\RequirePackage{fix-cm}
\documentclass[smallcondensed,draft,envcountsame]{svjour3}     %
\smartqed  %
\newcommand{\qedhere}{\qed}
\usepackage{graphicx}
\usepackage{mathptmx}      %
\PassOptionsToPackage{hyphens}{url}
\usepackage[T1]{fontenc}
\usepackage[misc]{ifsym}
\usepackage{amsmath}
\usepackage{amssymb}
\usepackage{bm}
\usepackage{booktabs}
\usepackage{bussproofs}
\usepackage{cite}
\usepackage{color}
\usepackage{enumerate}
\usepackage{stmaryrd}
\usepackage{url}
\usepackage{mathtools}
\usepackage{microtype}
\usepackage{array}
\usepackage{tabu}
\usepackage{prftree}
\usepackage{trimclip}
\usepackage{float}
\usepackage{bm}
\usepackage{tikz-cd}
\usepackage{tikz}
\usepackage{xr}

\usepackage[mathscr]{eucal}

\makeatletter
\g@addto@macro{\UrlBreaks}{\UrlOrds\do\=\do\_}
\makeatother

\usepackage{version}
\excludeversion{notyet}

\usepackage[
   a4paper,
   pdftitle={Superposition with Lambdas},
   pdfauthor={Alexander Bentkamp, Jasmin Blanchette, Sophie Tourret, Petar Vukmirovi\'c, and Uwe Waldmann},
   pdfkeywords={},
   pdfborder={0 0 0},
   draft=false,
   bookmarksnumbered,
   bookmarks,
   bookmarksdepth=2,
   bookmarksopenlevel=2,
   bookmarksopen]{hyperref}

\DeclareSymbolFont{letters}{OML}{txmi}{m}{it} %

\DeclareFontFamily{OT1}{pzc}{}
\DeclareFontShape{OT1}{pzc}{m}{it}{<-> s * [1.10] pzcmi7t}{}
\DeclareMathAlphabet{\mathcalx}{OT1}{pzc}{m}{it}

\def\negvthinspace{\kern-0.083333em}
\def\vthinspace{\kern+0.083333em}
\def\vvthinspace{\kern+0.0416667em}
\def\negvvthinspace{\kern-0.0416667em}
\def\hypsep{\hskip1.25em}

\newcommand\CORR{{(\vthinspace\Letter\vthinspace)}}

\spnewtheorem{lemmaxx}[theorem]{Lemma}{\bfseries}{\slshape}
\spnewtheorem{corollaryxx}[theorem]{Corollary}{\bfseries}{\slshape}
\spnewtheorem{theoremxx}[theorem]{Theorem}{\bfseries}{\slshape}
\spnewtheorem{conventionxx}[theorem]{Convention}{\bfseries}{\rmfamily}
\spnewtheorem{definitionxx}[theorem]{Definition}{\bfseries}{\rmfamily}
\spnewtheorem{notationxx}[theorem]{Notation}{\bfseries}{\rmfamily}
\spnewtheorem{examplexx}[theorem]{Example}{\bfseries}{\rmfamily}
\spnewtheorem{remarkxx}[theorem]{Remark}{\bfseries}{\rmfamily}

\newenvironment{lemmax}[1][]{\if\relax\detokenize{#1}\relax\begin{lemmaxx}\else\begin{lemmaxx}[#1]\,\fi\ignorespaces}{\end{lemmaxx}}

\newenvironment{theoremx}[1][]{\if\relax\detokenize{#1}\relax\begin{theoremxx}\else\begin{theoremxx}[#1]\,\fi\ignorespaces}{\end{theoremxx}}
\newenvironment{conventionx}[1][]{\if\relax\detokenize{#1}\relax\begin{conventionxx}\else\begin{conventionxx}[#1]\,\fi\ignorespaces}{\end{conventionxx}}
\newenvironment{definitionx}[1][]{\if\relax\detokenize{#1}\relax\begin{definitionxx}\else\begin{definitionxx}[#1]\,\fi\ignorespaces}{\end{definitionxx}}
\newenvironment{notationx}[1][]{\if\relax\detokenize{#1}\relax\begin{notationxx}\else\begin{notationxx}[#1]\,\fi\ignorespaces}{\end{notationxx}}
\newenvironment{examplex}[1][]{\if\relax\detokenize{#1}\relax\begin{examplexx}\else\begin{examplexx}[#1]\,\fi\ignorespaces}{\end{examplexx}}

\newcommand\ourparagraph[1]{\paragraph{\bfseries\upshape#1}}
\newcommand\oursubsection[1]{\bfseries\subsection{#1}\mdseries}

\newcommand\medrightarrow{\mathrel{{{\color{black}\relbar}\kern-0.9ex\rlap{\color{white}\ensuremath{\blacksquare}}\kern-0.9ex}\joinrel{\color{black}\rightarrow}}}
\newcommand\medleftarrow{\mathrel{{\color{black}\leftarrow}\kern-0.9ex\rlap{\color{white}\ensuremath{\blacksquare}}\kern-0.9ex\joinrel{{\color{black}\relbar}}}}
\newcommand\medleftrightarrow{\mathrel{\leftarrow\kern-1.685ex\rightarrow}}

\newcommand{\rewrite}{\medrightarrow}

\newcommand{\leftrightrewrite}{\medleftrightarrow}

\newcommand\Section{Sect.}

\definecolor{light-gray}{gray}{0.875}
\definecolor{darker-gray}{gray}{0.45}

\let\oldSigma=\Sigma
\renewcommand\Sigma{\mathrm{\oldSigma}}

\newcommand{\flooronly}{\mathcalx{F}}
\newcommand{\ceilonly}{\mathcalx{F}^{-1}}
\newcommand{\floor}[1]{\flooronly\!(#1)}
\newcommand{\ceil}[1]{\ceilonly\!(#1)}

\newcommand\eqIH{\overset{\smash{\scriptscriptstyle\text{IH}}}{=}}

\newcommand{\Sigmaty}{\Sigma_\mathsf{ty}}
\newcommand{\VV}{\mathscr{V}}
\newcommand{\Vty}{\VV_\mathsf{ty}}
\newcommand{\III}{\mathscr{I}}
\newcommand{\II}{\mathscr{J}}
\newcommand{\IIty}{\II_\mathsf{ty}}
\newcommand{\IIIty}{\III_\mathsf{ty}}
\newcommand{\DD}{\mathscr{D}}

\newcommand{\RfN}{R}

\newcommand{\UU}{\mathscr{U}}

\newcommand{\EE}{\mathscr{E}}

\newcommand{\IItyho}{\IIty^{\GH}}
\newcommand{\uho}{\UU^{\GH}}
\newcommand{\iho}{\II^{\GH}}

\newcommand{\dho}{\DD}
\newcommand{\IIIho}{\III^{\smash{\GH}}}

\newcommand{\ifo}{\II}

\newcommand{\ufo}{\UU}
\newcommand{\LL}{\mathscr{L}}
\newcommand{\lho}{\LL^{\GH}}

\newcommand{\infname}[1]{\textsc{#1}}

\newcommand{\leftsubterm}{[}
\newcommand{\rightsubterm}{]}
\newcommand{\subterm}[2]{#1\leftsubterm#2\rightsubterm}

\newcommand{\lang}{\begin{picture}(5,7)
\put(1.1,2.5){\rotatebox{45}{\line(1,0){6.0}}}
\put(1.1,2.5){\rotatebox{315}{\line(1,0){6.0}}}
\end{picture}}
\newcommand{\rang}{\begin{picture}(5,7)
\put(0,2.5){\rotatebox{135}{\line(1,0){6.0}}}
\put(0,2.5){\rotatebox{225}{\line(1,0){6.0}}}
\end{picture}}

\newcommand{\leftgreensubterm}{\lang\,}
\newcommand{\rightgreensubterm}{\rang}
\newcommand{\greensubterm}[2]{#1\leftgreensubterm #2\rightgreensubterm}

\newcommand{\leftyellowsubterm}{\lang\!\!\leftgreensubterm}
\newcommand{\rightyellowsubterm}{\rightgreensubterm\!\!\rang}
\newcommand{\yellowsubterm}[2]{#1\leftyellowsubterm #2\rightyellowsubterm}

\newcommand{\leftorangesubterm}{\leftyellowsubterm}
\newcommand{\rightorangesubterm}{\rightyellowsubterm}
\newcommand{\orangesubterm}[3]{#1\leftorangesubterm #2.\> #3\rightorangesubterm}
\newcommand{\orangesubtermeta}[3]{\orangesubterm{#1}{#2}{#3}_{\!\eta}}

\newcommand{\leftinterpret}{\llbracket}
\newcommand{\rightinterpret}{\rrbracket}
\newcommand{\interpret}[3]{\smash{\leftinterpret #1\rightinterpret_{#2}^{#3}}}

\newcommand{\interpretaxi}[1]{\interpret{#1}{\III}{\xi}}
\newcommand{\interpretfo}[2]{\interpret{#1}{\RfN}{#2}}
\newcommand{\interpretho}[2]{\interpret{#1}{\IIIho}{#2}}
\newcommand{\interpretfoxi}[1]{\interpretfo{#1}{\xi}}
\newcommand{\interprethoxi}[1]{\interpretho{#1}{\xi}}

\renewcommand{\doteq}{\mathrel{\dot\approx}}
\newcommand{\eq}{\approx}
\newcommand{\noteq}{\not\eq}

\newcommand{\eqR}[2]{#1\sim#2}

\newcommand{\namedinference}[3]{\prftree[r]{\relax{\infname{#1}}}{\strut#2}{\strut#3}}

\newcommand{\namedsimp}[3]{\prftree[d][r]{\relax{\infname{#1}}}{\strut#2}{\strut#3}}

\DeclareMathOperator{\csu}{CSU} %

\newcommand\UNIF{\mathrel{\smash{\stackrel{\lower.1ex\hbox{\ensuremath{\scriptscriptstyle ?}}}{=}}}}

\newcommand{\tuple}[1]{\bar{#1}}

\newcommand{\cst}[1]{{\mathsf{#1}}}
\newcommand{\var}[1]{{\mathit{#1}}}

\newcommand{\typ}[1]{{\mathit{#1}}}

\newcommand\defeq{=}

\newcommand\fun{\rightarrow}

\newcommand\foralltynospace[1]{\mathsf{\Pi}#1.}
\newcommand\forallty[1]{\foralltynospace{#1}\;}

\newcommand{\typeargs}[1]{{\langle#1\rangle\negvthinspace}}

\newcommand\oftype{:}
\newcommand\oftypedecl{:}

\newcommand{\diff}{\cst{diff}}

\newcommand{\db}{\cst{db}}

\newcommand{\llor}{\mathrel\lor}
\newcommand{\ccup}{\mathrel\cup}
\newcommand{\ccap}{\mathrel\cap}

\newcommand\betanf[1]{#1\vthinspace{\downarrow}_{\beta\eta}}

\newcommand{\TT}{\mathcalx{T}}
\newcommand{\Ty}{\mathcalx{Ty}}
\newcommand{\CC}{\mathcalx{C}}
\newcommand{\HH}{{\mathrm{H}}}
\newcommand{\GH}{{\mathrm{GH}}}
\newcommand{\GF}{{\mathrm{GF}}}
\newcommand{\THH}{\TT_\HH}
\newcommand{\TGH}{\TT_\GH}
\newcommand{\TGF}{\TT_\GF}
\newcommand{\TyHH}{\Ty_\HH}
\newcommand{\TyGH}{\Ty_\GH}
\newcommand{\TyGF}{\Ty_\GF}
\newcommand{\CHH}{\CC_\HH}
\newcommand{\CGH}{\CC_\GH}
\newcommand{\CGF}{\CC_\GF}

\newcommand{\Inf}{\mathit{Inf}}

\newcommand{\Gmodels}{\models_\gnd}
\newcommand{\HInf}{\mathit{HInf}}

\newcommand{\GFInf}{\mathit{GFInf}}
\newcommand{\GHInf}{\mathit{GHInf}}

\newcommand{\gnd}{{\mathcalx{G}}}
\newcommand{\concl}{\mathit{concl}}
\newcommand{\prem}{\mathit{prems}}
\newcommand{\mprem}{\mathit{mprem}}
\newcommand{\HSel}{\mathit{HSel}}
\newcommand{\GHSel}{\mathit{GHSel}}
\newcommand{\GFSel}{\mathit{GFSel}}
\newcommand{\RedI}{\mathit{Red}_{\mathrm{I}}}
\newcommand{\RedC}{\mathit{Red}_{\mathrm{C}}}
\newcommand{\HRedC}{{\mathit{HRed}}_{\mathrm{C}}}
\newcommand{\HRedI}{\mathit{HRed}_{\mathrm{I}}}
\newcommand{\GHRedC}{\mathit{GHRed}_{\mathrm{C}}}
\newcommand{\GHRedI}{\mathit{GHRed}_{\mathrm{I}}}
\newcommand{\GFRedC}{\mathit{GFRed}_{\mathrm{C}}}
\newcommand{\GFRedI}{\mathit{GFRed}_{\mathrm{I}}}
\newcommand{\wrt}{\hbox{w.r.t.}}

\newcommand{\soundmodels}{\mathrel{|\kern-.1ex}\joinrel\approx}
\newcommand{\issubsumedtilde}{\mathrel{\raisebox{+0.8pt}{\Large\rlap{\kern0.5pt$\cdot$}}}\gtrsim}

\newcommand\encOonly{\mathcalx{\negvvthinspace O}}
\newcommand{\encO}[1]{\encOonly(#1)}

\newcommand\encOsubonly[1]{\encOonly_{\vvthinspace\let\>=\vvthinspace#1}}
\newcommand{\encOsub}[2]{\encOsubonly{#1}(#2)}

\newcommand\encBonly{\mathcalx{B}}
\newcommand{\encB}[2]{\encBonly_{#1}(#2)}

\newcommand\fosubscript{{\mathsf{fo}}}
\newcommand\lsubscript{{\lambda}}

\newcommand\fosucc{\succ_{\negvthinspace\fosubscript}}
\newcommand\lsucc{\succ_{\negvthinspace\lsubscript}}
\newcommand\fosucceq{\succeq_{\negvvthinspace\fosubscript}}
\newcommand\lsucceq{\succeq_{\lsubscript}}
\newcommand\lsuccsim{\succsim_{\vvthinspace\lsubscript}}
\newcommand\lfprec{\prec_\fosubscript}
\newcommand\lprec{\prec_\lsubscript}

\newcommand\zof[1]{z_{\vvthinspace\let\>=\vvthinspace#1}}

\newcommand{\OK}[1]{#1}

\journalname{J.\ Autom.\ Reasoning}
\begin{document}

\title{Superposition with Lambdas%
}
\subtitle{}

\author{Alexander Bentkamp
  \and
  Jasmin Blanchette
  \and
  Sophie~Tourret
  \and
  Petar~Vukmirovi\'c
  \and
  Uwe~Waldmann
  }

\institute{%
Alexander Bentkamp \CORR
\and Jasmin Blanchette
\and Petar~Vukmirovi\'c\at
  Vrije Universiteit Amsterdam, Department of Computer Science, Section of Theoretical Computer Science,
  De Boelelaan 1111, 1081 HV Amsterdam, the Netherlands\\
  \email{\{a.bentkamp,j.c.blanchette,p.vukmirovic\}@vu.nl}
\and
Jasmin Blanchette \and Sophie~Tourret \and Uwe~Waldmann \at%
  Max-Planck-Institut f\"ur Informatik, Saarland Informatics Campus E1 4, 66123 Saarbr\"ucken, Germany
  \\
  \email{\{jblanche,stourret,uwe\}@mpi-inf.mpg.de}%
}

\date{Received: date / Accepted: date}

\maketitle

\begin{abstract}
We designed a superposition calculus for a clausal fragment of extensional
polymorphic higher-order logic that includes anonymous functions but excludes
Booleans. The inference rules work on $\beta\eta$-equivalence classes of
$\lambda$-terms and rely on higher-order unification to achieve refutational
completeness. We implemented the calculus in the Zipperposition prover and
evaluated it on TPTP and Isabelle benchmarks. The results suggest that
superposition is a suitable basis for higher-order reasoning.
\end{abstract}

\section{Introduction}
\label{sec:introduction}

Superposition \cite{bachmair-ganzinger-1994} is widely regarded as the
calculus par excellence for reasoning about first-order logic with equality.
To increase automation in proof assistants and other verification tools based
on higher-order formalisms, we propose to generalize superposition to an
extensional, polymorphic, clausal version of higher-order logic (also called
simple type theory). Our ambition is to achieve a \emph{graceful}
extension, which coincides with standard superposition on first-order problems
and smoothly scales up to arbitrary higher-order problems.

Bentkamp, Blanchette, Cruanes, and Waldmann~\cite{bentkamp-et-al-2018} designed a family of
superposition-like calculi for a $\lambda$-free clausal fragment of higher-order
logic, with currying and applied variables. We adapt their
extensional nonpurifying calculus to support \hbox{$\lambda$-terms}
(\Section~\ref{sec:the-calculus}).
Our calculus does not support interpreted Booleans; it is conceived as the
penultimate milestone towards a superposition calculus for full higher-order
logic. If desired, Booleans can be encoded in our logic fragment using an
uninterpreted type and uninterpreted ``proxy'' symbols corresponding to
equality, the connectives, and the quantifiers.

Designing a higher-order superposition calculus poses three main challenges:

\begin{enumerate}
\item Standard superposition is parameterized by a ground-total
  simplification order~$\succ$, but such orders do not exist for
  $\lambda$-terms equal up to $\beta$-conversion. The relations
  designed for proving termination of higher-order term rewriting systems, such
  as HORPO \cite{jouannaud-rubio-2007} and CPO \cite{blanqui-et-al-2015}, lack
  many of the desired properties (e.g., transitivity, stability under grounding
  substitutions).

\medskip

\item Higher-order unification is undecidable and may give rise to an infinite set of
  incomparable unifiers. For example, the constraint
  $\cst{f} \> (y \> \cst{a}) \UNIF y \> (\cst{f} \> \cst{a})$ admits
  infinitely many independent solutions of the form $\{ y \mapsto \lambda x.\;
  \cst{f}^n \, x \}.$

\pagebreak[2]
\medskip

\item In first-order logic, to rewrite into a term~$s$ using an
  oriented equation $t \eq t'$, it suffices to find a subterm of $s$ that is
  unifiable with $t$. In higher-order logic, this is insufficient. Consider
  superposition from $\cst{f} \> \cst{c} \eq \cst{a}$ into $y \> \cst{c}
  \noteq y \> \cst{b}$. The left-hand sides can obviously be unified by $\{y
  \mapsto \cst{f}\}$, but the more general $\{y \mapsto \lambda
  x.\> z\> x\> (\cst{f}\> x)\}$ also gives rise to a subterm $\cst{f} \>
  \cst{c}$ after $\beta$-reduction. The corresponding inference generates the
  clause $z\> \cst{c}\> \cst{a} \noteq z\> \cst{b} \>(\cst{f}\> \cst{b})$.
\end{enumerate}

To address the first challenge, we adopt the $\eta$-short $\beta$-normal form
to represent $\beta\eta$-equivalence classes of $\lambda$-terms.
In the spirit of Jouannaud and Rubio's early joint work
\cite{jouannaud-rubio-1998}, we state requirements on the term
order only for ground terms (i.e., closed monomorphic $\beta\eta$-equivalence
classes); the nonground case is connected to the ground case via stability under grounding
substitutions. Even on ground terms, we cannot obtain all desirable
properties. We sacrifice compatibility with arguments (the property that $s'
\succ s$ implies $s'\>t \succ s\>t$), compensating with an
\emph{argument congruence} rule (\infname{ArgCong}), as in
Bentkamp et al.\ \cite{bentkamp-et-al-2018}.

For the second challenge, we accept that there might be infinitely many
incomparable unifiers and enumerate a complete set (including the notorious
flex--flex pairs \cite{huet-1975}), relying on heuristics to postpone the
combinatorial explosion. The saturation loop must also be
adapted to interleave this enumeration with the theorem
prover's other activities (\Section~\ref{sec:implementation}).
Despite its reputation for explosiveness, higher-order unification is a
conceptual improvement over $\cst{SK}$ combinators, because it can often
\emph{compute} the right unifier.
Consider the conjecture $\exists z. \>
\forall x\> y.\> z\> x\> y \eq \cst{f}\> y\> x$. After negation,
clausification,
and skolemization (which are as for first-order logic),
the formula becomes
$z\> (\cst{sk}_\cst{x}\> z) \> (\cst{sk}_\cst{y}\> z) \noteq
\cst{f}\> (\cst{sk}_\cst{y}\> z) \> (\cst{sk}_\cst{x}\> z)$. Higher-order unification quickly
computes the unique unifier: $\{z \mapsto \lambda x\> y.\> \cst{f}\>y\>x\}$.
In contrast, an
encoding approach based on combinators, similar to the one implemented in
Sledgehammer \cite{meng-paulson-2008-trans}, would blindly enumerate all
possible $\cst{SK}$ terms for $z$ until the right one, $\cst{S}\>
(\cst{K}\> (\cst{S}\> \cst{f}))\> \cst{K}$, is found. Given the definitions
$\cst{S}\> z\> y\> x \eq z\> x\> (y\> x)$ and $\cst{K}\> x\> y \eq x$, the E
prover \cite{schulz-et-al-2019} in \emph{auto} mode needs to perform 3757 inferences
to derive the empty clause.

For the third challenge, the idea is that, when applying $t \eq t'$ to perform
rewriting inside a higher-order term $s$, we can encode an arbitrary context as
a fresh higher-order variable $z$, unifying $s$ with $z\>t$; the result is
$(z\>t')\sigma$, for some unifier~$\sigma$. This is performed by a dedicated
\emph{fluid subterm superposition} rule (\infname{FluidSup}).

Functional extensionality %
is also considered a quintessential higher-order challenge
\cite{benzmueller-kohlhase-1998}, although similar difficulties arise with
first-order sets and arrays \cite{gupta-et-al-2014}. Our approach
is to add extensionality as an axiom and provide optional rules as
optimizations (\Section~\ref{sec:extensions}). With this axiom, our
calculus is refutationally complete \wrt\ extensional Henkin
semantics (\Section~\ref{sec:refutational-completeness}).
Our proof employs the
new saturation framework by Waldmann et
al.~\cite{waldmann-et-al-2020-saturation} to derive
dynamic completeness of a given clause prover from ground static completeness.

We implemented the calculus in the Zipperposition prover \cite{cruanes-2017}
(\Section~\ref{sec:implementation}).
Our empirical evaluation includes benchmarks from
the TPTP~\cite{sutcliffe-2017-tptp} and interactive verification
problems exported from Isabelle/HOL \cite{boehme-nipkow-2010}
(\Section~\ref{sec:evaluation}). The results clearly demonstrate the
calculus's potential. 
The 2020 edition of
the CADE ATP System Competition (CASC) provides further confirmation:
Zipperposition finished 20~percentage points ahead of its closest rival.
This suggests that an implementation inside a
high-performance prover such as E \cite{schulz-et-al-2019} or
Vampire \cite{kovacs-voronkov-2013} could fulfill the promise of
strong proof automation for higher-order logic
(\Section~\ref{sec:discussion-and-related-work}).

An earlier version of this article was presented at CADE-27
\cite{bentkamp-et-al-2019-lamsup}. This article extends the conference paper
with more explanations, detailed soundness and completeness proofs, including
dynamic completeness, and new optional inference rules. We have also updated
the empirical evaluation and extended the coverage of related work.
Finally, we tightened side condition~4 of \infname{FluidSup}, making the rule
slightly less explosive.

\section{Logic}
\label{sec:logic}

Our \relax{extensional polymorphic clausal higher-order logic} is a
restriction of full TPTP THF \cite{benzmueller-paulson-2010} to rank-1
(top-level) polymorphism, as in TH1 \cite{kaliszyk-et-al-2016}.
In keeping with standard superposition, we consider only formulas in
conjunctive
normal form, without explicit quantifiers or Boolean type.
We use Henkin semantics
\cite{henkin-1950,benzmueller-miller-2014,fitting-2002}, as opposed
to the standard semantics 
that is commonly considered the foundation of the HOL systems \cite{gordon-melham-1993}.   
However, both of these semantics are compatible with the notion of provability employed by the HOL systems.
By admitting nonstandard models, Henkin
semantics is not subject to G\"odel's first incompleteness theorem, allowing us to
claim not only soundness but also refutational completeness of our calculus.

\ourparagraph{Syntax}

We fix a set $\Sigmaty$ of type constructors with arities and a set $\Vty$ of type variables.
We require at least one nullary type constructor and
a binary function type constructor ${\fun}$ to be present in $\Sigmaty$.
A \relax{type}~$\tau,\upsilon$ is either a type variable
$\alpha\in\Vty$ or has the form $\kappa(\tuple{\tau}_n)$
for an $n$-ary type constructor $\kappa\in\Sigmaty$ and types $\tuple{\tau}_n$.
We use the notation $\tuple{a}_n$ or $\tuple{a}$ to stand for the
tuple $(a_1,\dots,a_n)$ or product $a_1 \times \dots \times a_n$, where $n \ge 0$.
We write $\kappa$ for $\kappa()$ and $\tau\fun\upsilon$ for ${\fun}(\tau,\upsilon)$.
\relax{Type declarations} have
the form $\forallty{\tuple{\alpha}_m}\tau$ (or simply $\tau$
if $m = 0$), where all type variables occurring in $\tau$ belong to
$\tuple{\alpha}_m$.

We fix a set $\Sigma$ of (function) symbols $\cst{a}, \cst{b},
\cst{c}, \cst{f}, \cst{g}, \cst{h}, \dots$, with type declarations, written as
$\cst{f}\oftypedecl\forallty{\tuple{\alpha}_m}\tau$
or~$\cst{f}$, and a set $\VV$ of term variables with associated types, written as
$\var{x}\oftype\tau$ or~$\var{x}$. The notation $t \oftype\tau$ will also be used
to indicate the type of arbitrary terms $t$.
We require the presence of
a symbol %
of type $\forallty\alpha \alpha$
and of a symbol
$\diff\oftype\forallty{\alpha,\beta}(\alpha\fun\beta)\fun(\alpha\fun\beta)\fun{\alpha}$ in $\Sigma$.
We use $\diff$ to express the polymorphic functional extensionality axiom.
A \relax{signature} is a pair $(\Sigmaty,\Sigma)$.

In the following, we will define terms in three layers of abstraction: raw $\lambda$-terms, $\lambda$-terms, and terms;
where $\lambda$-terms will be $\alpha$-equivalence classes of raw $\lambda$-terms
and terms will be $\beta\eta$-equivalence classes of $\lambda$-terms.

The \emph{raw $\lambda$-terms} over a given signature and their associated types are
defined inductively as follows. Every $x \mathbin: \tau \in\VV$ is a raw $\lambda$-term of type $\tau$.
If $\cst{f}\oftypedecl\forallty{\tuple{\alpha}_m}\tau \in \Sigma$ and
$\tuple{\upsilon}_m$ is a tuple of types, called \emph{type arguments}, then
$\cst{f}\typeargs{\tuple{\upsilon}_m}$ (or %
$\cst{f}$ if $m = 0$) is a
raw $\lambda$-term of type $\tau\{\tuple{\alpha}_m \mapsto \tuple{\upsilon}_m\}$.
If $x\mathbin\oftype\tau$ and $t\oftype\upsilon$, then the \emph{$\lambda$-expression}
$\lambda x.\> t$ is a raw $\lambda$-term of type $\tau\fun\upsilon$.
If $s\oftype\tau\fun\upsilon$ and $t\oftype\tau$, then the
\emph{application} $s\>t$ is a raw $\lambda$-term of type $\upsilon$.

The function type constructor $\fun$ is right-associative;
application is left-associative.
Using the spine notation \cite{cervesato-pfenning-2003}, raw $\lambda$-terms can
be decomposed in a unique way as a nonapplication \emph{head} $t$ applied to
zero or more arguments: $t \> s_1\dots s_n$ or $t \> \tuple{s}_n$ (abusing
notation).

A raw $\lambda$-term $s$ is a \emph{subterm} of a raw $\lambda$-term $t$,
written $t = \subterm{t}{s}$,
if $t = s$,
if $t = (\lambda x.\>\subterm{u}{s})$,
if $t = (\subterm{u}{s})\>v$,
or if $t = u\>(\subterm{v}{s})$ for some raw $\lambda$-terms $u$ and $v$.
A \emph{proper} subterm of a  raw $\lambda$-term $t$ is any subterm of $t$
that is distinct from $t$ itself.

A variable occurrence is \emph{free} in a raw $\lambda$-term if it is not bound by
a \hbox{$\lambda$-expression}. A raw $\lambda$-term is \emph{ground}
if it is built without using type variables and contains no free term
variables.

The $\alpha$-renaming rule is defined as
$(\lambda x.\> t) \vthinspace\rewrite_\alpha\vthinspace (\lambda y.\> t\{x \mapsto y\})$,
where $y$ does not occur free in $t$ and is not captured by a $\lambda$-binder
in $t$. Raw $\lambda$-terms form equivalence classes modulo $\alpha$-renaming,
called \emph{$\lambda$-terms}. 
We lift the above notions on raw $\lambda$-terms to $\lambda$-terms.

A substitution $\rho$ is a function from type variables to types and from term
variables to $\lambda$-terms such that it maps all but finitely many variables to themselves.
We require that it is type-correct---i.e., for each $x\oftype\tau \in \VV$, 
$x\rho$ is of type $\tau\rho$.
The letters $\theta,\pi,\rho,\sigma$ are reserved for substitutions.
Substitutions $\alpha$-rename $\lambda$-terms to avoid capture; for
example, $(\lambda x.\> y)\{y \mapsto x\} = (\lambda x'\!.\> x)$.
The composition $\rho\sigma$ applies $\rho$ first: $t\rho\sigma = (t\rho)\sigma$.
The notation $\sigma[\tuple{x}_n \mapsto \tuple{s}_n]$ denotes the
substitution that replaces each $x_i$ by $s_i$ and that otherwise coincides
with $\sigma$.

The $\beta$- and $\eta$-reduction rules are specified on
$\lambda$-terms as
$(\lambda x.\> t)\> u \vthinspace\rewrite_\beta\vthinspace t\{x \mapsto u\}$ and
$(\lambda x.\> t\> x) \vthinspace\rewrite_\eta\vthinspace t$.
For $\beta$, bound variables in $t$ are implicitly renamed to avoid
capture; for $\eta$, the variable $x$ must not occur free in $t$.
The $\lambda$-terms form equivalence classes modulo $\beta\eta$-reduction,
called \emph{$\beta\eta$-equivalence classes} or simply \emph{terms}.

\begin{conventionx} \label{conv:beta-eta-normal-form}
When defining operations that need to analyze the structure of terms, we will
use the $\eta$-short $\beta$-normal form $\betanf{t}$, obtained by applying
$\rewrite_\beta$ and $\rewrite_\eta$ exhaustively, as a representative of
the equivalence class $t$. In particular, we lift the notions of subterms and
occurrences of variables to $\beta\eta$-equivalence classes via their
$\eta$-short $\beta$-normal representative.
\end{conventionx}

Many authors prefer the $\eta$-long $\beta$-normal form
\cite{huet-1975,mayr-nipkow-1998,jouannaud-rubio-1998}, but in a polymorphic
setting it has the\pagebreak[2] drawback that instantiating a type variable with a
functional type can lead to $\eta$-expansion. We reserve the letters
$s, t, u, v$ for terms and $x, y, z$ for variables.

An equation $s \eq t$ is formally an unordered pair of terms $s$ and $t$. A
literal is an equation or a negated equation, written $\lnot\; s \eq t$ or $s
\noteq t$. A clause $L_1 \lor \dots \lor L_n$ is a finite multiset of literals
$L_{\!j}$. The empty clause is written as $\bot$.

A \emph{complete set of unifiers} on a set~$X$ of variables for
two terms $s$~and~$t$ is a
set~$U$ of unifiers of $s$ and $t$ such that for every unifier $\theta$ of
$s$~and~$t$ there exists a member $\sigma \in U$ and a substitution~$\rho$ such that
$x\sigma\rho = x\theta$ for all $x \in X.$
We let $\csu_X(s,t)$ denote an arbitrary (preferably minimal)
complete set of unifiers on~$X$ for $s$~and~$t$.
We assume that all $\sigma \in \csu_X(s,t)$ are idempotent on $X$---i.e.,
$x\sigma\sigma = x\sigma$ for all $x \in X.$
The set~$X$ will
consist of the free variables of the clauses in which $s$ and $t$ occur and
will be left implicit.

Given a substitution $\sigma$, 
the $\sigma$-instance of a term $t$ or clause $C$ is the term $t\sigma$ or the clause $C\sigma$,
respectively.
If $t\sigma$ or $C\sigma$ is ground, we call it a $\sigma$-ground instance.

\ourparagraph{Semantics}

A \emph{type interpretation} $\IIIty = (\UU, \IIty)$ is defined as follows.
The \emph{universe} $\UU$ is a nonempty collection of nonempty sets, called
\emph{domains}.
The function $\IIty$ associates a function
$\IIty(\kappa) : \UU^n \rightarrow \UU$
with each $n$-ary type constructor~$\kappa$,
such that for all domains $\DD_1,\DD_2\in\UU$, the set $\IIty(\fun)(\DD_1,\DD_2)$
is a subset of the function space from $\DD_1$ to $\DD_2$.
The semantics is \emph{standard} if
$\IIty(\fun)(\DD_1,\DD_2)$ is the entire function space for all $\DD_1,\DD_2$.

A \emph{type valuation} $\xi$ is a function that maps every type variable to a domain.
The \emph{denotation} of a type for a type interpretation $\IIIty$
and a type valuation $\xi$ is defined by
$\interpret{\alpha}{\IIIty}{\xi}=\xi(\alpha)$ and
$\interpret{\kappa(\tuple{\tau})}{\IIIty}{\xi}=
\IIty(\kappa)(\interpret{\tuple{\tau}}{\IIIty}{\xi})$.
We abuse notation by applying an operation on a tuple when it must be applied elementwise;
thus, $\interpret{\tuple{\tau}_n}{\IIIty}{\xi}$ stands for $\interpret{\tau_1}{\IIIty}{\xi},\dots, \interpret{\tau_n}{\IIIty}{\xi}$.
A type valuation $\xi$ can be extended to be a \emph{valuation} by additionally
assigning an element $\xi(x)\in\interpret{\tau}{\IIIty}{\xi}$ to each variable $x \oftype \tau$.
An \emph{interpretation function} $\II$ for a type interpretation $\IIIty$ associates with each symbol
$\cst{f}\oftypedecl\forallty{\tuple{\alpha}_m}\tau$ and domain tuple $\tuple{\DD}_m\in\UU^m$
a value
$\II(\cst{f},\tuple{\DD}_m) \in
\interpret{\tau}{\IIIty}{\xi}$,
where $\xi$ is the type valuation that maps each $\alpha_i$ to $\DD_i$.

The comprehension principle states that every function designated by
a $\lambda$-expression is contained in the corresponding domain.
Loosely following Fitting~\cite[\Section~2.4]{fitting-2002}, we initially allow
$\lambda$-expressions to designate arbitrary elements of the domain, to be
able to define the denotation of a term. We impose restrictions afterwards
using the notion of a proper interpretation.
A \emph{$\lambda$-designation function} $\LL$
for a type interpretation $\IIIty$ is a function that maps
a valuation $\xi$ and a $\lambda$-expression of type $\tau$ to elements
of $\interpret{\tau}{\IIIty}{\xi}$.
A type interpretation, an interpretation function, and a $\lambda$-designation function form an
(\emph{extensional}) \emph{interpretation} $\III = (\IIIty,\II,\LL)$.
For an interpretation~$\III$ and a valuation~$\xi$, the \relax{denotation of a term} is defined
as
$\interpretaxi{x} \defeq \xi(x)$,
$\interpretaxi{\cst{f}\typeargs{\tuple{\tau}_m}} \defeq
  \II(\cst{f},\interpret{\tuple{\tau}_m}{\IIIty}{\xi})$,
$\interpretaxi{s\>t} \defeq \interpretaxi{s} (\interpretaxi{t})$, and
$\interpretaxi{\lambda x.\> t} \defeq \LL(\xi,\lambda x.\> t)$.
For ground terms $t$, the denotation does not depend on the choice of the valuation $\xi$,
which is why we sometimes write $\interpret{t}{\III}{}$ for $\interpretaxi{t}$.

An interpretation $\III$ is \emph{proper} if $\interpret{\lambda
x.\>t}{\III}{\xi}(a) = \interpret{t}{\III}{\xi[x\mapsto a]}$ for all
$\lambda$-expressions $\lambda x.\>t$, all valuations $\xi$, and all $a$. If a type interpretation
$\IIIty$ and an interpretation function $\II$ can be extended by a $\lambda$-designation function $\LL$ to a proper
interpretation $(\IIIty,\II,\LL)$, then this $\LL$ is unique
\cite[Proposition~2.18]{fitting-2002}.
Given an interpretation $\III$ and a valuation $\xi$, an equation $s\eq t$ is
\relax{true} if %
$\interpretaxi{s}$ and $\interpretaxi{t}$ are equal
and it is \relax{false} otherwise.
A disequation $s\noteq t$ is true if $s \eq t$ is false.
A clause is \relax{true} if at least one of its literals is true.
A clause set is \relax{true} if all its clauses are true.
A proper interpretation $\III$ is a \emph{model} of a clause set $N$,
written $\III \models N$, if $N$ is true in $\III$ for all valuations $\xi$.

\ourparagraph{Axiomatization of Booleans}

\looseness=-1
Our clausal logic lacks a Boolean type, but it can easily be
axiomatized as follows. We extend the signature with a nullary type
constructor $\typ{bool} \in \Sigmaty$ equipped with the proxy constants
$\cst{t}, \cst{f} : \typ{bool}$, $\cst{not} : \typ{bool} \fun \typ{bool}$,
$\cst{and}, \cst{or}, \cst{impl}, \cst{equiv} : \typ{bool} \fun \typ{bool} \fun \typ{bool}$,
$\cst{forall}, \cst{exists} : \forallty{\alpha} (\alpha \fun \typ{bool}) \fun \typ{bool}$,
$\cst{eq} : \forallty{\alpha} \alpha \fun \alpha \fun \typ{bool}$,
and $\cst{choice} : \forallty{\alpha} (\alpha \fun \typ{bool}) \fun \alpha$,
characterized by the axioms

\kern\abovedisplayskip

\noindent
\begin{minipage}[t]{.2\textwidth}
  \begin{center}
  $\cst{t} \noteq \cst{f}$ \\
  $x \eq \cst{t} \llor x \eq \cst{f}$ \\
  $\cst{not} \> \cst{t} \eq \cst{f}$ \\
  $\cst{not} \> \cst{f} \eq \cst{t}$ \\
  $\cst{and} \> \cst{t} \> x \eq x $ \\
  $\cst{and} \> \cst{f} \> x \eq \cst{f} $ \\
  \end{center}
  \end{minipage}%
  \begin{minipage}[t]{.3\textwidth}
  \begin{center}
  $\cst{or} \> \cst{t} \> x \eq \cst{t} $ \\
  $\cst{or} \> \cst{f} \> x \eq x $ \\
  $\cst{impl} \> \cst{t} \> x \eq x $ \\
  $\cst{impl} \> \cst{f} \> x \eq \cst{t} $ \\
  $x \noteq y \llor \cst{eq}\typeargs{\alpha}\;x\> y \eq \cst{t}$ \\
  $x \eq y \llor \cst{eq}\typeargs{\alpha}\;x\> y \eq \cst{f}$
  \end{center}
  \end{minipage}%
  \begin{minipage}[t]{.5\textwidth}
    \begin{center}
      $\cst{equiv} \> x \> y \eq \cst{and} \> (\cst{impl} \> x \> y) \> (\cst{impl} \> y \> x) $ \\
      $\cst{forall}\typeargs{\alpha}\; (\lambda x.\; \cst{t}) \eq \cst{t}$ \\
      $y \eq (\lambda x.\; \cst{t}) \llor \cst{forall}\typeargs{\alpha}\; y \eq \cst{f}$ \\
      $\cst{exists}\typeargs{\alpha}\; y \eq \cst{not} \> (\cst{forall}\typeargs{\alpha}\; (\lambda x. \> \cst{not} \> (y \> x)))$ \\
      $y \> x \eq \cst{f} \llor y \> (\textsf{choice}\typeargs{\alpha} \> y) \eq \cst{t}$ \\

    \end{center}
\end{minipage}

\kern\belowdisplayskip

This axiomatization of Booleans can be used in a prover to support full
higher-order logic with or without Hilbert choice, corresponding to the TPTP
THF format variants TH0 (monomorphic) \cite{sutcliffe-et-al-2009} and TH1
(polymorphic) \cite{kaliszyk-et-al-2016}. The prover's clausifier would
transform the outer first-order skeleton of a formula into a clause and use
the axiomatized Booleans within the terms. It would also add the proxy
axioms to the clausal problem.
As an alternative to this complete axiomatization,
Vukmirovi\'c and Nummelin \cite{vukmirovic-nummelin-2020-boolean} 
present a possibly refutationally incomplete calculus extension with dedicated
rules to support Booleans. This approach works better in practice and
contributed to Zipperposition's victory at CASC 2020.

\section{The Calculus}
\label{sec:the-calculus}

Our \emph{Boolean-free $\lambda$-superposition calculus} presented here is inspired by the
\relax{extensional nonpurifying} Boolean-free \hbox{$\lambda$-free} higher-order
superposition calculus described 
by Bentkamp et al.\ \cite{bentkamp-et-al-2018}.
The text of this and the next section is partly based on that paper
and the associated journal submission
\cite{bentkamp-et-al-lfhosup-arxiv}
(with Cruanes's~permission).
The central idea is that superposition inferences are restricted to
\emph{unapplied} subterms occurring in the first-order outer skeleton of
clauses---that is, outside $\lambda$-expressions and
outside the arguments of applied variables. We call these ``green subterms.''
Thus, $\cst{g} \eq (\lambda x.\> \cst{f}\>x\>x)$ cannot be used directly to
rewrite $\cst{g}\> \cst{a}$ to $\cst{f}\> \cst{a}\> \cst{a}$, because
$\cst{g}$ is applied in $\cst{g}\> \cst{a}$. A separate inference rule,
\infname{ArgCong}, takes care of deriving $\cst{g}\>x \eq \cst{f}\>x\>x$,
which can be oriented independently of its parent clause and used to rewrite
$\cst{g}\> \cst{a}$ or $\cst{f}\> \cst{a}\> \cst{a}$.
\begin{definitionx}[Green positions and subterms]
  The \emph{green positions} and \emph{green subterms} of a term (i.e., a
  $\beta\eta$-equivalence class) are defined inductively as follows.
  A green position is a tuple of natural numbers.
  For any term $t$, the empty tuple $\varepsilon$ is a green position of $t$,
  and $t$ is the green subterm of $t$ at position $\varepsilon$.
  For all symbols $\cst{f}\in\Sigma$, types $\tuple{\tau}$, and terms $\tuple{u}$,
  if $t$ is a green subterm of $u_i$ at some position $p$ for some~$i$, then $i.p$ is a 
  green position of $\cst{f}\typeargs{\tuple{\tau}}\vthinspace\> \tuple{u}$, and 
  $t$ is the green subterm of $\cst{f}\typeargs{\tuple{\tau}}\>\tuple{u}$ at position~$i.p$.
  We denote the green subterm of $s$ at the green position $p$ by $s|_p$.
\end{definitionx}
In $\cst{f}\> (\cst{g}\> \cst{a})\> (y\> \cst{b})\>
(\lambda x.\> \cst{h}\> \cst{c}\> (\cst{g}\> x))$,
the proper green subterms are $\cst{a}$, $\cst{g}\> \cst{a}$, $y\> \cst{b}$,
and $\lambda x.\> \cst{h}\> \cst{c}\> (\cst{g}\> x)$. The last two of these
do not look like first-order terms and hence their subterms are not green.
\begin{definitionx}[Green contexts]
  We write $t = \greensubterm{s}{u}_p$ to express that $u$ is a green subterm of~$t$ at the green position $p$ and call
  $\greensubterm{s}{\phantom{.}}_p$ a \emph{green context}.
  We omit the subscript $p$ if there are no ambiguities.
\end{definitionx}
In a $\beta\eta$-normal representative of a green context, the hole never occurs applied.
Therefore, inserting a $\beta\eta$-normal term into the context
produces another $\beta\eta$-normal term.

Another key notion is that of a fluid term:
\begin{definitionx}[Fluid terms]
A term $t$ is called \emph{fluid} if (1)~$\betanf{t}$ is of the form
$y\>\tuple{u}_n$ where $n \geq 1$, or (2)~$\betanf{t}$
is a $\lambda$-expression and there exists a substitution~$\sigma$ such that
$\betanf{t\sigma}$ is not a $\lambda$-expression (due to $\eta$-reduction).
\end{definitionx}
Case~(2) can arise only if $t$ contains an applied variable. 
Intuitively, fluid terms are terms whose
$\eta$-short $\beta$-normal form can change radically as a result of
instantiation. For example, 
$\lambda x.\> y\> \cst{a}\> (z\> x)$ is fluid because applying $\{z \mapsto \lambda
x.\>x\}$ makes the
$\lambda$ vanish:
$(\lambda x.\> y\> \cst{a}\> x) = y\> \cst{a}$.
Similarly, $\lambda x.\> \cst{f}\>(y\>x)\>x$ is fluid because 
$(\lambda x.\> \cst{f}\>(y\>x)\>x)\{y \mapsto \lambda
x.\>\cst{a}\} = (\lambda x.\> \cst{f}\>\cst{a}\>x) = \cst{f}\>\cst{a}$.

\oursubsection{The Core Inference Rules}
\label{ssec:the-core-inference-rules}

The calculus is parameterized by a strict and a nonstrict term order as well as
a selection function. These concepts are defined below.

\begin{definitionx}[Strict ground term order]
A \emph{strict ground term order} is a well-founded strict total order $\succ$ on
ground terms satisfying the following criteria, where $\succeq$ denotes the
reflexive closure of
$\succ$:
\begin{itemize}
\item \emph{green subterm property}:\enskip $\greensubterm{t}{s} \succeq s$;
\item \emph{compatibility with green contexts}:\enskip $s' \succ s$ implies
  $\greensubterm{t}{s'} \succ \greensubterm{t}{s}$.
\end{itemize}
Given a strict ground term order, we extend it to literals and clauses via the multiset extensions in the
standard way \cite[\Section~2.4]{bachmair-ganzinger-1994}. 
\label{def:strict-ground-term-order}
\end{definitionx}
Two properties that are not required are \emph{compatibility with
$\lambda$-expressions}
($s'\succ s$ implies $(\lambda x. \> s') \succ (\lambda
x.\> s)$) and \emph{compatibility with arguments} ($s' \succ s$ implies
$s'\>\relax{t} \succ s\vthinspace\>\relax{t}$). The latter would even be inconsistent
with totality. To see why, consider\pagebreak[2] the symbols $\cst{c} \succ
\cst{b} \succ \cst{a}$ and the terms $\lambda x.\> \cst{b}$ and $\lambda x.\>
x$. Owing to totality, one of the terms must be larger than the other, say,
$(\lambda x.\> \cst{b}) \succ (\lambda x.\> x)$. By compatibility
with arguments, we get $(\lambda x.\> \cst{b})\> \cst{c} \succ
(\lambda x.\>x)\> \cst{c}$, i.e., $\cst{b} \succ \cst{c}$,
a contradiction. A similar line of reasoning applies if $(\lambda x.\>
\cst{b}) \prec (\lambda x.\> x)$, using $\cst{a}$ instead of~$\cst{c}$.

\begin{definitionx}[Strict term order]
  A \emph{strict term order} is a relation $\succ$ on terms, literals, and clauses such that
  restriction to ground entities is a strict ground term order %
  and such that it is
  stable under grounding substitutions (i.e., $t \succ s$ implies $t\theta
  \succ s\theta$ for all substitutions~$\theta$ grounding the entities $t$ and $s$).
  \label{def:strict-term-order}
\end{definitionx}

\begin{definitionx}[Nonstrict term order]
  Given a strict term order $\succ$ and its reflexive closure $\succeq$, a \emph{nonstrict term order} is a relation $\succsim$ on terms, literals, and clauses such
  that $t \succsim s$ implies $t\theta \succeq s\theta$ for all $\theta$ grounding the entities $t$ and $s$.
  \label{def:nonstrict-term-order}
\end{definitionx}
Although we call them orders, a strict term order $\succ$ is not required to be transitive on nonground entities,
and a nonstrict term order $\succsim$ does not need to be transitive at all.
Normally, $t \succeq s$ should imply $t \succsim s$, but this is not
required either. A nonstrict term order~$\succsim$ allows us to be more precise than the
reflexive closure $\succeq$ of %
$\succ$. For example,
we cannot have $y\>\cst{b} \succeq y\>\cst{a}$, because
$y\>\cst{b} \not= y\>\cst{a}$ and $y\>\cst{b} \not\succ y\>\cst{a}$
by stability under grounding substitutions (with $\{y \mapsto \lambda x.\>\cst{c}\}$).
But we can have $y\>\cst{b} \succsim y\>\cst{a}$ if $\cst{b} \succ \cst{a}$.
In practice, the strict and the nonstrict term order should be chosen so that they can compare
as many pairs of terms as possible while being computable and reasonably
efficient.

\begin{definitionx}[Maximality] \label{def:maximal}
  An element $x$ of a multiset $M$ is $\unrhd$-\emph{maximal} for some relation $\unrhd$
  if for all $y \in M$ with $y \unrhd x$, we have $y \unlhd x$.
  It is \emph{strictly $\unrhd$-maximal} if it is $\unrhd$-maximal and occurs only once in $M$.
\end{definitionx}

\begin{definitionx}[Selection function]
  A \emph{selection function} is a function that maps each clause to a subclause consisting of
  negative literals, which we call the \emph{selected} literals of that clause. 
  A literal $\greensubterm{L}{\,y}$ must not be selected if $y\>
  \tuple{u}_n$, with $n > 0$, is a $\succeq$-maximal term of the clause.
  \label{def:sel}
\end{definitionx}
The restriction on the selection function is needed for our proof, but it is an open question
whether it is actually necessary for refutational completeness.

Our calculus is parameterized by a strict term order $\succ$, a nonstrict term order $\succsim$, and a selection function $\HSel$.
The calculus rules depend on the following auxiliary notions.

\begin{definitionx}[Eligibility]
A literal $L$ is (\emph{strictly}) $\unrhd$-\emph{eligible} \wrt\ a substitution $\sigma$ in $C$ 
for some relation $\unrhd$ if it is
selected in $C$ or there are no selected literals in $C$ and $L\sigma$ is
(strictly) $\unrhd$-maximal in $C\sigma.$
If $\sigma$ is the identity substitution, we leave it implicit.
\end{definitionx}
\begin{definitionx}[Deep occurrences]
A variable \emph{occurs deeply} in a clause $C$ if it occurs
inside a $\lambda$-expression or inside an argument of an applied variable. 
\end{definitionx}
For example, $x$ and $z$ occur deeply in $\cst{f}\vvthinspace x \> y \eq y \> x \llor z \noteq (\lambda w.\>z\>\cst{a})$,
whereas $y$ does not occur deeply.
The purpose of this definition is to capture
all variables with an occurrence that
corresponds to a position inside a $\lambda$-expression in some ground instances of $C$.

The first rule of our calculus is the superposition rule. We regard positive and negative superposition as two cases of a single
rule
\[\namedinference{Sup}
{\overbrace{D' \llor { t \eq t'}}^{\vphantom{\cdot}\smash{D}} \hypsep
 \overbrace{C' \llor s\leftgreensubterm u\rightgreensubterm \doteq s'}^{\smash{C}}}
{(D' \llor C' \llor s\leftgreensubterm t'\rightgreensubterm \doteq s')\sigma}\]
where $\doteq$ denotes either $\eq$ or $\noteq$. The following side conditions
apply:
\begin{enumerate}
  \item[1.] $u$ is not fluid;%
  \hfill 2.\enskip $u$ is not a variable deeply occurring in $C$;%
  \hfill\hbox{}
  \hfill\hbox{}
  \item[3.] \emph{variable condition}: if
    $u$ is a variable~$y$,
    there must exist a grounding substitution~$\theta$
    such that $t\sigma\theta \succ
    t'\negvthinspace\sigma\theta$ and
    $C\sigma\theta \prec C''\sigma\theta$, where $C'' = C\{y\mapsto t'\}$;
  \item[4.] $\sigma\in\csu(t,u)$;%
  \hfill 5.\enskip $t\sigma \not\precsim t'\negvthinspace\sigma$;%
  \hfill 6.\enskip $s\leftgreensubterm u\rightgreensubterm\sigma \not\precsim s'\sigma$;%
  \hfill 7.\enskip $C\sigma \not\precsim D\sigma$;%
  \hfill\hbox{}
  \item[8.] $t \eq t'$ is strictly $\succsim$-eligible in $D$ \wrt\ $\sigma$;
  \item[9.] $s\leftgreensubterm u\rightgreensubterm \doteq s'$ is $\succsim$-eligible in $C$ \wrt\ $\sigma$,
        and strictly $\succsim$-eligible if it is positive.
\end{enumerate}
There are four main differences with the statement of the
standard superposition rule:
Contexts $\subterm{s}{~}$ are replaced by green contexts
  $\greensubterm{s}{\phantom{.}}$.
The standard condition $u \notin \VV$ is generalized
by conditions 2~and~3.
Most general unifiers are replaced by complete sets of unifiers.
And $\not\preceq$ is replaced by the more precise $\not\precsim$.

The second rule is a variant of \infname{Sup} that focuses on fluid green subterms:
\[\namedinference{FluidSup}
{\overbrace{D' \llor t \eq t'}^{\phantom{\cdot}\smash{D}} \hypsep
\overbrace{C' \llor \greensubterm{s}{u} \doteq s'}^{\smash{C}}}
{(D' \llor C' \llor \greensubterm{s}{z\>t'} \doteq s') \sigma}\]
with the following side conditions, in addition to \infname{Sup}'s conditions
5~to~9:

\begin{enumerate}
  \item[1.] $u$ is either a fluid term or a variable deeply occurring in $C$;
  \item[2.] $z$ is a fresh variable;%
  \hfill 3. $\sigma\in\csu(z\>t{,}\;u)$;%
  \hfill 4.\enskip $(z\>t')\sigma \not= (z\>t)\sigma$.\hfill\hbox{}
\end{enumerate}

The equality resolution and equality factoring rules are
almost identical to their standard counterparts:
\begin{align*}
 &\namedinference{ERes}
{\overbrace{C' \llor {u \noteq u'}}^C}
{C'\sigma}
&&\namedinference{EFact}
{\overbrace{C' \llor {u'} \eq v' \llor {u} \eq v}^C}
{(C' \llor v \noteq v' \llor u \eq v')\sigma}
\end{align*}
For \infname{ERes}: $\sigma\in\csu(u,u')$
and $u \noteq u'$ is $\succsim$-eligible in $C$ \wrt\ $\sigma$.
For~\infname{EFact}: $\sigma\in\csu(u,u')$,
$u\sigma \not\precsim v\sigma$,
and $u \eq v$ is $\succsim$-eligible in $C$ \wrt\ $\sigma$.

Argument congruence, a higher-order concern, is embodied by the rule
\[\namedinference{ArgCong}
{\overbrace{C' \llor s \eq s'}^C}
{C'\sigma \llor s\sigma\>\tuple{x}_n \eq s'\sigma\>\tuple{x}_n}\]
where $\sigma$ is the most general type substitution that ensures
well-typedness of the conclusion. In particular, if the result type of $s$ is
not a type variable, $\sigma$ is the identity substitution; and if the result
type is a type variable, it is instantiated with $\alpha_1 \fun \cdots \fun \alpha_m \fun \beta$,
where $\tuple{\alpha}_m$ and $\beta$ are fresh. This yields
infinitely many conclusions, one for each $m$.
The literal $s \eq s'$ must be strictly $\succsim$-eligible in $C$ \wrt\ $\sigma$,
and $\tuple{x}_n$ is a nonempty tuple of distinct fresh variables.

The rules are complemented by the polymorphic functional
extensionality axiom:
\[y\>(\diff\typeargs{\alpha,\beta}\> y\> z)\noteq z\>(\diff\typeargs{\alpha,\beta}\> y\> z) \llor y \eq z
\tag*{\text{(\infname{Ext})}}
\]
From now on, we will omit the type arguments to $\diff$ since they can be
inferred from the term arguments.

\oursubsection{Rationale for the Rules}
\label{ssec:rationale-for-the-rules}

The calculus realizes the following division of labor: \infname{Sup} and
\infname{FluidSup} are responsible for green subterms, which are outside
$\lambda$s, \infname{ArgCong} effectively gives access to the remaining
positions outside $\lambda$s, and the extensionality axiom takes care of
subterms inside $\lambda$s.

\begin{examplex}
Prefix subterms such as $\cst{g}$ in the term $\cst{g}\>\cst{a}$ are not green subterms
and thus cannot be superposed into.
\infname{ArgCong} gives us access to those positions. Consider the clauses
$\cst{g}\>\cst{a} \noteq \cst{f}\>\cst{a}$ and $\cst{g} \eq \cst{f}$.
An \infname{ArgCong} inference from $\cst{g} \eq \cst{f}$ generates
$\cst{g}\>x \eq \cst{f}\>x$.
This clause can be used for a $\infname{Sup}$ inference into the first clause,
yielding $\cst{f}\>\cst{a} \noteq \cst{f}\>\cst{a}$ and thus $\bot$ by \infname{ERes}.
\end{examplex}

\begin{examplex}
\label{ex:wsup-1}
Applied variables give rise to subtle situations with no counterparts in
first-order logic. Consider the clauses
$\cst{f}\>\cst{a} \eq \cst{c}$ and
$\cst{h}\>(y\>\cst{b})\>(y\>\cst{a}) \noteq \cst{h}\>(\cst{g}\>(\cst{f}\>\cst{b}))\>(\cst{g}\>\cst{c})$,
where $\cst{f}\>\cst{a} \succ \cst{c}$. It is easy to see that the clause set
is unsatisfiable, by grounding the second clause with $\theta = \{y \mapsto \lambda
x.\> \cst{g}\>(\cst{f}\>x)\}$. However, to mimic the superposition inference
that can be performed at the ground level, it is necessary to superpose at an
imaginary position \emph{below} the applied variable $y$ and yet \emph{above}
its argument~$\cst{a}$, namely, into the subterm $\cst{f}\>\cst{a}$ of
$\cst{g}\>(\cst{f}\>\cst{a})
= (\lambda x.\> \cst{g}\>(\cst{f}\>x))\>\cst{a}
= (y\>\cst{a})\theta$.
\infname{FluidSup}'s $z$~variable effectively transforms
$\cst{f}\>\cst{a} \eq \cst{c}$ into $z\>(\cst{f}\>\cst{a}) \eq
z\>\cst{c}$, whose left-hand side can be unified with $y\>\cst{a}$
by taking $\{y \mapsto \lambda x.\> z\>(\cst{f}\>x)\}$.
The resulting clause is
$\cst{h}\>(z\>(\cst{f}\>\cst{b}))\>(z\>\cst{c}) \noteq
\cst{h}\>(\cst{g}\>(\cst{f}\>\cst{b}))\>(\cst{g}\>\cst{c})$,
from which $\bot$ follows by \infname{ERes}.
\end{examplex}

\begin{examplex}
\label{ex:wsup-2}
The clause set consisting of
$\cst{f}\>\cst{a} \eq \cst{c}$,
$\cst{f}\>\cst{b} \eq \cst{d}$, and
$\cst{g}\>\cst{c} \noteq y\>\cst{a} \llor \cst{g}\>\cst{d} \noteq y\>\cst{b}$
has a similar flavor.
\infname{ERes} is applicable on either literal of the third clause,
but the computed unifier, $\{y \mapsto \lambda x.\> \cst{g}\>\cst{c}\}$
or $\{y \mapsto \lambda x.\> \cst{g}\>\cst{d}\}$, is not the right one.
Again, we need \infname{FluidSup}.
\end{examplex}

\begin{examplex}
\label{ex:wsup-3}
Third-order clauses containing subterms of the form $y\>(\lambda x.\> t)$ can
be even more stupefying. The clause set consisting of
$\cst{f}\> \cst{a} \eq \cst{c}$ and
$\cst{h}\> (y\> (\lambda x.\> \cst{g}\> (\cst{f}\> x))\> \cst{a})\> y
  \noteq \cst{h}\> (\cst{g}\> \cst{c})\> (\lambda w\>x.\> w\>x)$
is unsatisfiable. To see why, apply
$\theta = \{y \mapsto \lambda w\>x.\> w\>x\}$
to the second clause, yielding
$\cst{h}\> (\cst{g}\> (\cst{f}\> \cst{a}))\> (\lambda w\>x.\> w\>x)
  \noteq \cst{h}\> (\cst{g}\> \cst{c})\> (\lambda w\>x.\> w\>x)$.
Let $\cst{f}\> \cst{a} \succ \cst{c}$. A \infname{Sup} inference is
possible between the
first clause and this ground instance of the second one.
But at the nonground level,
the subterm $\cst{f}\> \cst{a}$ is not clearly localized:
$\cst{g}\> (\cst{f}\> \cst{a})
= (\lambda x.\> \cst{g}\> (\cst{f}\> x))\> \cst{a}
= (\lambda w\>x.\> w\>x)\> (\lambda x.\> \cst{g}\> (\cst{f}\> x))\> \cst{a}
= (y\> (\lambda x.\> \cst{g}\> (\cst{f}\> x))\> \cst{a})\theta$.
The \infname{FluidSup} rule can cope with this. One of the unifiers of
$z\> (\cst{f}\> \cst{a})$ and $y\> (\lambda x.\> \cst{g}\> (\cst{f}\> x))\> \cst{a}$
will be $\{y \mapsto \lambda w\>x.\> w\>x{,}\; z \mapsto \cst{g}\}$,
yielding the clearly unsatisfiable clause
$\cst{h}\> (\cst{g}\> \cst{c})\> (\lambda w\>x.\> w\>x) \noteq
  \cst{h}\> (\cst{g}\> \cst{c})\> (\lambda w\>x.\> w\>x)$.
\end{examplex}

\begin{examplex}
The \infname{FluidSup} rule is concerned not only with applied variables but also with
$\lambda$-expressions that, after substitution, may be $\eta$-reduced to
reveal new applied variables or green subterms. Consider the clauses
$\cst{g}\> \cst{a} \eq \cst{b}$,
$\cst{h}\> (\lambda y.\> x\> y\> \cst{g}\> z) \eq \cst{c}$, and
$\cst{h}\>(\cst{f}\> \cst{b}) \noteq \cst{c}$.
Applying
$\{x \mapsto \lambda y'\> w\> z'.\> \cst{f}\> (w\> \cst{a})\> y' \}$
to the second clause yields
$\cst{h}\> (\lambda y.\> (\lambda y'\> w\> z'.\;
    \cst{f}\> (w\> \cst{a})\> y')\> y\> \cst{g}\> z) \eq \cst{c}$,
which $\beta$-reduces to
$\cst{h}\;(\lambda y.\> \cst{f}\> (\cst{g}\> \cst{a})\> y) \eq \cst{c}$
and $\beta\eta$-reduces to
$\cst{h}\;(\cst{f}\> (\cst{g}\> \cst{a})) \eq \cst{c}
$.
A \infname{Sup} inference is possible between the first clause and this new ground clause,
generating the clause $\cst{h}\>(\cst{f}\> \cst{b}) \eq \cst{c}$. By also
considering $\lambda$-expressions, the \infname{FluidSup} rule is applicable at
the nonground level to derive this clause.
\end{examplex}

\begin{examplex} \label{ex:prod-div}
  Consider the clause set consisting of the facts
  $C_{\text{succ}} = \cst{succ}\>x \noteq \cst{zero}$,
  $C_{\text{div}} = n \eq \cst{zero} \llor \cst{div}\;n\;n \eq \cst{one}$,
  $C_{\text{prod}} = \cst{prod}\; K\;(\lambda k.\>\cst{one}) \eq \cst{one}$,
  and the negated conjecture
  $C_{\text{conj}} = \cst{prod}\; K\;(\lambda k.\> \cst{div}\; (\cst{succ}\; k)\; (\cst{succ}\; k)) \noteq \cst{one}$.
  Intuitively, the term $\cst{prod}\;K\;(\lambda k.\; u)$ is intended to denote
  the product $\smash{\prod_{k\in K} u}$, where $k$ ranges over a finite set~$K$
  of natural numbers.
  The calculus derives the empty clause as follows:
  \[
    \prftree[r]{\infname{ERes}}{
    \prftree[r]{\infname{Sup}}
    {C_{\text{prod}}\!}
    {
  \prftree[r]{
  \infname{Sup}}{C_{\text{conj}}}
  {\trimbox{3.2em 0pt 0em 0em}{%
  {\prftree[r]{\infname{ERes}}{
    \prftree[r]{\infname{Sup}}
    {\prftree[r]{\infname{ERes}}{
    \prftree[r]{\infname{FluidSup}}
    {C_{\text{div}}}{
      \prftree[r]
      {\infname{Ext}}{}{y\>(\diff\typeargs{\alpha,\beta}\> y\> z)\noteq z\>(\diff\typeargs{\alpha,\beta}\> y\> z) \llor y \eq z}
    }{
    \begin{aligned}
      w&\>(\diff\typeargs{\alpha,\iota}\>(\lambda k.\>\cst{div}\>(w\>k)\>(w\>k))\>z) \eq \cst{zero} \\
      &\llor 
    \cst{one} \noteq z\>(\diff\typeargs{\alpha,\iota}\> (\lambda k.\>\cst{div}\>(w\>k)\>(w\>k))\> z) \llor
    (\lambda k.\>\cst{div}\>(w\>k)\>(w\>k)) \eq z
    \end{aligned}
    }
    }
    {{\begin{aligned}\\C_{\text{succ}}\end{aligned}}\quad
      \begin{aligned} 
    w&\>(\diff\typeargs{\alpha,\iota}\>(\lambda k.\>\cst{div}\>(w\>k)\>(w\>k))\>(\lambda k.\>\cst{one})) \eq \cst{zero}\\
    &\llor 
    (\lambda k.\>\cst{div}\>(w\>k)\>(w\>k)) \eq (\lambda k.\>\cst{one})
    \end{aligned}
    }}
  {\cst{zero} \noteq \cst{zero}
  \llor 
  (\lambda k.\>\cst{div}\>(\cst{succ}\>k)\>(\cst{succ}\>k)) \eq (\lambda k.\>\cst{one})}
  }
  {(\lambda k.\>\cst{div}\>(\cst{succ}\>k)\>(\cst{succ}\>k)) \eq (\lambda k.\>\cst{one})}}}}
  {\cst{prod}\; K\;(\lambda k.\>\cst{one}) \noteq \cst{one}}}
  {\cst{one} \noteq \cst{one}}}
  {\bot}
  \]
  Since the calculus does not superpose into $\lambda$-expressions, we
  need to 
  use the extensionality axiom to refute this clause set. We perform a
  \infname{FluidSup} inference
  into the extensionality axiom with the unifier
  $\{
    \beta \mapsto \iota,\>\allowbreak
    z' \mapsto \lambda x.\>x,\>\allowbreak
    n \mapsto w\>(\diff\typeargs{\alpha,\iota}\> (\lambda k.\>\cst{div}\>(w\>k)\>(w\>k))\> z),\> \allowbreak
    y \mapsto \lambda k.\>\cst{div}\>(w\>k)\>(w\>k)
  \} 
  \in \csu(z'\>(\cst{div}\>n\>n){,}\; y\>(\diff\typeargs{\alpha,\beta}\> y\> z))$.
  Then we apply \infname{ERes}
  with the unifier 
  $\{z \mapsto \lambda k.\>\cst{one}\} \in 
  \csu(\cst{one}{,}\; z\>(\diff\typeargs{\alpha,\iota}\> (\lambda k.\>\cst{div}\>(w\>k)\>(w\>k))\> z))$
  to eliminate the negative literal.
  Next, we perform a $\infname{Sup}$ inference 
  into $C_{\text{succ}}$
  with the unifier 
  $\{
    \alpha \mapsto \iota,\>\allowbreak
    w \mapsto \cst{succ},\>\allowbreak
    x \mapsto \diff\typeargs{\alpha,\iota}\>\allowbreak(\lambda k.\>\cst{div}\>(w\>k)\>(w\>k))\>(\lambda k.\>\cst{one})\} \in
  \csu(w\>(\diff\typeargs{\alpha,\iota}\>(\lambda k.\>\cst{div}\>(w\>k)\>(w\>k))\>(\lambda k.\>\cst{one})),\allowbreak
  \>\cst{succ}\>x)$. To eliminate the trivial literal, we apply \infname{ERes}.
  We then apply a \infname{Sup} inference into $C_{\text{conj}}$ and superpose into the resulting clause with $C_{\text{prod}}$.
  Finally we derive the empty clause by \infname{ERes}.
  The unifiers in this example were chosen to keep the clauses reasonably small.
\end{examplex}

Because it gives rise to flex--flex pairs, which are unification constraints
where both sides are variable-headed, \infname{FluidSup} can be very
prolific. With variable-headed terms on both sides of its maximal literal, the
extensionality axiom is another prime source of flex--flex pairs. Flex--flex
pairs can also arise in the other rules (\infname{Sup}, \infname{ERes}, and
\infname{EFact}).
Due to order restrictions and fairness, we cannot postpone solving flex--flex
pairs indefinitely. Thus, we cannot use Huet's pre-unification procedure
\cite{huet-1975} and must instead choose a full unification procedure such as Jensen
and Pietrzykowski's \cite{jensen-pietrzykowski-1976}, Snyder and Gallier's
\cite{snyder-gallier-1989}, or the procedure
that has recently been developed by Vukmirovi\'c, Bentkamp, and Nummelin
\cite{vukmirovic-et-al-2020-unif}.
On the positive side, optional inference rules can efficiently cover many
cases where \infname{FluidSup} or the extensionality axiom would otherwise be
needed (\Section~\ref{sec:extensions}), and heuristics can help
postpone the explosion.
Moreover, flex--flex pairs are not always as bad as their reputation; for
example, $y\> \cst{a}\> \cst{b} \UNIF z\> \cst{c}\> \cst{d}$ admits a most general unifier:
$\{y \mapsto \lambda w\> x.\> y'\,w\> x\> \cst{c}\>\cst{d}{,}\; z \mapsto y'\, \cst{a}\>\cst{b}\}$.

The calculus is a graceful generalization of standard superposition, except
for the extensionality axiom. 
From simple first-order clauses, the axiom can be used
to derive clauses containing $\lambda$-expressions, which are useless if the problem is
first-order.
For instance, the clause $\cst{g}\>x \eq \cst{f}\>x\>x$
can be used for a \infname{FluidSup} inference into the axiom (\infname{Ext})
yielding the clause
$w\>t\>(\cst{f}\>t\>t)\noteq z\>t
\llor (\lambda u.\>w\>u\>(\cst{g} u)) \eq z$
via the unifier 
$\{
\alpha \mapsto \iota,\>\allowbreak
\beta\mapsto \iota,\>\allowbreak
x \mapsto t,\>\allowbreak
v \mapsto \lambda u.\> w\>t\>u,\>\allowbreak
y \mapsto \lambda u.\>w\>u\>(\cst{g}\>u)
\} \in \csu(v\>(\cst{g}\>x),\>y\>(\diff\typeargs{\alpha,\beta}\>y\>z))$
where $t = \diff\typeargs{\iota,\iota}\>(\lambda u.\>w\>u\>(\cst{g}\>u))\>z$,
the variable $w$ is freshly introduced by unification, and
$v$ is the fresh variable introduced by \infname{FluidSup} (named $z$ in the definition of the rule).
By \infname{ERes}, 
with the unifier
$\{
  z \mapsto \lambda u.\> w\>u\>(\cst{f}\>u\>u)
\}\in \csu(w\>t\>(\cst{f}\>t\>t),\>z\>t)$,
we can then derive
$(\lambda u.\> w\>u\>(\cst{g}\>u)) \eq
(\lambda u.\> w\>u\>(\cst{f}\>u\>u))$,
an equality of two $\lambda$-expressions, although we started with a simple first-order clause.
This could be avoided if we could find a way to make the positive
literal $y \eq z$ of (\infname{Ext}) larger than the other literal,
or to select $y \eq z$ without losing refutational
completeness. The literal $y \eq z$ interacts only with green subterms of functional type,
which do not arise in first-order clauses.

\oursubsection{Soundness}
\label{ssec:soundness}

To show soundness of the inferences, we need the substitution lemma for our logic:
\begin{lemmax}[Substitution lemma]
  Let $\III = (\IIIty,\II,\LL)$ be a proper interpretation. Then
  \[\interpret{\tau\rho}{\IIIty}{\xi} = \interpret{\tau}{\IIIty}{\xi'}%
  \text{\quad and\quad}%
  \interpret{t\rho}{\III}{\xi} = \interpret{t}{\III}{\xi'}\]
  for all terms $t$, all types $\tau$, and all substitutions
  $\rho$,
  where $\xi'(\alpha) = \interpret{\alpha\rho}{\IIIty}{\xi}$ for all type
  variables~$\alpha$ and $\xi'(x) = \interpret{x\rho}{\III}{\xi}$ for all term
  variables $x$.
  \label{lem:subst-lemma-general}
\end{lemmax}
\begin{proof}
  First, we prove that $\interpret{\tau\rho}{\IIIty}{\xi} =
  \interpret{\tau}{\IIIty}{\xi'}$ by induction on the structure of $\tau$.
  If $\tau = \alpha$ is a type variable,
  \[\interpret{\alpha\rho}{\IIIty}{\xi} = \xi'(\alpha) = \interpret{\alpha}{\IIIty}{\xi'}\]
  If $\tau = \kappa(\tuple{\upsilon})$ for some type constructor $\kappa$ and types $\tuple{\upsilon}$,
  \[\interpret{\kappa(\tuple{\upsilon})\rho}{\IIIty}{\xi}
  = \IIty(\kappa)(\interpret{\tuple{\upsilon}\rho}{\IIIty}{\xi})
  \eqIH
    \IIty(\kappa)(\interpret{\tuple{\upsilon}}{\IIIty}{\xi'})
    = \interpret{\kappa(\tuple{\upsilon})}{\IIIty}{\xi'}\]
  Next, we prove $\interpret{t\rho}{\III}{\xi} = \interpret{t}{\III}{\xi'}$
  by induction on the structure of a $\lambda$-term representative of $t$,
  allowing arbitrary substitutions $\rho$ in the induction hypothesis.	
  If~$t = y$, then by the definition of the denotation of a variable
  \[\interpret{y\rho}{\III}{\xi} = \xi'(y) = \interpret{y}{\III}{\xi'}\]
  If $t = \cst{f}\typeargs{\tuple{\tau}}$, then by the definition of the term denotation
  \[\interpret{\cst{f}\typeargs{\tuple{\tau}}\rho}{\III}{\xi} =
  \II(\cst{f},\interpret{\tuple{\tau}\rho}{\IIIty}{\xi})
  \eqIH
  \II(\cst{f},\interpret{\tuple{\tau}}{\IIIty}{\xi'}) =
  \interpret{\cst{f}\typeargs{\tuple{\tau}}}{\III}{\xi'}\]
  If $t = u\>v$, then by the definition of the term denotation
  \[\interpret{(u\>v)\rho}{\III}{\xi}
  = \interpret{u\rho}{\III}{\xi}(\interpret{v\rho}{\III}{\xi})
  \eqIH  \interpret{u}{\III}{\xi'}(\interpret{v}{\III}{\xi'})
  =  \interpret{u\>v}{\III}{\xi'}\]
  If $t = \lambda z.\>u$, let $\rho'(z)=z$ and $\rho'(x)=\rho(x)$ for $x\neq z$. Using properness of $\III$ in the second and the last step, we have
\[\interpret{(\lambda z.\>u)\rho}{\III}{\xi}(a)
= \interpret{(\lambda z.\>u\rho')}{\III}{\xi}(a)
= \interpret{u\rho'}{\III}{\xi[z\mapsto a]}
\eqIH \interpret{u}{\III}{\xi'[z\mapsto a]}
= \interpret{\lambda z.\>u}{\III}{\xi'}(a)
\tag*{\qedhere}\]
\end{proof}

\begin{lemmax}\label{lem:apply-subst}
If $\III\models C$ for some interpretation $\III$ and some clause $C$, then $\III\models C\rho$ for all substitutions $\rho$.
\end{lemmax}
\begin{proof}
We have to show that $C\rho$ is true in $\III$ for all valuations $\xi$. Given a valuation $\xi$,
define $\xi'$ as in Lemma~\ref{lem:subst-lemma-general}.
Then, by Lemma~\ref{lem:subst-lemma-general}, a literal in $C\rho$ is true in $\III$ for $\xi$ if and
only if the corresponding literal in $C$ is true in $\III$ for $\xi'$.
There must be at least one such literal because $\III \models C$ and hence $C$ is in particular true in $\III$ for $\xi'$.
Therefore, $C\rho$ is true in $\III$ for $\xi$. \qedhere
\end{proof}

\begin{theoremx}[Soundness] 
The inference rules \infname{Sup}, \infname{FluidSup}, \infname{ERes},
\infname{EFact}, and \infname{Arg\-Cong} are sound
{\upshape(}even without the variable condition and the side conditions on fluidity, deeply occurring variables, order, and eligibility{\upshape)}.
\label{lem:soundness}
\end{theoremx}
\begin{proof}
We fix an inference and an interpretation $\III$
that is a model of the premises.
We need to show that it is also a model of the conclusion.

From the definition of the denotation of a term, it is obvious that
congruence holds in our logic, at least for subterms that are not inside a $\lambda$-expression.
In particular, it holds for green subterms and for the left subterm $t$ of an application $t\>s$.

By Lemma~\ref{lem:apply-subst}, $\III$ is a model of the $\sigma$-instances of the premises as well,
where $\sigma$ is the substitution used for the inference. Let $\xi$ be a valuation.
By making case distinctions on the truth under $\III,\xi$ of the literals
of the $\sigma$-instances of the premises, using the conditions that $\sigma$ is a unifier, and applying congruence,
it follows that the conclusion is true under $\III,\xi$. Hence, $\III$ is a model of the conclusion. \qedhere
\end{proof}

As in the $\lambda$-free higher-order logic of
Bentkamp et al.~\cite{bentkamp-et-al-lfhosup-arxiv},
skolemization is unsound in our logic. As a consequence, axiom (\infname{Ext}) does not hold
in all interpretations, but the axiom is consistent with our logic, i.e., there exist models of (\infname{Ext}).

\oursubsection{The Redundancy Criterion}
\label{ssec:the-redundancy-criterion}

A redundant clause is usually defined as a clause whose ground
instances are entailed by smaller (\negvvthinspace$\prec$) ground instances of
existing clauses. This would be too strong for our calculus, as it
would make most clauses produced by \infname{ArgCong}
redundant. The solution is to base the redundancy criterion on a weaker ground
logic---ground monomorphic first-order logic---in which argument congruence and
extensionality do not hold. The resulting notion of redundancy gracefully
generalizes the standard first-order notion.
We employ an encoding $\flooronly$ to translate
ground higher-order terms into ground first-order terms. 
$\flooronly$ indexes
each symbol occurrence with the type arguments and the number of term
arguments. For example,
$\floor{\cst{f}\>\cst{a}}=\cst{f}_1(\cst{a}_0)$
and $\floor{\cst{g}\typeargs{\kappa}} = \cst{g}\vvthinspace{}^{\kappa}_{0}$.
In addition, $\flooronly$ conceals $\lambda$-expressions by replacing them with
fresh symbols. These measures effectively disable argument congruence and
extensionality.
For example, the clause sets
$\{\cst{g} \eq \cst{f}{,}\; \cst{g}\>\cst{a} \noteq \cst{f}\> \cst{a}\}$
and
$\{\cst{b} \eq \cst{a}{,}\; (\lambda x.\; \cst{b}) \noteq (\lambda x.\; \cst{a})\}$
are unsatisfiable in higher-order logic,
but the encoded clause sets
$\{\cst{g}_0 \eq \cst{f}_0{,}\allowbreak\; \cst{g}_1(\cst{a}_0) \noteq \cst{f}_1(\cst{a}_0)\}$
and
$\{\cst{b}_0 \eq \cst{a}_0{,}\allowbreak\; \cst{lam}_{\lambda x.\; \cst{b}} \noteq \cst{lam}_{\lambda x.\; \cst{a}}\}$
are satisfiable in first-order logic, where $\cst{lam}_{\lambda x.\>t}$ is a family of fresh symbols.

Given a higher-order signature ($\Sigmaty,\Sigma)$, we define
a ground first-order signature ($\Sigmaty,\allowbreak\Sigma_\GF)$ as follows.
The type constructors $\Sigmaty$ are the same in both signatures, but ${\fun}$
is uninterpreted in first-order logic. For each ground
instance $\cst{f}\typeargs{\tuple{\upsilon}}
: \tau_1\fun\cdots\fun\tau_n\fun\tau$ of a symbol $\cst{f} \in
\Sigma$, %
we introduce a
first-order symbol $\smash{\cst{f}^{\tuple{\upsilon}}_{\!j}} \in
\Sigma_\GF$ with argument types~$\tuple{\tau}_{\!j}$ and return
type~$\tau_{\!j+1} \fun \cdots \fun \tau_n \fun \tau$, for
each $j$.
Moreover, for each ground term $\lambda x.\>t$, we introduce a
symbol $\cst{lam}_{\lambda x.\>t} \in \Sigma_\GF$ of the same type.

Thus, we 
consider three levels of logics:\
the higher-order level $\HH$ over a given signature ($\Sigmaty,\Sigma)$,
the ground higher-order level $\GH$, which is the ground fragment of $\HH$,
and the ground monomorphic first-order level $\GF$ over the signature ($\Sigmaty,\Sigma_\GF)$ defined above.
We use $\THH$, $\TGH$, and $\TGF$ to denote the respective sets of terms,
$\TyHH$, $\TyGH$, and $\TyGF$ to denote the respective sets of types,
and $\CHH$, $\CGH$, and $\CGF$ to denote the respective sets of clauses.
Each of the three levels has an entailment relation $\models$.
A clause set $N_1$ entails a clause set $N_2$, denoted $N_1 \models N_2$,
if every model of $N_1$ is also a model of $N_2$. For $\HH$ and $\GH$,
we use higher-order models; for $\GF$, we use first-order models.
This machinery may seem excessive, but it is essential to define redundancy of
clauses and inferences properly, and it will play an important role in the
refutational completeness proof (\Section~\ref{sec:refutational-completeness}).
The 
three levels are connected by two functions
$\gnd$ and $\flooronly$:
\begin{definitionx}[Grounding function $\boldsymbol{\gnd}$ on terms and clauses]
The grounding function $\gnd$ maps
terms $t \in \THH$ to the set of their ground instances---i.e.,
the set of all $t\theta \in \TGH$ where $\theta$ is a substitution.
It also maps clauses $C\in\CHH$ to the set of their ground instances---i.e.,
the set of all $C\theta \in \CGH$ where $\theta$ is a substitution.
\end{definitionx}
\begin{definitionx}[Encoding $\boldsymbol{\flooronly}$ on terms and clauses]
  The encoding $\flooronly : \TGH \rightarrow \TGF$ is recursively defined
  as
  \begin{align*}
    \floor{\lambda x.\>t} = \cst{lam}_{\lambda x.\>t}
    &&&
  \floor{\cst{f}\typeargs{\tuple{\upsilon}}\> \tuple{s}_{\!j}}
  = \cst{f}^{\tuple{\upsilon}}_{\!j} (\floor{\tuple{s}_{\!j}})
  \end{align*}
  using $\eta$-short $\beta$-normal representatives of terms.
  The encoding $\flooronly$ is
  extended to map from $\CGH$ to $\CGF$ 
  by mapping each literal and each side of a literal individually.
\end{definitionx}

Schematically, the three levels are connected as follows:
\[
\begin{tikzpicture}[level distance=13em]
  \node[align=center]{$\HH$\\higher-order}[grow'=right]
    child {
      node[align=center]{$\GH$\\ground higher-order}
        child {
          node[align=center]{$\GF$\\ground first-order}
          edge from parent[->]
            node[above] {$\flooronly$}
        }
      edge from parent[->]
        node[above] {$\gnd$}
    };
\end{tikzpicture}
\]

The mapping $\flooronly$ is clearly bijective.
Using the inverse mapping, the
order $\succ$ can be transferred from $\TGH$ to $\TGF$ and from $\CGH$ to $\CGF$ by defining $t \succ s$
as $\ceil{t} \succ \ceil{s}$ and $C \succ D$
as $\ceil{C} \succ \ceil{D}$. The property that $\succ$ on clauses
is the multiset extension of
$\succ$ on literals, which in turn is the
multiset extension of
$\succ$ on terms, is maintained because
$\ceilonly$ maps the multiset representations elementwise.

For example, let $C = y\>\cst{b} \eq y\>\cst{a} \lor y \noteq \cst{f}\>\cst{a}\in\CHH$.
Then $\gnd(C)$ contains, among many other clauses,
$C\theta = 
\cst{f}\>\cst{b}\>\cst{b} \eq \cst{f}\>\cst{a}\>\cst{a} \lor (\lambda x.\>\cst{f}\>x\>x) \noteq \cst{f}\>\cst{a}\in\CGH$,
where $\theta = \{y \mapsto \lambda x.\>\cst{f}\>x\>x\}$.
On the $\GF$ level, this clause corresponds to 
$\floor{C\theta} = 
\cst{f}_2(\cst{b}_0,\cst{b}_0) \eq \cst{f}_2(\cst{a}_0,\cst{a}_0) \lor \cst{lam}_{\lambda x.\>\cst{f}\>x\>x} \noteq \cst{f}_1(\cst{a}_0)\in\CGF$.

A key property of $\flooronly$ is that green subterms in $\TGH$ correspond to
subterms in $\TGF$. This allows us to show that well-foundedness, totality on
ground terms, compatibility with contexts, and the subterm property hold for
$\succ$ on $\TGF$. 

\begin{lemmax}\label{lem:subterm-correspondence1}
  Let $s,t\in\TGH$.
  We have $\floor{\greensubterm{t}{s}_p} = \subterm{\floor{t}}{\floor{s}}_p$.
  In other words, $s$ is a green subterm of $t$ at position $p$ if and only if $\floor{s}$ is
  a subterm of $\floor{t}$ at position $p$.
\end{lemmax}
\begin{proof}
Analogous to 
Lemma~3.13
of Bentkamp et al.~\cite{bentkamp-et-al-lfhosup-arxiv}.
\qedhere
\end{proof}

\begin{lemmax}\label{lem:order-prop-transfer}
  Well-foundedness,
  totality,
  compatibility with
  contexts, and
  the subterm property
  hold for $\succ$ in $\TGF$.
\end{lemmax}
\begin{proof}
Analogous to 
Lemma~3.15
of Bentkamp et al.~\cite{bentkamp-et-al-lfhosup-arxiv}, 
using Lemma~\ref{lem:subterm-correspondence1}.
\qedhere
\end{proof}

The saturation procedures of superposition provers
aggressively delete clauses that are strictly subsumed by other clauses. A
clause $C$ \emph{subsumes}~$D$ if there exists a substitution
$\sigma$ such that $C\sigma \subseteq D$. A clause $C$ \emph{strictly subsumes}~$D$ if
$C$ subsumes $D$ but $D$ does not subsume $C$. For example, $x \eq \cst{c}$
strictly subsumes both $\cst{a} \eq \cst{c}$ and $\cst{b} \noteq \cst{a} \llor
x \eq \cst{c}$. The proof of refutational completeness of resolution and
superposition provers relies on the well-foundedness of the strict subsumption
relation. 
Unfortunately, this
property does not hold for 
higher-order logic, 
where $\cst{f}\>x\>x \eq
\cst{c}$ is strictly subsumed by $\cst{f}\>(x\>\cst{a})\>(x\>\cst{b}) \eq
\cst{c}$, which is strictly subsumed by
$\cst{f}\>(x\>\cst{a}\>\cst{a}')\>(x\>\cst{b}\>\cst{b}') \eq \cst{c}$, and so
on. To prevent such infinite chains,
we use a well-founded partial order $\sqsupset$ on $\CHH$.
We can define $\sqsupset$ as
${\issubsumedtilde} \ccap {>_\text{size}}$,
where $\issubsumedtilde$ stands for ``subsumed by'' and
$D >_\text{size} C$ if
either $\mathit{size}(D) > \mathit{size}(C)$
or $\mathit{size}(D) = \mathit{size}(C)$
and $D$ contains fewer distinct variables than $C$;
the $\mathit{size}$ function is some notion of syntactic size, such as
the number of constants and variables contained in a clause.
This yields for instance $\cst{a} \mathbin\eq \cst{c} \sqsupset x \mathbin\eq \cst{c}$
and $\cst{f}\>(x\>\cst{a}\>\cst{a}) \mathbin\eq \cst{c} \sqsupset \cst{f}\>(y\>\cst{a}) \mathbin\eq \cst{c}$.
To justify the deletion of subsumed clauses, we set up our redundancy criterion
to cover subsumption, following Waldmann et al.~\cite{waldmann-et-al-2020-saturation}.

We define the sets of redundant clauses \wrt\ a given clause set as follows:
\begin{itemize}
\item Given $C\in\CGF$ and $N\subseteq\CGF$, let $C\in\GFRedC(N)$ if
$\{D \in N \mid D \prec C\}\models C$.
\item Given $C\in\CGH$ and $N\subseteq\CGH$, let $C\in\GHRedC(N)$ if
$\floor{C} \in \GFRedC(\floor{N})$.
\item
Given $C\in\CHH$ and $N\subseteq\CHH$, let $C\in\HRedC(N)$ if
for every $D \in \gnd(C)$,
we have $D \in \GHRedC(\gnd(N))$ or
there exists $C' \in N$ such that $C \sqsupset C'$ and $D \in \gnd(C')$.
\end{itemize}
For example, 
$(\cst{h}\>\cst{g})\>x \eq (\cst{h}\>\cst{f})\>x$ is redundant \wrt\ $\cst{g}\eq \cst{f}$, 
but $\cst{g}\>x \eq \cst{f}\>x$ and $(\lambda x.\>\cst{g}) \eq (\lambda x.\>\cst{f})$ are not,
because $\flooronly$ translates an unapplied $\cst{g}$ to $\cst{g}_0$, whereas
an applied $\cst{g}$ is translated to $\cst{g}_1$ and 
the expression $\lambda x.\>\cst{g}$ is translated to $\cst{lam}_{\lambda x.\>\cst{g}}$.
These different translations prevent entailment on the $\GF$ level.
For an example of subsumption, we assume that
$\cst{a} \mathbin\eq \cst{c} \sqsupset x \mathbin\eq \cst{c}$ holds, 
for instance using the above definition of $\sqsupset$. Then
$\cst{a} \mathbin\eq \cst{c}$ is redundant \wrt\ $x \mathbin\eq \cst{c}$.

Along with the three levels of logics, we consider three inference systems%
:\ %
$\HInf$, $\GHInf$, and $\GFInf$. $\HInf$ is the inference system described in
\Section~\ref{ssec:the-core-inference-rules}. For uniformity, we regard the
extensionality axiom as a premise-free inference rule \infname{Ext} whose
conclusion is axiom~(\infname{Ext}). The rules of $\GHInf$
include \infname{Sup}, \infname{ERes}, and \infname{EFact} from $\HInf$,
but with the restriction that premises and conclusion are ground
and with all references to $\succsim$ replaced by $\succeq$.
In addition, $\GHInf$ contains a premise-free rule \infname{GExt}
whose infinitely many conclusions are the ground instances of (\infname{Ext}),
and the following ground variant of \infname{ArgCong}:
\[\namedinference{GArgCong}
{C' \llor s \eq s'}
{C' \llor s\>\tuple{u}_n \eq s'\>\tuple{u}_n}\]
where $s \eq s'$ is strictly $\succeq$-eligible in $C' \llor s \eq s'$
and $\tuple{u}_n$ is a nonempty tuple of ground terms.

$\GFInf$ contains all \infname{Sup}, \infname{ERes}, and \infname{EFact} inferences
from $\GHInf$ translated by $\flooronly$.
 It coincides with standard first-order
superposition. 

\looseness=-1
Each of the three inference systems is parameterized by a selection function.
For $\HInf$, we globally fix one selection function $\HSel$.
For $\GHInf$ and $\GFInf$, we need to consider different selection functions.
We write $\GHInf^\GHSel$ for $\GHInf$ and $\GFInf^\GFSel$ for $\GFInf$
to make the dependency on the respective selection functions $\GHSel$ and $\GFSel$ explicit.
Let $\gnd(\HSel)$ denote the set of all selection functions on~$\CGH$
such that for each clause
in $C\in\CGH$, there exists a clause $D\in\CHH$ with $C\in\gnd(D)$ and corresponding selected literals.
For each selection function $\GHSel$ on $\CGH$, via the bijection $\flooronly$, we obtain a corresponding selection function
on $\CGF$, which we denote by $\floor{\GHSel}$.

We extend the functions $\flooronly$ and $\gnd$ to inferences:
\begin{notationx}
  Given an inference $\iota$, we write $\prem(\iota)$ for the tuple of premises,
  $\mprem(\iota)$ for the main (i.e., rightmost) premise, and $\concl(\iota)$ for
  the conclusion.
\end{notationx}

\begin{definitionx} [Encoding $\flooronly$ on inferences]\,
  Given a \infname{Sup}, \infname{ERes}, or \infname{EFact}
  inference $\iota \in \GHInf$, let $\floor{\iota}\in\GFInf$ denote 
  the inference defined by $\prem(\floor{\iota}) = \floor{\prem(\iota)}$
  and $\concl(\floor{\iota}) = \floor{\concl(\iota)}$.
\end{definitionx}

\begin{definitionx} [Grounding function $\gnd$ on inferences]\,
  Given an inference $\iota\in\HInf$, and a selection function $\GHSel\in\gnd(\HSel)$,
  we define the set $\gnd^\GHSel(\iota)$ of ground instances of $\iota$
  to be all inferences $\iota'\in\GHInf^\GHSel$ such that $\prem(\iota') = \prem(\iota)\theta$
  and $\concl(\iota') = \concl(\iota)\theta$ for some grounding substitution $\theta$.
\end{definitionx}
This will map \infname{Sup} and \infname{FluidSup} to \infname{Sup},
\infname{EFact} to \infname{EFact},
\infname{ERes} to \infname{ERes},
\infname{Ext} to \infname{GExt},
and \infname{Arg\-Cong} to \infname{GArgCong} inferences, but it
is also possible that $\gnd^\GHSel(\iota)$ is the empty set for some inferences $\iota$.

We define the sets of redundant inferences \wrt\ a given clause
set as follows:
\begin{itemize}
\item Given $\iota\in\GFInf^\GFSel$ and $N\subseteq\CGF$, let $\iota\in\GFRedI^\GFSel(N)$
  if 
  $\prem(\iota) \ccap \GFRedC(N) \not= \varnothing$ or
  $\{D \in N \mid D \prec \mprem(\iota)\} \models \concl(\iota)$.
\item Given $\iota\in\GHInf^\GHSel$ and $N\subseteq\CGH$, let $\iota\in\GHRedI^\GHSel(N)$ if
\begin{itemize}
\item $\iota$ is not a \infname{GArgCong} or \infname{GExt} inference and
  $\floor{\iota}\in\smash{\GFRedI^{\floor{\GHSel}}}(\floor{N})$; or
\item $\iota$ is a \infname{GArgCong} or \infname{GExt} inference
  and $\concl(\iota)\in N\ccup\GHRedC(N)$.
\end{itemize}
\item Given $\iota\in\HInf$ and $N\subseteq\CHH$, let $\iota\in\HRedI(N)$ if
$\gnd^\GHSel(\iota)\subseteq\GHRedI(\gnd(N))$ for all $\GHSel\in\gnd(\HSel)$.
\end{itemize}
Occasionally, we omit the selection function in the notation when it is irrelevant.
A clause set $N$ is \emph{saturated} \wrt\ an inference system and the inference
component $\RedI$ of a redundancy criterion if every inference from clauses
in $N$ is in~$\RedI(N).$

\oursubsection{Simplification Rules}
\label{ssec:simplification-rules}
The redundancy criterion $(\HRedI, \HRedC)$ is strong enough to support
most of the simplification rules implemented in Schulz's first-order prover E
\cite[Sections 2.3.1~and~2.3.2]{schulz-2002-brainiac},
some only with minor adaptions.
Deletion of duplicated literals, 
deletion of resolved literals, 
syntactic tautology deletion,  
negative simplify-reflect, and
clause subsumption
adhere to our redundancy criterion.

Positive simplify-reflect and equality subsumption are supported
by our criterion if they are applied in green contexts
$\greensubterm{u}{\phantom{.}}$ instead of arbitrary contexts $u[\phantom{.}]$.
Semantic tautology deletion can be applied as well, but we must use the
entailment relation of the GF level---i.e., only rewriting in green contexts can
be used to establish the entailment.
Similarly, rewriting of positive and negative literals (demodulation) 
can only be applied in green contexts.
Moreover, for positive literals, the rewriting clause must be smaller than the rewritten
clause---a condition that is also necessary with the standard first-order redundancy criterion
but not always fulfilled by Schulz's rule.
As for destructive equality resolution, even in first-order logic the rule cannot be justified with the standard redundancy criterion,
and it is unclear whether it preserves refutational completeness.

\oursubsection{A Derived Term Order}
\label{ssec:a-derived-term-order}

We stated some requirements on the term orders $\succ$ and
$\succsim$ in \Section~\ref{ssec:the-core-inference-rules} but have not shown
how to fulfill them.
To derive a suitable strict term order $\succ$, we propose to encode
$\eta$-short $\beta$-normal forms into untyped first-order
terms and apply an order~$\fosucc$ of first-order terms such as the Knuth--Bendix
order \cite{knuth-bendix-1970} or the lexicographic path order
\cite{kamin-levy-1980-cannotfind}.

The encoding, denoted by $\encOonly$, 
indexes symbols with their number of term arguments, similarly to the $\flooronly$ encoding.
Unlike the $\flooronly$ encoding, $\encOonly$ translates $\lambda x \mathbin:\nobreak
\tau.\; t$ to $\cst{lam}(\encO{\tau},\encO{t})$ and uses
De Bruijn \cite{de-bruijn-1972} symbols to represent bound variables.
The $\encOonly$ encoding replaces fluid terms~$t$ by fresh variables~$\zof{t}$
and maps type arguments to term arguments, while erasing any other
type information. 
For example, 
$\encO{\lambda x \mathbin: \kappa.\> \cst{f}\>(\cst{f}\>(\cst{a}\typeargs{\kappa}))\>(y\>\cst{b}) } =
\cst{lam}(\kappa, \cst{f}_2(\cst{f}_1(\cst{a}_0(\kappa)), \zof{y \> \cst{b}}))$.
The use of De Bruijn indices and the monolithic encoding of fluid terms
ensure stability under both $\alpha$-renaming and substitution. 
\begin{definitionx}[Encoding $\boldsymbol{\encOonly}$]
Given a signature $(\Sigmaty,\allowbreak\Sigma)$,
$\encOonly$ encodes types and terms as terms over the untyped first-order
signature $\Sigmaty \uplus \{\cst{f}_k \mid \cst{f}\in\Sigma,\>k\in\nobreak\mathbb{N}\} \uplus \{\cst{lam}\}\uplus \{\smash{\db^i_k}\mid
i,k\in\nobreak\mathbb{N}\}$.
We reuse higher-order type variables
as term variables in the target untyped first-order logic.
Moreover, let $\zof{t}$ be an untyped first-order variable for each higher-order term $t$.
The auxiliary function $\encB{x}{t}$ replaces each free occurrence of the
variable~$x$ by a symbol $\db^i$, where $i$ is the number of
$\lambda$-expressions surrounding the variable occurrence.
The type-to-term version of $\encOonly$ is defined
by $\encO{\alpha} = \alpha$ and $\encO{\kappa(\tuple{\tau})} = \kappa(\encO{\tuple{\tau}})$.
The term-to-term version is defined by
\[\encO{t} =
\begin{cases}
\zof{t}
  & \text{if $t = x$ or $t$ is fluid} \\
\cst{lam}(\encO{\tau}, \encO{\encB{x}{u}})
  & \text{if $t = (\lambda x \mathbin: \tau.\; u)$ and $t$ is not fluid} \\
\cst{f}_k(\encO{\tuple{\tau}}, \encO{\tuple{u}_k})
  & \text{if $t = \cst{f}\typeargs{\tuple{\tau}}\>\tuple{u}_k$}
\end{cases}\]%
\end{definitionx}
For example, let
$s = \lambda y.\>\cst{f}\> y\> (\lambda w.\> \cst{g}\>(y\> w))$
where $y$ has type $\kappa \fun \kappa$ and $w$ has type $\kappa$.
We have
$\encB{y}{\cst{f}\> y\> (\lambda w.\> \cst{g}\>(y\> w))}
 = \cst{f}\> \db^0 (\lambda w.\> \cst{g}\> (\db^1 \> w))$
and
 $\encB{w}{\cst{g}\> (\db^1 \> w)}
 = \cst{g}\> (\db^1 \> \db^0)$.
Neither $s$ nor $\lambda w.\> \cst{g}\>(y\> w)$ are fluid.
Hence, 
we have 
$\encO{s} = \cst{lam}({\fun}(\kappa,\kappa),\allowbreak \cst{f}_2( \db^0_0, \cst{lam}(\kappa, \cst{g}_1(\db^1_1(\db^0_0))))$.

\begin{definitionx}[Derived strict term order]
Let the strict term order derived from $\fosucc$ be $\lsucc$ where $t \lsucc s$ if $\encO{t} \fosucc \encO{s}$.
\end{definitionx}
We will show that the derived $\lsucc$ fulfills all properties
of a strict term order (Definition \ref{def:strict-term-order})
if $\fosucc$ fulfills the corresponding properties on first-order terms.
For the nonstrict term order $\succsim$, we can use the reflexive closure
$\lsucceq$ of $\lsucc$.

\begin{lemmax}
\label{lem:lsucc-preserves-lfsucc-ground-properties}
Let $\fosucc$ be a strict partial order on first-order terms and $\lsucc$
the derived term order on $\beta\eta$-equivalence classes. If the restriction of
$\fosucc$ to ground terms enjoys well-foundedness, totality, the subterm
property, and compatibility with contexts
{\upshape(}\wrt\
first-order terms{\upshape)},
the restriction of $\lsucc$ to
ground terms enjoys 
well-foundedness, totality, the green subterm
property, and compatibility with green contexts
{\upshape(}\wrt\
$\beta\eta$-equivalence classes{\upshape)}.
\end{lemmax}

\begin{proof}
Transitivity and irreflexivity of $\fosucc$
imply transitivity and irreflexivity of $\lsucc$.

\medskip

\noindent
\textsc{Well-foundedness}:\enskip
If there existed an infinite %
chain $t_1 \lsucc t_2 \lsucc \cdots$ of ground terms, there would also be the
chain $\encO{t_1} \fosucc \encO{t_2} \fosucc \cdots$,
contradicting the well-foundedness of $\fosucc$ on ground $\lambda$-free terms.

\medskip

\noindent
\textsc{Totality:}\enskip
By ground totality of $\fosucc$, for any ground terms $t$ and $s$ we have
$\encO{t} \fosucc \encO{s}$,
$\encO{t} \lfprec \encO{s}$,
or $\encO{t} = \encO{s}$.
In the first two cases, it follows that $t \lsucc s$ or $t\lprec s$.
In the last case, it follows that $t = s$
because $\encOonly$ is clearly injective.

\medskip

\noindent
\textsc{Green subterm property:}\enskip
Let $s$ be a term. We show that $s \lsucceq s|_p$ by induction on $p$, where
$s|_p$ denotes the green subterm at position $p$. If $p = \varepsilon$, this is trivial.
If $p = p'.i$, we have $s \lsucceq s|_{p'}$ by the induction hypothesis.
Hence, it suffices to show that $s|_{p'} \lsucceq s|_{p'.i}$.
From the existence of the position $p'.i$,
we know that $s|_{p'}$ must be of the form
$s|_{p'} = \cst{f}\typeargs{\tuple{\tau}}\>\tuple{u}_k$.
Then $s|_{p'.i} = u_i$.
The encoding yields
$\encO{s|_{p'}} = \cst{f}_k(\encO{\tuple{\tau}},\encO{\tuple{u}_k})$
and hence $\encO{s|_{p'}} \fosucceq \encO{s|_{p'.i}}$
by the ground subterm property of $\fosucc$. Hence,
$s|_{p'} \lsucceq s|_{p'.i}$ and thus $s \lsucceq s|_{p}$.

\medskip

\noindent
\textsc{Compatibility with green contexts:}\enskip
By induction on the depth of the context,
it suffices to show that $t \lsucc s$ implies
$\cst{f}\typeargs{\tuple{\tau}}\>\tuple{u}\>t\>\tuple{v}
\lsucc
 \cst{f}\typeargs{\tuple{\tau}}\>\tuple{u}\>s\>\tuple{v}$
for all $t$, $s$, $\cst{f}$, $\tuple{\tau}$, $\tuple{u}$, and $\tuple{v}$.
This amounts to showing that
$\encO{t} \fosucc \encO{s}$ implies
$\encO{\cst{f}\typeargs{\tuple{\tau}}\>\tuple{u}\>t\>\tuple{v}}
=\cst{f}_k(\encO{\tuple{\tau}},\encO{\tuple{u}},\encO{t},\encO{\tuple{v}})
\fosucc
\cst{f}_k(\encO{\tuple{\tau}},\encO{\tuple{u}},\encO{s},\encO{\tuple{v}})
=\encO{\cst{f}\typeargs{\tuple{\tau}}\>\tuple{u}\>s\>\tuple{v}}$,
which follows directly from ground compatibility of $\fosucc$ with
contexts and the induction hypothesis.
\qedhere
\end{proof}

\begin{lemmax}
\label{lem:lsucc-preserves-lfsucc-stability-under-ground-subst}
Let $\fosucc$ be a strict partial order on first-order terms. If $\fosucc$
is stable under grounding substitutions {\upshape(}\wrt\ first-order
terms{\upshape)}, the derived term order $\lsucc$ is
stable under grounding substitutions {\upshape(}\wrt\
$\beta\eta$-equivalence classes{\upshape)}.
\end{lemmax}

\begin{proof}
Assume $s \lsucc s'$ for some terms $s$ and $s'$.
Let $\theta$ be a higher-order substitution grounding $s$ and $s'$. 
We must show $s\theta \lsucc s'\theta$.
We will define a first-order substitution $\rho$
grounding $\encO{s}$ and $\encO{s'}$
such that $\encO{s}\rho = \encO{s\theta}$ and $\encO{s'}\rho = \encO{s'\theta}$.
Since $s \lsucc s'$, we have $\encO{s} \fosucc \encO{s'}$.
By stability of $\fosucc$ under grounding substitutions,
$\encO{s}\rho \fosucc \encO{s'}\rho$.
It follows that
$\encO{s\theta} \fosucc \encO{s'\theta}$
and hence $s\theta \lsucc s'\theta$.

We define the first-order substitution $\rho$ as $\alpha\rho = \alpha\theta$ for
type variables $\alpha$ and $\zof{u}\rho = \encO{u\theta}$ for terms $u$.
Strictly speaking, the domain of a substitution must be finite, so we restrict
this definition of $\rho$ to the finitely many variables that occur
in the computation of $\encO{s}$ and $\encO{s'}$.

Clearly $\encO{\tau}\rho = \encO{\tau\theta}$ for all types $\tau$ occurring
in the computation of $\encO{s}$ and $\encO{s'}$.
Moreover, $\encO{t}\rho = \encO{t\theta}$ for all $t$ occurring
in the computation of $\encO{s}$ and $\encO{s'}$, which we show by induction on the
definition of the encoding.
If $t=x$ or if $t$ is fluid, $\encO{t}\rho = \zof{t}\rho = \encO{t\theta}$.
If $t = \cst{f}\typeargs{\tuple{\tau}}\>\tuple{u}$,
then
$\encO{t}\rho = \cst{f}_k(\encO{\tuple{\tau}}\rho,\encO{\tuple{u}}\rho)
\eqIH \cst{f}_k(\encO{\tuple{\tau}\theta},\encO{\tuple{u}\theta})
 =  \encO{\cst{f}\typeargs{\tuple{\tau}\theta}\>(\tuple{u}\theta)}
 = \encO{t\theta}$.
 If $t = (\lambda x\mathbin\oftype\tau.\;u)$ and $t$ is not fluid, then
 $\encO{t}\rho = \cst{lam}(\encO{\tau}\rho,\allowbreak\encO{\encB{x}{u}}\rho)
\eqIH \cst{lam}(\encO{\tau\theta},\allowbreak\encO{\encB{x}{u}\theta})
 = \cst{lam}(\encO{\tau\theta},\allowbreak\encO{\encB{x}{u}\theta[x\mapsto x]})
 = \encO{\lambda x\mathbin\oftype\tau\theta.\;u\theta[x\mapsto x]}
 = \encO{(\lambda x\mathbin\oftype\tau.\;u)\theta}
  = \encO{t\theta}$. \qedhere
\end{proof}

\begin{notyet}
Starting with a $\lambda$-free higher-order term order $\fosucc$, we can also
derive a relation $\lsuccsim$ which, unlike the reflexive closure $\lsucceq$
of $\lsucc$,
is precise enough to orient pairs of terms such as $y\>\cst{b} \lsuccsim
y\>\cst{a}$. The idea is that if $\cst{b} \lsuccsim \cst{a}$ and their type is
neither functional nor a type variable (which could be instantiated by a
functional type), then $\cst{b}$ and $\cst{a}$ will appear unapplied in
$(y\>\cst{b})\theta$ and $(y\>\cst{a})\theta$, as green subterms, regardless
of~$\theta$. Intuitively, in higher-order logic, $y\>\cst{b}$ can be construed
as an arbitrary term with zero or more occurrences of $\cst{b}$, and
$y\>\cst{a}$ is the same term but with zero or more occurrences of $\cst{a}$
instead.

Like $\lsucc$, the relation $\lsuccsim$ is defined via an encoding:
$t \lsuccsim s$ if and only if $\encO{t} \fosucceq \encOsub{t}{s}$.
In turn, $\encOsub{t}{s} = \encZsub{\encB{XXX}{t}}{\encB{XXX}{s}}$, where $\encBonly$
encodes the bound variables as above and
\[\encZsub{t}{s} =
\begin{cases}
\zof{y\>\tuple{t}}
  & \text{if $t = y\>\tuple{t}_n$, $s = y\>\tuple{s}_n$, and each $(t_i, s_i)$ satisfies ($*$) below} \\
\cst{lam}\>\encZ{\tau}\> \encZsub{v}{u}
  & \text{if $t = (\lambda x \mathbin: \tau.\; v)$, $s = (\lambda x \mathbin: \tau.\> u)$, and $t$ and $s$ are not fluid} \\
\cst{f}\>\encZ{\tuple{\tau}}\> \tuple{u}'_n
  & \text{if $t = \cst{f}\typeargs{\tuple{\tau}}\>\tuple{v}_n$, $s = \cst{f}\typeargs{\tuple{\tau}}\>\tuple{u}_n$, and $u'_i = \encZsub{v_i}{u_i}$ for all $i$} \\
\encZ{s}
  & \text{otherwise}
\end{cases}\]%
The condition ($*$) on $(t_i, s_i)$ is that (1) $t_i
\lsuccsim s_i$ and $t_i$'s type is neither functional nor a type
variable or (2) $t_i = s_i$. Notice the (well-founded) mutual dependency
between $\lsuccsim$ and~$\encZsubonly{t}$.

The difference between $\encOsubonly{t}$ and $\encOonly$, and hence between
$\lsuccsim$ and $\lsucceq$, concerns applied variables.
If the subterm $y\>\cst{b}$ occurs in $t$ where $s$ has $y\>\cst{a}$, and it
can be determined that $\cst{b} \lsuccsim \cst{a}$, we proceed as if
$y\>\cst{a}$ had been $y\>\cst{b}$, using the same variable $\zof{y\>\cst{b}}$ to
represent both $y\>\cst{b}$ and~$y\>\cst{a}$. For example,
we have $\cst{g}\> (y\> \cst{b}\> \cst{f}) \lsuccsim \cst{g}\> (y\> \cst{a}\>
\cst{f})$ and $\cst{h}\> \cst{b}\> (y\> \cst{b}) \lsuccsim \cst{h}\> \cst{a}\>
(y\> \cst{a})$ because
\[\begin{array}{r@{}c@{}r@{\>}c@{\>}l@{}c@{}l}
\encO{\cst{g}\> (y\> \cst{b}\> \cst{f})}
& {} = {} &
\cst{g}\> \zof{y\> \cst{b}\> \cst{f}}
& = &
\cst{g}\> \zof{y\> \cst{b}\> \cst{f}}
& {} = {} &
\encOsub{\cst{g}\> (y\> \cst{b}\> \cst{f})}{\cst{g}\> (y\> \cst{a}\> \cst{f})}
\\
\encO{\cst{h}\> \cst{b}\> (y\> \cst{b})}
& {} = {} &
\cst{h}\> \cst{b}\> \zof{y\> \cst{b}}
& {} \fosucc\negvthinspace {} &
\cst{h}\> \cst{a}\> \zof{y\> \cst{b}}
& {} = {} &
\encOsub{\cst{h}\> \cst{b}\> (y\> \cst{b})}{\cst{h}\> \cst{a}\> (y\> \cst{a})}
\end{array}\]
On the other hand,
$\cst{g}\> (y\> \cst{b}\> \cst{f}) \not\lsucceq \cst{g}\> (y\> \cst{a}\>
\cst{f})$ and $\cst{h}\> \cst{b}\> (y\> \cst{b}) \not\lsucceq \cst{h}\> \cst{a}\>
(y\> \cst{a})$.

\begin{lemmax}
\label{lem:lsuccsim-preserves-lfsucc-stability-under-ground-subst}
Let $\fosucc$ be a strict partial order on $\lambda$-free terms that
is stable under grounding substitutions
and whose ground restriction is compatible with green contexts. If $t \lsuccsim
s$, then $t\theta \lsucceq s\theta$ for all grounding substitutions $\theta$.
\end{lemmax}

\begin{proof}
\looseness=-1
By induction on $t$.
Let $\rho$ be the grounding substitution based on $\theta$ and $\{t, s\}$
defined in the proof of
Lemma~\ref{lem:lsucc-preserves-lfsucc-stability-under-ground-subst}.
From the hypothesis $t \lsuccsim s$ (i.e., $\encO{t} \fosucceq \encOsub{t}{s}$),
we have $\encO{t}\rho \fosucceq \encOsub{t}{s}\rho$ by stability of $\fosucc$
under grounding substitutions. The crux is to show that
$\encOsub{t}{s}\rho \fosucceq \encO{s}\rho$. The rest follows
easily: From $\encO{t}\rho \fosucceq \encOsub{t}{s}\rho \fosucceq \encO{s}\rho$,
$\encO{t\theta} \fosucceq \encO{s\theta}$ follows by construction of $\rho$,
and thus $t\theta \lsucceq s\theta$ by injectivity of $\encOonly$.

We show $\encOsub{t}{s}\rho \fosucceq \encO{s}\rho$ by exploiting
ground compatibility of $\fosucc$ with green contexts.
The difference between $\encOsub{t}{s}$ and $\encO{s}$ is that the former may
have $\zof{y\>\tuple{t}}$ where the latter has $\zof{y\>\tuple{s}}$. We call
each such pair $(t_i, s_i)$ with $t_i \not= s_i$ a \emph{mismatch}.

After applying $\rho$, the terms $\encOsub{t}{s}\rho$ and
$\encO{s}\rho$ have the form $\subterm{u}{t'_1, \ldots, t'_n}$ and
$\subterm{u}{s'_1, \ldots, s'_n}$, respectively, where the holes in
$\subterm{u}{\phantom{i}}$ correspond to positions at or below variables
$\zof{y\>\tuple{t}}$ in $\encOsub{t}{s}$ and, in parallel, $\zof{y\>\tuple{s}}$ in
$\encO{s}$. Clearly, each pair $(t'_{\!j}, s'_{\!j})$ must be equal to a pair
$(\encO{t_i\theta},\allowbreak \encO{s_i\theta})$ of argument tuples of some
applied variable.
By~($*$), all mismatches are of nonfunctional, nonvariable types, and this
property is preserved by $\theta$ and $\encOonly$. Hence, all the holes in
$\subterm{u}{\phantom{i}}$ are located at green positions.

For each $i$, we have $t_i \lsuccsim s_i$ by ($*$) and hence $t_i\theta
\lsucceq s_i\theta$ by the induction hypothesis. Thus, $t'_{\!j} \fosucceq
s'_{\!j}$ for all $j$. Using these inequalities in turn together with
ground compatibility with green contexts, we form the following transitive
chain, thereby resolving the crux:

\vskip\abovedisplayskip

\noindent\hbox{}%
\phantom{\squareforqed}\hfill
\raise1.35\baselineskip\hbox{$\underbrace{\greensubterm{u}{t'_1, \ldots, t'_n}}_{\encOsub{t}{s}\rho}
\>\fosucceq\>
\greensubterm{u}{s'_1, t'_2, \ldots, t'_n}
\>\fosucceq\,
\cdots
\,\fosucceq\>
\greensubterm{u}{s'_1, \ldots, s'_{n-1}, t'_n}
\>\fosucceq\>
\underbrace{\greensubterm{u}{s'_1, \ldots, s'_n}}_{\encO{s}\rho}$}%
\hfill\squareforqed
\end{proof}
\end{notyet}

\section{Refutational Completeness}
\label{sec:refutational-completeness}

Besides soundness, the most important property of the Boolean-free
$\lambda$-superposition calculus introduced in \Section~\ref{sec:the-calculus}
is refutational completeness.
We will prove static and dynamic refutational completeness of $\HInf$ \wrt\ $(\HRedI, \HRedC)$, which is defined as follows:
\begin{definitionx}[Static refutational completeness]
  Let $\Inf$ be an inference system and let $(\RedI, \RedC)$ be a redundancy criterion.
  The inference system $\Inf$ is \emph{statically refutationally complete} \wrt\ $(\RedI, \RedC)$ if
  we have $N \models \bot$ if and only if $\bot \in N$
  for every clause set $N$ that is saturated \wrt\ $\Inf$ and $\RedI$.
\end{definitionx}
\begin{definitionx}[Dynamic refutational completeness]
Let $\Inf$ be an inference system and let $(\RedI, \RedC)$ be a redundancy criterion.
Let $(N_i)_i$ be a finite or infinite sequence over sets of clauses. 
Such a sequence is
a \emph{derivation} if
$N_i \setminus N_{i+1} \subseteq \RedC(N_{i+1})$ for all $i$.
It is \emph{fair} if all $\Inf$-inferences from clauses in
the limit inferior $\bigcup_i \bigcap_{\!j \geq i} N_{\!j}$ are contained in
$\bigcup_i \RedI(N_i)$.
The inference system $\Inf$ is \emph{dynamically refutationally complete} \wrt\ $(\RedI, \RedC)$ if
for every fair derivation $(N_i)_i$ such that $N_0 \models \bot$,
we have $\bot \in N_i$ for some $i$.
\label{def:dyn-complete}
\end{definitionx}

\oursubsection{Outline of the Proof}

The proof proceeds in three steps, corresponding to the three levels $\GF$, $\GH$, and $\HH$ introduced in \Section~\ref{ssec:the-redundancy-criterion}:
\begin{enumerate}
\item We use Bachmair and Ganzinger's work on the refutational completeness of standard
(first-order) superposition~\cite{bachmair-ganzinger-1994}
to prove static refutational completeness of $\GFInf$.
\item From the first-order model constructed in Bachmair and Ganzinger's proof,
we derive a clausal higher-order model and thus prove static refutational completeness of $\GHInf$.
\item We use the saturation framework by Waldmann et al.~\cite{waldmann-et-al-2020-saturation} to lift the static refutational completeness of $\GHInf$
to static and dynamic refutational completeness of $\HInf$.
\end{enumerate}

In the first step,
since the inference system $\GFInf$ is standard ground superposition,
we can make use of Bachmair and Ganzinger's
results. %
Given a saturated clause set $N\subseteq\CGF$ with $\bot\not\in N$,
Bachmair and Ganzinger prove refutational completeness by constructing
a term rewriting system $R_N$ and showing that it can be
viewed as an interpretation that is a model of $N$.
This first step deals exclusively with ground first-order clauses.

In the second step, we derive refutational completeness of $\GHInf$.
Given a saturated clause set $N\subseteq\CGH$ with $\bot\not\in N$,
we use the first-order model $R_{\floor{N}}$ of $\floor{N}$ constructed in the first step
to derive a clausal higher-order interpretation that is a model of $N$.
Under the encoding $\flooronly$,
occurrences of the same symbol with different numbers of arguments are
regarded as different symbols---e.g.,
$\floor{\cst{f}}=\cst{f}_0$
and
$\floor{\cst{f}\>\cst{a}}=\cst{f}_1(\cst{a}_0)$.
All $\lambda$-expressions $\lambda x.\>t$ are regarded as uninterpreted
symbols $\cst{lam}_{\lambda x.\>t}$.
The difficulty is to
construct a higher-order interpretation
that merges the first-order denotations of all $\cst{f}_i$ into a single
higher-order denotation of $\cst{f}$
and to show that the symbols $\cst{lam}_{\lambda x.\>t}$ behave like
$\lambda x.\>t$. This step relies on saturation
\wrt\ the \infname{GArgCong} rule---which connects a
term of functional type with its value when applied to an argument~$x$---and on
the presence of the extensionality rule \infname{GExt}.

In the third step,
we employ the saturation framework by Waldmann et al.~\cite{waldmann-et-al-2020-saturation}%
,
which is based on Bachmair and Ganzinger's framework~\cite[\Section~4]{bachmair-ganzinger-2001-resolution},
to prove refutational completeness of $\HInf$.
Both saturation frameworks help calculus designers prove static and dynamic refutational completeness of nonground calculi.
In addition, the framework by Waldmann et al.\ explicitly supports the redundancy criterion defined in \Section~\ref{ssec:the-redundancy-criterion},
which can be used to justify the deletion of subsumed clauses.
Moreover, their saturation framework provides completeness theorems for prover architectures, such as the DISCOUNT loop.

The main proof obligation we must discharge to use the framework is that there
should exist nonground inferences in $\HInf$ corresponding to all nonredundant inferences in $\GHInf$.
We face two specifically higher-order difficulties.
First, in standard superposition, we can avoid \infname{Sup} inferences into
  variables~$x$ by exploiting the clause order's compatibility with contexts:
  If $t' \prec t$, we have $C\{x \mapsto\nobreak t'\} \prec C\{x \mapsto t\}$,
  which allows us to show that \infname{Sup} inferences into
  variables are redundant. This technique fails for higher-order variables~$x$
  that occur applied in~$C$, because the order lacks compatibility
  with arguments. This is why our \infname{Sup} rule must perform some inferences
  into variables.
The other difficulty also concerns applied variables. We must show that any
nonredundant \infname{Sup} inference in level $\GH$ into a position corresponding
  to a fluid term or a deeply occurring variable in level $\HH$ can be
  lifted to a \infname{FluidSup} inference. This involves showing that the $z$
  variable in \infname{FluidSup} can represent arbitrary contexts around a
  term~$t$.

For the entire proof of refutational completeness, $\beta\eta$-normalization is the
proverbial dog that did not bark. On level $\GH$, the rules
\infname{Sup}, \infname{ERes}, and \infname{EFact} preserve
$\eta$-short $\beta$-normal form, and so does first-order term rewriting.
Thus, we can completely ignore $\rewrite_\beta$ and
$\rewrite_\eta$.
On level $\HH$, instantiation can cause $\beta$- and $\eta$-reduction, but
this poses no difficulties thanks to the clause order's stability
under grounding substitutions.

\oursubsection{The Ground First-Order Level}

We use Bachmair and Ganzinger's results on standard superposition~\cite{bachmair-ganzinger-1994}
to prove refutational completeness of $\GF$. In the subsequent steps, we will also make use of specific properties
of the model Bachmair and Ganzinger construct.
The basis of Bachmair and Ganzinger's proof is that
a term rewriting system $R$ defines an interpretation $\TGF/R$
such that for every ground equation $s
\eq t$, we have $\TGF/R \models s \eq t$ if and only if $s
\leftrightrewrite_R^* t$.
Formally, $\TGF/R$ denotes the
monomorphic first-order interpretation
whose universes $\ufo_\tau$ consist of the $R$-equivalence
classes over $\TGF$ containing terms of type $\tau$.
The interpretation $\TGF/R$ is term-generated---that is,
for every element $a$ of the universe of this interpretation and for any
valuation $\xi$, there exists
a ground term~$t$ such that $\interpret{t}{\TGF/R}{\xi} = a$.
To lighten notation, we will write $R$ to refer to both the term rewriting
system $R$ and the interpretation $\TGF/R$.

The term rewriting system is constructed as follows:
\begin{definitionx}
Let $N\subseteq\CGF$.
We first define sets of rewrite rules $E_N^C$ and $R_N^C$ for all $C\in N$ by induction on the clause order.
Assume that $E_N^D$ has already been defined for all $D \in N$
such that $D \prec C.$ Then $R_N^C = \bigcup_{D \prec C} E_N^D.$
Let $E_N^C=\{s \rewrite t\}$ if the following conditions are met:\
\begin{enumerate}[(a)]
	\item $C = C' \lor s \eq t$; \label{cond:C-eq-C'-st}
	\item $s \eq t$ is $\succsim$-maximal in $C$; \label{cond:st-strictly-max}
	\item $s \succ t$; \label{cond:s-gt-t}
	\item $C'$ is false in $R_N^C$; \label{cond:C'-false}
	\item $s$ is irreducible \wrt\ $R_N^C.$ \label{cond:s-irred}
\end{enumerate}
Then $C$ is said to \emph{produce} $s \rewrite t$. %
Otherwise, $E_N^C = \emptyset$.
Finally, $R_N = \bigcup_{D} E_N^D.$
\end{definitionx}
Based on Bachmair and Ganzinger's work, 
Bentkamp et al.\ \cite[Lemma~4.2 %
and Theorem~4.3] %
{bentkamp-et-al-lfhosup-arxiv} prove the following properties of $R_N$:

\begin{lemmax} \label{lem:productive-clauses}
  Let $\bot\not\in N$ and $N\subseteq\CGF$ be saturated \wrt\ $\GFInf$
  and $\GFRedI$. If $C = C' \lor s \eq t \in N$ produces $s \rewrite t$,
  then $s \eq t$ is strictly $\succeq$-eligible in $C$ and $C'$ is false in $R_N$.
\end{lemmax}

\begin{theoremx}[Ground first-order static refutational completeness]
  The inference system $\GFInf$ is statically refutationally complete \wrt\ $(\GFRedI, \GFRedC)$.
  More precisely, if $N\subseteq\CGF$ is a clause set saturated \wrt\ $\GFInf$
  and $\GFRedI$ such that $\bot\not\in N$,
  then $R_N$ is a model of $N$.
  \label{thm:GF-refutational-completeness}
\end{theoremx}

\oursubsection{The Ground Higher-Order Level}
\label{ssec:the-ground-higher-order-level}

In this subsection, let $\GHSel$ be a selection function on $\CGH$,
let $N\subseteq\CGH$ be a clause set saturated \wrt\
$\GHInf^\GHSel$ and $\GHRedI^\GHSel$ such that $\bot\not\in N$. Clearly, $\floor{N}$ is then saturated \wrt\
$\smash{\GFInf^{\floor{\GHSel}}}$ and $\smash{\GFRedI^{\floor{\GHSel}}}$.

We abbreviate $R_{\floor{N}}$ as $\RfN$. Given two terms $s,t\in\TGH$, we write $\eqR{s}{t}$ to abbreviate
$\RfN\models \floor{s}\eq\floor{t}$,
which is equivalent to
$\interpretfo{\floor{s}}{} = \interpretfo{\floor{t}}{}$.

\begin{lemmax}\label{lem:arg-cong-ext}
	For all terms $t,s\oftype\tau\fun\upsilon$ in $\TGH$, the following statements are equivalent:
	\begin{enumerate}
		\item[\upshape 1.]
		  $\eqR{t}{s}$;
		\item[\upshape 2.]
		  $\eqR{t\>(\diff\>t\>s)}{s\>(\diff\>t\>s)}$;
		\item[\upshape 3.]
		  $\eqR{t\>u}{s\>u}$ for all $u\in\TGH$.
	\end{enumerate}
\end{lemmax}

\begin{proof}	
	(3)\,$\Rightarrow$\,(2): Take $u := \diff\>t\>s$.

	\medskip\noindent
  (2)\,$\Rightarrow$\,(1):
  Since $N$ is saturated, the \infname{GExt} inference that generates
  the clause $C = t\>(\diff\>t\>s) \noteq s\>(\diff\>t\>s) \llor t \eq s$ is
  redundant---i.e., $C \in N \ccup \GHRedC(N)$---and hence
  $\RfN\models\floor{C}$ by 
  Theorem~\ref{thm:GF-refutational-completeness} 
  and
  the assumption that $\bot\not\in N$.
	Therefore, it follows from
	$\eqR{t\>(\diff\>t\>s)}{s\>(\diff\>t\>s)}$ that $\eqR{t}{s}$.

	\medskip\noindent
	(1)\,$\Rightarrow$\,(3):
	We assume that $\eqR{t}{s}$---i.e., $\floor{t} \leftrightrewrite^*_{\RfN}\floor{s}$.
	By induction on the number of rewrite steps between $\floor{t}$ and $\floor{s}$
	and by transitivity of $\eqR{}{}$,
	it suffices to show
	that $\floor{t} \rewrite_{\RfN} \floor{s}$ implies $\eqR{t\>u}{s\>u}$.
  If the rewrite step $\floor{t} \rewrite_{\RfN} \floor{s}$ is not at the
  top level, then neither $\betanf{s}$ nor $\betanf{t}$ can be $\lambda$-expressions.
  Therefore, $(\betanf{s})\>(\betanf{u})$ and $(\betanf{t})\>(\betanf{u})$ are in
  $\eta$-short $\beta$-normal form,
  and there is an analogous rewrite step  $\floor{t\>u} \rewrite_{\RfN} \floor{s\>u}$
  using the same rewrite rule.
  It follows that $\eqR{t\>u}{s\>u}$.
  If the rewrite step $\floor{t} \rewrite_{\RfN} \floor{s}$ is at the
	top level,
	$\floor{t}\rewrite \floor{s}$ must be a rule of $\RfN$.
	This rule must originate from a productive clause of the form
  $\floor{C} = \floor{C' \llor t \eq s}$.
  By Lemma~\ref{lem:productive-clauses},
  $\floor{t \eq s}$ is
  strictly $\succeq$-eligible in $\floor{C}$ \wrt\ $\floor{\GHSel}$,
  and hence $t \eq s$ is
  strictly $\succeq$-eligible in $C$ \wrt\ $\GHSel$.
  Thus, the following
  \infname{GArgCong} inference $\iota$ is applicable:
	\[
	\namedinference{GArgCong}
	{C' \llor t \eq s}
	{C' \llor t\>u \eq s\>u}
  \]
  By saturation,
  $\iota$ is redundant \wrt\ $N$---i.e., $\concl(\iota)\in N \ccup \GHRedC(N)$.
  By 
Theorem~\ref{thm:GF-refutational-completeness} 
  and the assumption
  that $\bot\not\in N$,
  $\floor{\concl(\iota)}$ is then true in $\RfN$.
  By Lemma~\ref{lem:productive-clauses}, 
  $\floor{C'}$ is false in $\RfN$. Therefore,
  $\floor{t\>u \eq s\>u}$ must be true in $\RfN$. \qedhere
\end{proof}

\begin{lemmax} \label{lem:subst-congruence}
	Let $s\in\THH$ and $\theta$, $\theta'$ grounding
	substitutions such that $\eqR{x\theta}{x\theta'}$ for all variables~$x$
	and $\alpha\theta = \alpha\theta'$ for all type variables $\alpha$.
	Then $\eqR{s\theta}{s\theta'}$.
\end{lemmax}
\begin{proof}
  In this proof, we work directly on $\lambda$-terms. To prove the
  lemma, it suffices to prove it
  for any $\lambda$-term $s$.
  Here, for $\lambda$-terms $t_1$ and $t_2$, the notation
  $\eqR{t_1}{t_2}$ is to be read as $\eqR{\betanf{t_1}}\betanf{{t_2}}$
  because $\flooronly$ is only defined on $\eta$-short $\beta$-normal terms.

	\newcommand{\choice}{\oplus}%
  \medskip\noindent
	\textsc{Definition}\enskip  %
  We extend the syntax of $\lambda$-terms with a new
  polymorphic function symbol $\choice\oftype\forallty{\alpha}\alpha\fun\alpha\fun\alpha$.
  We will omit its type argument. It is equipped with two reduction rules:
  $\choice\>t\>s \rewrite t$ and $\choice\>t\>s \rewrite s$.
  A \emph{$\beta\choice$-reduction step} 
  is either a rewrite step following one of these rules or a $\beta$-reduction step.
  
  \medskip\noindent
  The computability path order $\succ_\cst{CPO}$ \cite{blanqui-et-al-2015} guarantees that
  \begin{itemize}
    \item $\choice\>t\>s \succ_\cst{CPO} s$ by applying rule $@\rhd$;
    \item $\choice\>t\>s \succ_\cst{CPO} t$ by applying rule $@\rhd$ twice;
    \item $(\lambda x.\>t)\>s \succ_\cst{CPO} t[x\mapsto s]$ by applying rule $@\beta$.
 \end{itemize}
  Since this order is moreover monotone, it decreases with $\beta\choice$-reduction steps.
  The order is also well founded; thus, $\beta\choice$-reductions terminate.
  And since the $\beta\choice$-reduction steps describe a finitely branching term rewriting
  system, by K\H{o}nig's lemma \cite{koenigs-lemma-1927}, there is a maximal number of
  $\beta\choice$-reduction steps
  from each $\lambda$-term.

  \medskip\noindent
	\textsc{Definition}\enskip %
  A $\lambda$-term is \emph{term-ground} if it does not contain free term variables.
	It may contain polymorphic type arguments.

	\newcommand{\choicesubst}{\sigma}%
	\newcommand{\livesize}{\mathscr{S}}%
  \medskip\noindent
	\textsc{Definition}\enskip %
  We introduce an auxiliary function $\livesize$
	that essentially measures the size of a $\lambda$-term but assigns a size of $1$ to
  term-ground $\lambda$-terms.
\[\livesize(s) =
\begin{cases}
1
  & \text{if $s$ is term-ground or is a bound or free variable or a symbol} \\
1 + \livesize(t)
  & \text{if $s$ is not term-ground and has the form $\lambda x.\>t$} \\
\livesize(t) + \livesize(u)
  & \text{if $s$ is not term-ground and has the form $t\>u$}
\end{cases}\]%
    We prove $\eqR{s\theta}{s\theta'}$ by well-founded
    induction on $s$, $\theta$, and $\theta'$ using
    the left-to-right lexicographic order on the triple
	$\bigl(n_1(s), n_2(s), n_3(s)\bigr)\in\mathbb{N}^3$, where
	\begin{itemize}
		\item \label{mea:bi-red} $n_1(s)$ is the maximal number of
            $\beta\choice$-reduction steps starting from $s\choicesubst$, where
            $\choicesubst$ is the substitution mapping each term variable $x$
            to  $\choice\>x\theta\>x\theta'$;
        \item $n_2(s)$ is the number of free term variables occurring more than once in $s$;
		\item $n_3(s) = \livesize(s)$.
	\end{itemize}

	\medskip\noindent
	\textsc{Case 1:}\enskip
	The $\lambda$-term $s$ is term-ground. Then the lemma is trivial.

	\medskip\noindent
	\textsc{Case 2:}\enskip
	The $\lambda$-term $s$ contains $k \geq 2$ free term variables. Then we can apply the induction
	hypothesis twice and use the transitivity of $\eqR{}{}$ as follows. Let $x$
	be one of the free term variables in $s$. Let $\rho = \{x \mapsto x\theta\}$ the
	substitution that maps $x$ to $x\theta$ and ignores all other variables. Let
	$\rho' = \theta'[x\mapsto x]$.
	
	We want to invoke the induction hypothesis on $s\rho$ and $s\rho'$. This is
	justified because $s\choicesubst$ $\choice$-reduces to $s\rho\choicesubst$ and to
  $s\rho'\choicesubst$. 
  These $\choice$-reductions have at least one step because $x$ occurs in $s$ and $k \geq 2$.
  Hence, $n_1(s)>n_1(s\rho)$ and $n_1(s)>n_1(s\rho')$.
	
	This application of the induction hypothesis gives us
	$\eqR{s\rho\theta}{s\rho\theta'}$ and $\eqR{s\rho'\theta}{s\rho'\theta'}$.
  Since $s\rho\theta = s\theta$ and $s\rho'\theta' = s\theta'$, 
  this is equivalent to 
  $\eqR{s\theta}{s\rho\theta'}$ and $\eqR{s\rho'\theta}{s\theta'}$.
  Since moreover $s\rho\theta' = s\rho'\theta$, we have $\eqR{s\theta}{s\theta'}$ by
  transitivity of $\eqR{}{}$. 
  The following illustration visualizes the above argument:
	\[
	\begin{tikzpicture}
		\matrix (m) [matrix of math nodes,row sep=1.5em,column sep=0.1em,minimum width=1em]
		{
		   		   & s\rho &    &    &    & s\rho' &          \\
		   s\theta & \underset{\scriptscriptstyle\text{IH}}{\eqR{}{}} & s\rho\theta' & = & s\rho'\theta & \underset{\scriptscriptstyle\text{IH}}{\eqR{}{}} & s\theta' \\};
		\draw[-stealth] (m-1-2) edge node [left] {$\theta$\,} (m-2-1);
		\draw[-stealth] (m-1-2) edge node [right] {\,\,$\theta'$} (m-2-3);
		\draw[-stealth] (m-1-6) edge node [left] {$\theta$\,} (m-2-5);
		\draw[-stealth] (m-1-6) edge node [right] {\,\,$\theta'$} (m-2-7);
	\end{tikzpicture}
	\]

  \vskip-\baselineskip %
  \medskip\noindent
	\textsc{Case 3:}\enskip
	The $\lambda$-term $s$ contains a free term variable that occurs more than once. Then we rename
	variable occurrences apart by replacing each occurrence of each free term variable $x$ by
	a fresh variable $x_i$, for which we define $x_i\theta = x\theta$ and
	$x_i\theta' = x\theta'$. Let $s'$ be the resulting $\lambda$-term. Since $s\choicesubst
    = s'\choicesubst$, we have $n_1(s)=n_1(s')$. All free term variables occur only once in $s'$.
    Hence, $n_2(s)>0=n_2(s')$. Therefore, we can invoke the induction
	hypothesis on $s'$ to obtain $\eqR{s'\theta}{s'\theta'}$. Since $s\theta =
	s'\theta$ and $s\theta' = s'\theta'$, it follows that
	$\eqR{s\theta}{s\theta'}$.
	
	\medskip\noindent
	\textsc{Case 4:}\enskip
	The $\lambda$-term $s$ contains only one free term variable $x$, which occurs exactly once.
	
	\medskip\noindent
	\textsc{Case 4.1:}\enskip
	The $\lambda$-term $s$ is of the form $\cst{f}\typeargs{\tuple{\tau}}\>\tuple{t}$
	for some symbol~$\cst{f}$, some types $\tuple{\tau}$, and some
    $\lambda$-terms~$\tuple{t}$. Then let $u$ be the $\lambda$-term in
    $\tuple{t}$ that contains $x$.
    We want to apply the induction hypothesis to $u$, which can be
	justified as follows. Consider the longest sequence of
	$\beta\choice$-reductions from $u\choicesubst$. This sequence can be
	replicated inside $s\choicesubst=(\cst{f}\typeargs{\tuple{\tau}}\>\tuple{t})\choicesubst$.
	Therefore,
	the longest sequence of $\beta\choice$-reductions from $s\choicesubst$ is at
	least as long---i.e., $n_1(s)\geq n_1(u)$. Since both $s$ and $u$ have
	only one free term variable occurrence, we have $n_2(s) = 0 =
	n_2(u)$. But $n_3(s) > n_3(u)$ because $u$ is a term-nonground subterm of
	$s$.
	
	Applying the induction hypothesis gives us $\eqR{u\theta}{u\theta'}$. By
	definition of $\flooronly$, we have
  $\floor{(\cst{f}\typeargs{\tuple{\tau}}\>\tuple{t})\theta}
  = \cst{f}_m^{\smash{\tuple{\tau}\theta}}\>\floor{\tuple{t}\theta}$
  and analogously for $\theta'$,
	where $m$ is the length of $\tuple{t}$.
	By congruence of $\eq$ in first-order logic, it follows that $\eqR{s\theta}{s\theta'}$.
	
	\medskip\noindent
	\textsc{Case 4.2:}\enskip
	The $\lambda$-term $s$ is of the form $x\>\tuple{t}$ for some $\lambda$-terms $\tuple{t}$. Then we
	observe that, by assumption, $\eqR{x\theta}{x\theta'}$. By applying
	Lemma~\ref{lem:arg-cong-ext} repeatedly, we have
	$\eqR{x\theta\>\tuple{t}}{x\theta'\>\tuple{t}}$. Since $x$ occurs only once,
	$\tuple{t}$ is term-ground and hence $s\theta = x\theta\>\tuple{t}$ and $s\theta'
	= x\theta'\>\tuple{t}$. Therefore, $\eqR{s\theta}{s\theta'}$.

	\medskip\noindent
	\textsc{Case 4.3:}\enskip
	The $\lambda$-term $s$ is of the form $\lambda z.\>u$ for some $\lambda$-term $u$. Then we observe that to prove
	$\eqR{s\theta}{s\theta'}$, it suffices to show that
	$\eqR{s\theta\>(\diff\>s\theta\>s\theta')}{s\theta' \>(\diff\>s\theta\>s\theta')}$
	by Lemma~\ref{lem:arg-cong-ext}. Via $\beta\eta$-conversion, this is equivalent to
	$\eqR{u\rho\theta}{u\rho\theta'}$ where $\rho = \{z\mapsto
	\diff\>(\betanf{s\theta})\>(\betanf{s\theta'})\}$.
	To prove $\eqR{u\rho\theta}{u\rho\theta'}$, we apply the
	induction hypothesis on $u\rho$.
	
	It remains to show that the induction hypothesis is applicable on $u\rho$.
	Consider the longest sequence of $\beta\choice$-reductions from
	$u\rho\choicesubst$. Since $z\rho$ starts with the $\diff$ symbol, $z\rho$
	will not cause more $\beta\choice$-reductions than $z$. Hence, the same
	sequence of $\beta\choice$-reductions can be applied inside $s\choicesubst =
    (\lambda z.\>u)\choicesubst$, proving that $n_1(s) \geq n_1(u\rho)$.
    Since both $s$ and $u\rho$ have	only one free term variable occurrence,
	$n_2(s) = 0 = n_2(u\rho)$. But $n_3(s) = \livesize(s) = 1 + \livesize(u)$
	because $s$ is term-nonground. Moreover,
	$\livesize(u)\geq\livesize(u\rho)=n_3(u\rho)$ because $\rho$ replaces a
	variable by a ground $\lambda$-term. Hence, $n_3(s)
	> n_3(u\rho)$, which justifies the application of the induction hypothesis.
	
	\medskip\noindent
	\textsc{Case 4.4:}\enskip
	The $\lambda$-term $s$ is of the form $(\lambda z.\>u)\>t_0\>\tuple{t}$ for some $\lambda$-terms $u$,
	$t_0$, and $\tuple{t}$.
	We apply the induction hypothesis on $s' = u\{z \mapsto t_0\}\>\tuple{t}$.
	To justify it, consider the longest sequence of $\beta\choice$-reductions from
	$s'\choicesubst$. Prepending the reduction $s\choicesubst \rewrite_\beta
	s'\choicesubst$ to it gives us a longer sequence from $s\choicesubst$. Hence,
	$n_1(s) > n_1(s')$.
	The induction hypothesis gives us $\eqR{s'\theta}{s'\theta'}$. Since
	$\eqR{}{}$ is invariant under $\beta$-reductions, it follows that
	$\eqR{s\theta}{s\theta'}$.	
	\qedhere
\end{proof}

We proceed by defining a higher-order interpretation
$\IIIho=(\uho,\IItyho,\iho,\allowbreak\lho)$ derived from~$\RfN$.
The interpretation $\RfN$ is
an interpretation in monomorphic first-order logic.
Let $\ufo_\tau$ be its universe for type $\tau$
and $\ifo$ its interpretation function.

To illustrate the construction, we will employ the following running example.
Let the higher-order signature be $\Sigmaty = \{\iota, \fun\}$ %
and $\Sigma = \{\cst{f}\oftype \iota \fun \iota,\> \cst{a} \oftype \iota,\> \cst{b} \oftype \iota\}$.
The first-order signature accordingly consists of $\Sigmaty$ and 
$\Sigma_\GF = \{\cst{f}_0, \cst{f}_1, \cst{a}_0, \cst{b}_0\} \cup \{ \cst{lam}_{\lambda x.\>t} \mid \lambda x.\>t \in \TGH\}$.
We write $[t]$ for the equivalence class of $t\in\TGF$ modulo $\RfN$.
We assume that $[\cst{f}_0] = [\cst{lam}_{\lambda x.\>x}]$,
$[\cst{a}_0] = [\cst{f}_1(\cst{a}_0)]$,
$[\cst{b}_0] = [\cst{f}_1(\cst{b}_0)]$,
and that $\cst{f}_0$, $\cst{lam}_{\lambda x.\>\cst{a}}$,
$\cst{lam}_{\lambda x.\>\cst{b}_0}$, $\cst{a}_0$, and $\cst{b}_0$
are in disjoint equivalence classes.
Hence, $\UU_{\iota\fun\iota} = 
\{
  [\cst{f}_0],
  [\cst{lam}_{\lambda x.\>\cst{a}}],
  [\cst{lam}_{\lambda x.\>\cst{b}}],
  \dots
\}$ and $\UU_{\iota} = \{
  [\cst{a}_0],
  [\cst{b}_0]
\}$.

When defining the universe $\uho$ of the higher-order interpretation,
we need to ensure that it contains subsets of function spaces,
since $\IItyho(\fun)(\DD_1,\DD_2)$ must be a subset of the function space
from $\DD_1$ to $\DD_2$ for all $\DD_1,\DD_2\in\uho$.
But the first-order universes $\ufo_\tau$ consist of equivalence classes of terms from $\TGF$
\wrt\ the rewriting system $\RfN$, not of functions.

To repair this mismatch, we will define a family of functions $\EE_\tau$ that give a meaning
to the elements of the first-order universes $\ufo_{\tau}$.
We will define a domain $\dho_\tau$ for each ground type $\tau$
and then let $\uho$ be the set of all these domains $\dho_\tau$.
Thus, there will be a one-to-one
correspondence between ground types and domains.
Since the higher-order and first-order type signatures are identical
(including ${\fun}$, which is uninterpreted in first-order logic),
we can identify higher-order and first-order types.

We define $\EE_\tau$ and
$\dho_{\tau}$ in a mutual recursion and prove that $\EE_\tau$ is a bijection
simultaneously. We start with nonfunctional types $\tau$:
Let $\dho_\tau = \ufo_{\tau}$ and
let $\EE_{\tau} : \ufo_{\tau} \medrightarrow \dho_\tau$
be the identity.
We proceed by defining $\EE_{\tau\fun\upsilon}$ and $\dho_{\tau\fun\upsilon}$.
We assume that $\EE_\tau$, $\EE_\upsilon$,
$\dho_{\tau}$, and $\dho_{\upsilon}$ have already been defined and that
$\EE_\tau$, $\EE_\upsilon$ are bijections. To ensure that
$\EE_{\tau\fun\upsilon}$ will be bijective, we first define an injective
function
$\EE^0_{\tau\fun\upsilon}:(\ufo_{\tau\fun\upsilon}\medrightarrow
\dho_{\tau})\medrightarrow\dho_{\upsilon}$,
define $\dho_{\tau\fun\upsilon}$ as its image
$\EE^0_{\tau\fun\upsilon}(\ufo_{\tau\fun\upsilon})$, and finally
define $\EE_{\tau\fun\upsilon}$ as $\EE^0_{\tau\fun\upsilon}$ with its codomain
restricted to $\dho_{\tau\fun\upsilon}$:
\begin{align*}
	&\EE^0_{\tau\fun\upsilon}:\ufo_{\tau\fun\upsilon}\medrightarrow\dho_{\tau}\medrightarrow\dho_{\upsilon}\\
    &\EE^0_{\tau\fun\upsilon}(\interpretfo{\floor{s}}{})
    \bigl(\EE_{\tau}\bigl(\interpretfo{\floor{u}}{}\bigr)\bigr) =
    \EE_{\upsilon}\bigl(\interpretfo{\floor{s\>u}}{}\bigr)
\end{align*}
This is a valid definition because each element of
$\smash{\ufo_{\tau\fun\upsilon}}$ is of the form $\interpretfo{\floor{s}}{}$ for
some $s$ and each element of $\dho_{\tau}$  is of the form
$\EE_{\tau}\bigl(\interpretfo{\floor{u}}{}\bigr)$ for some $u$. This function is
well defined if it does not depend on the choice of $s$ and $u$. To show this,
we assume that there are other ground terms $t$ and $v$ such that
$\interpretfo{\floor{s}}{} = \interpretfo{\floor{t}}{}$ and
$\EE_{\tau}\bigl(\interpretfo{\floor{u}}{}\bigr) =
\EE_{\tau}\bigl(\interpretfo{\floor{v}}{}\bigr)$. Since $\EE_{\tau}$ is
bijective, we have $\interpretfo{\floor{u}}{} = \interpretfo{\floor{v}}{}$.
Using the $\eqR{}{}$-notation, we can write this as $\eqR{u}{v}$.
Applying Lemma~\ref{lem:subst-congruence} to the term $x\>y$ and the
substitutions $\{x\mapsto s, y\mapsto u\}$ and $\{x\mapsto t, y\mapsto v\}$,
we obtain $\eqR{s\>u}{t\>v}$---i.e.,
$\interpretfo{\floor{s\>u}}{} = \interpretfo{\floor{t\>v}}{}$.
Thus, $\EE^0_{\tau\fun\upsilon}$ is well defined.
It remains to show that $\EE^0_{\tau\fun\upsilon}$ is injective as a function
from $\smash{\ufo_{\tau\fun\upsilon}}$ to
$\dho_{\tau}\medrightarrow\dho_{\upsilon}$. Assume two
terms $s, t \in \TGH$ such that for all $u \in\TGH$, we have
$\interpretfo{\floor{s\>u}}{} = \interpretfo{\floor{t\>u}}{}$.
By Lemma~\ref{lem:arg-cong-ext}, it follows that  $\interpretfo{\floor{s}}{} =
\interpretfo{\floor{t}}{}$, which concludes the proof that
$\EE^0_{\tau\fun\upsilon}$ is injective.

We define $\dho_{\tau\fun\upsilon} = \EE^0_{\tau\fun\upsilon}(\ufo_{\tau\fun\upsilon})$ and
$\EE_{\tau\fun\upsilon}(a) = \EE^0_{\tau\fun\upsilon}(a)$.
This ensures that $\EE_{\tau\fun\upsilon}$ is bijective and concludes the
inductive definition of $\dho$ and $\EE$. In the following, we will usually
write $\EE$ instead of $\EE_\tau$, since the type $\tau$ is determined by the first
argument of $\EE_\tau$.

In our running example, we thus have 
$\DD_\iota = \UU_\iota = \{
  [\cst{a}_0],
  [\cst{b}_0]
\}$ 
and $\EE_\iota$ is the identity $\UU_\iota \medrightarrow \DD_\iota,\ c \mapsto c$.
The function $\EE^0_{\iota\fun\iota}$ maps $[\cst{f}_0]$ to the identity $\DD_\iota \medrightarrow \DD_\iota,\ c \mapsto c$;
it maps $[\cst{lam}_{\lambda x.\>\cst{a}}]$ to the constant function $\DD_\iota \medrightarrow \DD_\iota,\ c \mapsto [\cst{a}_0]$; and
it maps $[\cst{lam}_{\lambda x.\>\cst{b}}]$ to the constant function $\DD_\iota \medrightarrow \DD_\iota,\ c \mapsto [\cst{b}_0]$.
The swapping function $[\cst{a}_0] \mapsto [\cst{b}_0], [\cst{b}_0] \mapsto [\cst{a}_0]$ is not in the image of $\EE^0_{\iota\fun\iota}$.
Therefore, $\DD_{\iota\fun\iota}$ contains only the identity and the two constant functions, but not this swapping function.

We define the higher-order universe as $\uho= \{\dho_{\tau}\mid \tau \text{ ground}\}$.
Moreover, we define
$\IItyho(\kappa)(\dho_{\tuple{\tau}}) =
\ufo_{\kappa(\tuple{\tau})}$
for all $\kappa \in \Sigmaty$,
completing the type interpretation $\IIIty^\GH = (\uho,\IItyho)$.
We define the interpretation function as
$\iho(\cst{f},\dho_{\tuple{\upsilon}_m})
\defeq \EE(\ifo(\cst{f}_0^{\tuple{\upsilon}_m}))$
for all $\cst{f}\oftypedecl\forallty{\tuple{\alpha}_m}\tau$.

In our example, we thus have $\iho(\cst{f}) = \EE([\cst{f}_0])$, 
which is the identity on $\DD_\iota \medrightarrow \DD_\iota$.

Finally, we need to define the designation function $\lho$, which takes a
valuation $\xi$ and a $\lambda$-expression as arguments. Given a valuation
$\xi$, we choose a grounding substitution~$\theta$ such that
$\dho_{\alpha\theta}=\xi(\alpha)
\text{ and }
\EE(\interpretfo{\floor{x\theta}}{}) = \xi(x)$
for all type variables $\alpha$ and all variables $x$.
Such a substitution can be constructed as follows:
We can fulfill the first equation in a unique way
because there is a one-to-one correspondence between ground types and domains.
Since $\EE^{-1}(\xi(x))$ is an element of a first-order universe and $\RfN$ is term-generated, there exists
a ground term $t$ such that $\interpretfoxi{t}=\EE^{-1}(\xi(x))$.
Choosing one such $t$ and defining $x\theta = \ceil{t}$ gives us a grounding
substitution $\theta$ with the desired property.

We define
$\lho(\xi,(\lambda x.\>t)) =
\EE(\interpretfo{\floor{(\lambda x.\>t)\theta}}{})$. To prove that
this is well defined, we assume that there exists another substitution~$\theta'$
with the properties $\smash{\dho_{\alpha\theta'}}=\xi(\alpha)$ for all $\alpha$
and $\EE(\interpretfo{\floor{x\theta'}}{}) = \xi(x)$ for
all $x$. Then
we have $\alpha\theta = \alpha\theta'$ for all $\alpha$
due to the one-to-one correspondence between domains and ground types.
We have $\interpretfo{\floor{x\theta}}{} =
\interpretfo{\floor{x\theta'}}{}$ for all $x$ because $\EE$ is
injective.
By Lemma~\ref{lem:subst-congruence} it follows that
$\interpretfo{\floor{(\lambda x.\>t)\theta}}{} =
\interpretfo{\floor{(\lambda x.\>t)\theta'}}{}$, which proves that $\lho$ is
well defined.

In our example, for all $\xi$ we have $\lho(\xi,\lambda x.\> x) = \EE([\cst{lam}_{\lambda x.\>x}]) = \EE([\cst{f}_0])$, which is the identity.
If $\xi(y) = [\cst{a}_0]$, then $\lho(\xi,\lambda x.\> y) =  \EE([\cst{lam}_{\lambda x.\>\cst{a}}])$, which is the constant function $c \mapsto [\cst{a}_0]$.
Similarly, if $\xi(y) = [\cst{b}_0]$, then $\lho(\xi,\lambda x.\> y)$ is the constant function $c \mapsto [\cst{b}_0]$.

This concludes the definition of the interpretation
$\IIIho=(\uho,\IItyho,\iho,\lho)$. It remains to show that $\smash{\IIIho}$
is proper. In a proper interpretation, the denotation $\interpretho{t}{}$ of a
term $t$ does not depend on the representative of $t$ modulo $\beta\eta$, but
since we have not yet shown $\IIIho$ to be proper, we cannot rely on
this property. For this reason, we use $\lambda$-terms in the following three
lemmas and mark all $\beta\eta$-reductions explicitly.

The higher-order interpretation $\IIIho$ relates to the first-order
interpretation $\RfN$ as follows:
\begin{lemmax}\label{lem:ceil-floor-correspondence}
	Given a ground $\lambda$-term $t$, we have
  $\interpretho{t}{} = \EE(\interpretfo{\floor{\betanf{t}}}{})$
\end{lemmax}

\begin{proof}
	By induction on $t$.
	Assume that $\interpretho{s}{} = \EE(\interpretfo{\floor{\betanf{s}}}{})$ for all proper subterms $s$ of~$t$.
	If $t$ is of the form $\cst{f}\typeargs{\tuple{\tau}}$, then
	\begin{align*}
	\interpretho{t}{} &= \iho(\cst{f},\dho_{\tuple{\tau}})\\[-.5\jot]
	&=\EE(\ifo(\cst{f}_0,\ufo_{\floor{\tuple{\tau}}}))\\[-.5\jot]
	&=\EE(\interpretfo{\cst{f}_0\typeargs{\floor{\tuple{\tau}}}}{})\\[-.5\jot]
	&=\EE(\interpretfo{\floor{{\cst{f}\typeargs{\tuple{\tau}}}}}{})\\[-.5\jot]
	&=\EE(\interpretfo{\floor{\betanf{\cst{f}\typeargs{\tuple{\tau}}} }}{})
	=\EE(\interpretfo{\floor{\betanf{t}}}{})
	\end{align*}
  If $t$ is an application $t = t_1\>t_2$, where $t_1$ is of type $\tau\fun\upsilon$, then
	\begin{align*}
	\interpretho{t_1\>t_2}{}
	&= \interpretho{t_1}{} (\interpretho{t_2}{}) \\[-.5\jot]
	&\overset{\!\scriptscriptstyle\text{IH}\!}{=}
	\EE_{\tau\fun\upsilon}(\interpretfo{\floor{\betanf{t_1}}}{}) (\EE_\tau(\interpretfo{\floor{\betanf{t_2}}}{}))\\[-.5\jot]
	&\overset{\kern-10mm\text{Def }\EE\kern-10mm}{=}
	\enskip\EE_\upsilon(\interpretfo{\floor{\betanf{(t_1\>t_2)}}}{})
	\end{align*}
	If $t$ is a $\lambda$-expression, then
	\begin{align*}
	\interprethoxi{\lambda x.\>u}
	&= \lho (\xi, (\lambda x.\>u)) \\[-.5\jot]
	& = \EE(\interpretfo{\floor{\betanf{(\lambda x.\>u)\theta}}}{}) \\[-.5\jot]
	& = \EE(\interpretfo{\floor{\betanf{(\lambda x.\>u)}}}{})
  \end{align*}
  where $\theta$
  is a substitution such that $\dho_{\alpha\theta}=\xi(\alpha)$ and
  $\EE(\interpretfo{\floor{x\theta}}{}) = \xi(x)$.
	\qedhere
\end{proof}

We need to show that the interpretation $\IIIho=(\uho,\IItyho,\iho,\lho)$ is
proper. In the proof, we will need to employ the following lemma, which is
very similar to the substitution lemma (Lemma~\ref{lem:subst-lemma-general}),
but we
must prove it here for our particular interpretation $\IIIho$ because we have not
shown that $\IIIho$ is proper yet.

\begin{lemmax}[Substitution lemma]
	$\interpret{\tau\rho}{\IIIty^\GH}{\xi} = \interpret{\tau}{\IIIty^\GH}{\xi'}$
	and
  $\vphantom{(_{(_(}}\interpretho{t\rho}{\xi} = \interpretho{t}{\xi'}$
  for all $\lambda$-terms $t$,
  all~$\tau\in\TyHH$ and all grounding substitutions $\rho$,
	where $\xi'(\alpha) = \vphantom{(_{(_(}}\interpret{\alpha\rho}{\IIIty^\GH}{\xi}$ for all type variables $\alpha$
	and $\xi'(x) = \interpretho{x\rho}{\xi}$ for all term variables $x$.
  \label{lem:subst-lemma-special}
\end{lemmax}

\begin{proof}
  We proceed by induction on the structure of $\tau$ and $t$.
  The proof is identical to the one of Lemma~\ref{lem:subst-lemma-general},
  except for the last step, which uses properness of the interpretation, a
  property we cannot assume here.
  However, here, we have the assumption that $\rho$ is a grounding substitution.
  Therefore, if $t$ is a $\lambda$-expression, we argue as follows:
\begin{align*}
\interpretho{(\lambda z.\>u)\rho}{\xi}
&=\interpretho{(\lambda z.\>u\rho')}{\xi}&&\text{ where $\rho'(z)=z$ and $\rho'(x)=\rho(x)$ for $x\neq z$}\\
&= \lho(\xi,(\lambda z.\>u\rho'))
&&\text{ by the definition of the term denotation}\\
&= \EE(\interpretfo{\floor{\betanf{(\lambda z.\>u)\rho\theta}}}{\xi})
&&\text{ for some $\theta$ by the definition of $\lho$}\\
&= \EE(\interpretfo{\floor{\betanf{(\lambda z.\>u)\rho}}}{\xi})
&&\text{ because $(\lambda z.\>u)\rho$ is ground}\\
&\overset{\smash{*}}{=} \lho(\xi',\lambda z.\>u)
&&\text{ by the definition of $\lho$ and Lemma~\ref{lem:ceil-floor-correspondence}}\\
&= \interpretho{\lambda z.\>u}{\xi'}
&&\text{ by the definition of the term denotation}
\end{align*}
The step $*$ is justified as follows:
We have
$\lho(\xi',\lambda z.\>u) = \EE(\interpretfo{\floor{\betanf{(\lambda z.\>u)\theta'}}}{\xi})$
by the definition of $\lho$,
if $\theta'$ is a substitution such that
$\smash{\dho_{\alpha\theta'}}=\xi'(\alpha)$ for all $\alpha$ and
$\EE(\interpretfo{\floor{\betanf{x\theta'}}}{\xi}) = \xi'(x)$
for all $x$.
By the definition of $\xi'$ and by Lemma~\ref{lem:ceil-floor-correspondence},
$\rho$ is such a substitution.
Hence,
$\lho(\xi',\lambda z.\>u) = \EE(\interpretfo{\floor{\betanf{(\lambda z.\>u)\rho}}}{\xi})$.
\qedhere
\end{proof}

\begin{lemmax} \label{lem:proper}
	The interpretation $\IIIho$ is proper.
\end{lemmax}
\begin{proof}
	We must show that
	$\interprethoxi{(\lambda x.\>t)}(a) = \interpretho{t}{\xi[x\mapsto a]}$
	for all $\lambda$-expressions $\lambda x.\>t$, all valuations $\xi$, and all values $a$.
	\begin{align*}
	\interprethoxi{\lambda x.\>t}(a) &= \lho(\xi,\lambda x.\>t)(a)
	&&\text{by the definition of $\interpretho{\phantom{\cdot}}{}$}\\
	&= \EE(\interpretfo{\floor{\betanf{(\lambda x.\>t)\theta}}}{})(a)
	&&\text{by the definition of $\lho$ for some $\theta$}\\[-\jot]
	&&&\text{such that $\EE(\interpretfo{\floor{z\theta}}{}) = \xi(z)$ for all $z$}\\[-\jot]
	&&&\text{and $\smash{\dho}_{\alpha\theta} = \xi(\alpha)$ for all $\alpha$}\\
	&= \EE(\interpretfo{\floor{\betanf{((\lambda x.\>t)\theta\;s)}}}{})
	&&\text{by the definition of $\EE$}\\[-\jot]
	&&&\text{where $\EE(\interpretfo{\floor{s}}{})=a$}\\
	&= \EE(\interpretfo{\floor{\betanf{t(\theta[x\mapsto s])}}}{})
	&&\text{by $\beta$-reduction}\\
	&= \interpretho{t(\theta[x\mapsto s])}{}
  &&\text{by Lemma~\ref{lem:ceil-floor-correspondence}}\\
  &= \interpretho{t}{\xi[x\mapsto a]}
	&&\text{by Lemma~\ref{lem:subst-lemma-special}}
	\end{align*}

\vskip-\baselineskip
\vskip-\belowdisplayskip

~\qedhere  %
\end{proof}

  \begin{lemmax}\label{lem:B-inverse-model}
     $\IIIho$ is a model of $N$.
  \end{lemmax}
  \begin{proof}
    By Lemma \ref{lem:ceil-floor-correspondence},
    we have
    $\interpretho{t}{} = \EE(\interpretfo{\floor{t}}{})$
    for all $t \in \TGH$.
    Since $\EE$ is a bijection,
    it follows that any (dis)equation $s \doteq t \in \CGH$ is true in $\IIIho$
    if and only if $\floor{s \doteq t}$ is true in $\RfN$.
    Hence, a clause $C \in \CGH$ is true in $\IIIho$
    if and only if $\floor{C}$ is true in $\RfN$.
    By 
Theorem~\ref{thm:GF-refutational-completeness} 
    and the assumption
    that $\bot \notin N$, $\RfN$ is a model of $\floor{N}$---%
    that is, for all clauses $C\in N$, $\floor{C}$ is
    true in $\RfN$.
    Hence, all clauses $C\in N$ are true in $\IIIho$ and therefore
    $\IIIho$ is a model of $N$.
    \qedhere
  \end{proof}

We summarize the results of this subsection in the following theorem:

\begin{sloppypar}
\begin{theoremx}[Ground static refutational completeness]
  Let $\GHSel$ be a selection function on $\CGH$.
  Then the inference system $\GHInf^\GHSel$ is statically refutationally complete
  \wrt\ $(\GHRedI, \GHRedC)$.
  In other words, if $N \subseteq \CGH$ is a clause set saturated \wrt\ $\GHInf^\GHSel$ and $\GHRedI^\GHSel$,
  then $N \models \bot$ if and only if $\bot \in N$.
  \label{thm:GH-refutational-completeness}
\end{theoremx}
\end{sloppypar}

The construction of $\IIIho$ relies on specific properties of $\RfN$. It
would not work with an arbitrary first-order interpretation. Transforming a
higher-order interpretation into a first-order interpretation is easier:

\begin{lemmax} \label{lem:gf-interpretation-from-gh}
  Given a clausal higher-order interpretation $\III$
  on $\GH$,
  there exists a first-order interpretation $\III^\GF$ on $\GF$
  such that for any clause $C\in\CGH$ the truth values of
  $C$ in $\III$ and of $\floor{C}$ in $\III^\GF$ coincide.
  \end{lemmax}
  \begin{proof}
  Let $\III = (\IIIty,\II,\LL)$ be a clausal higher-order interpretation.
  Let $\UU^\GF_\tau = \interpret{\tau}{\IIIty}{}$ be the first-order type universe for the ground type $\tau$.
  For a symbol $\smash{\cst{f}^{\tuple{\upsilon}}_{\!j}} \in
  \Sigma_\GF$, let $\II^\GF (\smash{\cst{f}^{\tuple{\upsilon}}_{\!j}}) =
  \interpret{\cst{f}\typeargs{\tuple{\upsilon}}}{\III}{}$ (up to currying).
  For a symbol $\cst{lam}_{\lambda x.\>t} \in \Sigma_\GF$,
  let $\II^\GF (\cst{lam}_{\lambda x.\>t}) = \interpret{\lambda x.\>t}{\III}{}$.
  This defines a first-order interpretation $\III^\GF = (\UU^\GF,\II^\GF)$.
  
  We need to show that for any $C\in\CGH$, $\III \models C$ if and only if $\III^\GF \models \floor{C}$.
  It suffices to show that $\interpret{t}{\III}{} = \interpret{\floor{t}}{\III^\GF}{}$
  for all terms $t\in\TGH$.
  We prove this by induction on the structure of the $\eta$-short $\beta$-normal form of $t$.
  If $t$ is a $\lambda$-expression, this is obvious.
  If $t$ is of the form $\cst{f}\typeargs{\tuple{\upsilon}}\>\tuple{s}_{\!j}$, then
  $\floor{t} = \smash{\cst{f}^{\tuple{\upsilon}}_{\!j}}(\floor{\tuple{s}_{\!j}})$
  and hence
  $\interpret{\floor{t}}{\III^\GF}{}
  = \II^\GF (\smash{\cst{f}^{\tuple{\upsilon}}_{\!j}})(\interpret{\floor{\tuple{s}_{\!j}}}{\III^\GF}{})
  = \interpret{\cst{f}\typeargs{\tuple{\upsilon}}}{\III}{}(\interpret{\floor{\tuple{s}_{\!j}}}{\III^\GF}{})
  \eqIH \interpret{\cst{f}\typeargs{\tuple{\upsilon}}}{\III}{}(\interpret{\tuple{s}_{\!j}}{\III}{})
  = \interpret{t}{\III}{}$.
  \qedhere
\end{proof}

\oursubsection{The Nonground Higher-Order Level}

To lift the result to the nonground level, we employ the saturation framework of
Waldmann et al.~\cite{waldmann-et-al-2020-saturation}.
It is easy to see that the entailment relation $\models$ on $\GH$ is a consequence relation in the sense of the framework.
We need to show that our redundancy criterion on $\GH$ is a redundancy criterion in the sense of the framework and that
$\gnd$ is a grounding function in the sense of the framework:

\begin{lemmax} \label{lem:redundancy-criterion}
  The redundancy criterion for $\GH$ is a redundancy criterion 
  in the sense of \Section~2 of the saturation framework.
\end{lemmax}
\begin{sloppypar}
\begin{proof}
  We must prove the conditions (R1) to (R4) of the saturation framework.
  Adapted to our context, they state the following
  for all clause sets $N,N' \subseteq \CGH$:
  \begin{enumerate}[(R1)]
    \item if $N \models \bot$, then $N \setminus \GHRedC(N) \models \bot$;
    \item if $N \subseteq N'$, then $\GHRedC(N) \subseteq \GHRedC(N')$ 
        and $\GHRedI(N) \subseteq \GHRedI(N')$;
    \item if $N' \subseteq \GHRedC(N)$, then $\GHRedC(N) \subseteq \GHRedC(N \setminus N')$
        and $\GHRedI(N) \subseteq \GHRedI(N \setminus N')$;
    \item if $\iota \in \GHInf$ and $\concl(\iota) \in N$, then $\iota \in \GHRedI(N)$.
  \end{enumerate}
  The proof is analogous to the proof of Lemma~4.10 %
  of Bentkamp et al.~\cite{bentkamp-et-al-lfhosup-arxiv}, 
  using Lemma~\ref{lem:gf-interpretation-from-gh}. \qedhere
\end{proof}
\end{sloppypar}

\begin{lemmax} \label{lem:grounding-function}
  The grounding functions $\gnd^\GHSel$ for $\GHSel\in\gnd(\HSel)$
  are grounding functions in the sense of \Section~3 of the saturation framework.
\end{lemmax}
\begin{proof}
We must prove the conditions (G1), (G2), and (G3) of the saturation framework.
Adapted to our context, they state the following:
\begin{enumerate}[(G1)]
\item $\gnd(\bot) = \{ \bot \}$;
\item for every $C \in \CHH$, if $\bot \in \gnd(C)$, then $C = \bot$;
\item for every $\iota \in \HInf$, $\gnd^\GHSel(\iota) \subseteq \GHRedI^\GHSel(\gnd(\concl(\iota)))$.
\end{enumerate}
Clearly, $C = \bot$ if and only if $\bot \in \gnd(C)$ if and only if $\gnd(C) =
\{\bot\}$, proving (G1) and (G2).
For every $\iota\in\HInf$, by the definition of $\gnd^\GHSel$,
we have $\concl(\gnd^\GHSel(\iota))\subseteq\gnd(\concl(\iota))$, and thus (G3) by (R4). \qedhere
\end{proof}

To lift the completeness result of the previous subsection to
the nonground calculus $\HInf$,
we employ Theorem~14 of the saturation framework, which, 
adapted to
our context, is stated as follows.
The theorem uses the notation $\Inf(N)$ 
to denote the set of $\Inf$-inferences
whose premises are in $N$,
for an inference system $\Inf$ and a clause set $N$.
Moreover, it uses 
Herbrand entailment $\Gmodels$ on $\CHH$, which is
defined so that $N_1 \Gmodels N_2$ if and only if $\gnd(N_1) \models \gnd(N_2)$.

\begin{theoremx}[Lifting theorem]
If $\GHInf^\GHSel$ is statically refutationally complete
\wrt\ $(\GHRedI^\GHSel, \GHRedC)$
for every $\GHSel \in \gnd(\HSel)$,
and if for every $N\subseteq \CHH$ that is
saturated \wrt\ $\HInf$ and $\HRedI$
there exists a $\GHSel \in \gnd(\HSel)$
such that
$\GHInf^\GHSel(\gnd(N)) 
\subseteq \gnd^\GHSel(\HInf(N)) \cup \GHRedI^\GHSel(\gnd(N))$,
then also
$\HInf$ is statically refutationally complete \wrt\ $(\HRedI, \HRedC)$
and $\Gmodels$.
\label{thm:lifting-theorem}
\end{theoremx}
\begin{proof}
  This is almost an instance of Theorem~14 %
  of the saturation framework.
  We take $\CHH$ for $\mathbf{F}$, $\CGH$ for $\mathbf{G}$, and $\gnd(\HSel)$ for $Q$.
  It is easy to see that the entailment relation $\models$ on $\GH$ is a consequence relation in the sense
  of the framework. By Lemma~\ref{lem:redundancy-criterion} and~\ref{lem:grounding-function},
  $(\GHRedI^\GHSel,\allowbreak\GHRedC)$ is a redundancy criterion
  in the sense of the framework,
  and $\gnd^\GHSel$ are grounding functions
  in the sense of the framework,
  for all $\GHSel\in\gnd(\HSel)$.
  The redundancy criterion $(\HRedI,\HRedC)$
  matches exactly the intersected lifted redundancy criterion
  $\mathit{Red}^{\ccap,\sqsupset}$
  of the saturation framework.
  Their Theorem~14 %
  states the theorem only for ${\sqsupset} = \varnothing$.
  By their Lemma~16, %
  it also holds if ${\sqsupset} \not= \varnothing$. \qedhere
\end{proof}

Let $N\subseteq\CHH$ be a clause set saturated \wrt\ $\HInf$ and $\HRedI$.
We assume that $\HSel$ fulfills the selection restriction that
a literal $\greensubterm{L}{\,y}$ must not be selected if $y\>
\tuple{u}_n$, with $n > 0$, is a $\succeq$-maximal term of the clause,
as required in Definition~\ref{def:sel}.
For the above theorem to apply,
we need to show that there exists a selection
function $\GHSel\in\gnd(\HSel)$ such that
all inferences $\iota\in\GHInf^\GHSel$ with $\prem(\iota)\in\gnd(N)$
are liftable or redundant.
Here, for $\iota$ to be \emph{liftable} means that $\iota$ is a $\smash{\gnd^\GHSel}$-ground instance
of a $\smash{\HInf}$-inference from $N$;
for $\iota$ to be \emph{redundant} means that $\iota\in\smash{\GHRedI^\GHSel(\gnd(N))}$.

To choose the right selection function $\GHSel\in\gnd(\HSel)$, we observe that each ground clause
$C\in\gnd(N)$ must have at least one corresponding clause $D\in N$ such that $C$ is a ground instance of $D$.
We choose one of them for each $C\in\gnd(N)$, which we denote by $\gnd^{-1}(C)$. Then let $\GHSel$ select those literals in $C$ that 
correspond to literals selected by $\HSel$ in $\gnd^{-1}(C)$. With respect to this selection function
$\GHSel$, we can show that all inferences from $\gnd(N)$ are liftable or redundant:

\begin{lemmax} \label{lem:eligibility-lifting}
Let $\gnd^{-1}(C) = D \in N$ and $D\theta = C$. 
Let $\sigma$ and $\rho$ be substitutions such that 
$x\sigma\rho = x\theta$ for all variables $x$ in $D$.
Let $L$ be a (strictly) $\succeq$-eligible literal in $C$ \wrt\ $\GHSel$.
Then there exists a (strictly) $\succsim$-eligible literal $L'$ in $D$ \wrt\ $\sigma$ and $\HSel$ such that $L'\theta = L$.
\end{lemmax}
\begin{proof}
If $L \in \GHSel(C)$, then there exists $L'$ such that $L'\theta = L$ and $L' \in \HSel(D)$ by the definition of $\gnd^{-1}$.
Otherwise, $L$ is $\succeq$-maximal in $C$.
Since $C = D\sigma\rho$, there are literals $L'$ in $D\sigma$ such that $L'\rho = L$.
Choose $L'$ to be a $\succsim$-maximal among them. 
Then $L'$ is $\succsim$-maximal in $D\sigma$ because for any literal $L''\in D$ with $L'' \succsim L'$,
we have $L''\rho \succeq L'\rho = L$ and hence $L''\rho = L$ by $\succeq$-maximality of $L$.

If $L$ is \relax{strictly} $\succeq$-maximal in $C$,
$L'$ is also \relax{strictly} $\succsim$-maximal in $D\sigma$ because a duplicate of $L'$ in $D\sigma$ would imply a duplicate of $L$ in $C$.
\qedhere
\end{proof}

\begin{lemmax}[Lifting of \infname{ERes}, \infname{EFact}, \infname{GArgCong}, and \infname{GExt}]
  All \infname{ERes}, \infname{EFact}, \infname{GArgCong}, and \infname{GExt} inferences from $\gnd(N)$ are liftable.
  \label{lem:lifting1}
\end{lemmax}
\begin{proof}
  \infname{ERes}: Let $\iota\in\GHInf^\GHSel$ be an \infname{ERes} inference with $\prem(\iota)\in\gnd(N)$.
  Then $\iota$ is of the form
  \[\namedinference{ERes}{C\theta~=~C'\theta \llor s\theta \noteq s'\theta}{C'\theta}\]
  where $\gnd^{-1}(C\theta) = C = C' \llor s \noteq s'$
  and the literal $s\theta \noteq s'\theta$
  is $\succeq$-eligible \wrt\ $\GHSel$.
  Since $s\theta$ and $s'\theta$ are unifiable and ground,
  we have $s\theta = s'\theta$.
  Thus, there exists an idempotent $\sigma \in \csu(s,s')$
  such that for some substitution~$\rho$ and for all variables $x$ in $C$, 
  we have $x\sigma\rho = x\theta$.
  By Lemma~\ref{lem:eligibility-lifting}, we may assume without loss of generality that
  $s \noteq s'$ is $\succsim$-eligible in $C$ \wrt\ $\sigma$ and $\HSel$.
  Hence, the following inference $\iota'\in\HInf$ is applicable:
  \[\namedinference{ERes}{C' \llor s \noteq s'}{C'\sigma}\]
  Then $\iota$ is the $\sigma\rho$-ground instance of $\iota'$
  and is therefore liftable.

  \medskip

  \noindent
  \infname{EFact}:\enskip Analogously, if $\iota\in\GHInf^\GHSel$
  is an \infname{EFact} inference with $\prem(\iota)\in\gnd(N)$,
  then $\iota$ is of the form
  \[\namedinference{EFact}{C\theta~=~C'\theta \llor s'\theta \eq t'\theta \llor s\theta \eq t\theta}
  {C'\theta \llor t\theta \noteq t'\theta \llor s\theta \eq t'\theta}\]
  where $\gnd^{-1}(C\theta) = C =  C' \llor s' \eq t' \llor s \eq t$,
  the literal $s\theta \eq t\theta$ is $\succeq$-eligible in $C$ \wrt\ $\GHSel$,
  and $s\theta\not\prec t\theta$. Then $s\not\prec t$.
  Moreover, $s\theta$ and $s'\theta$ are unifiable and ground. 
  Hence,
  $s\theta = s'\theta$ and 
  there exists an idempotent $\sigma \in \csu(s,s')$
  such that for some substitution~$\rho$ and for all variables $x$ in $C$, 
  we have $x\sigma\rho = x\theta$.
  By Lemma~\ref{lem:eligibility-lifting}, we may assume without loss of generality that
  $s\eq t$ is $\succsim$-eligible in $C$ \wrt\ $\sigma$ and $\HSel$.
  It follows that the following inference $\iota'\in\HInf$ is applicable:
  \[\namedinference{EFact}{C' \llor s' \eq t' \llor s \eq t}
  {(C' \llor t \noteq t' \llor s \eq t')\sigma} \]
  Then $\iota$ is the $\sigma\rho$-ground instance of $\iota'$
  and is therefore liftable.

  \medskip

  \noindent
  \infname{GArgCong}:
  Let $\iota\in\GHInf^\GHSel$ be a \infname{GArgCong} inference with $\prem(\iota)\in\gnd(N)$.
  Then $\iota$ is of the form
  \[\namedinference{GArgCong}{C\theta~=~C'\theta \llor s\theta \eq s'\theta}
  {C'\theta \llor s\theta\>\tuple{u}_n \eq s'\theta\>\tuple{u}_n}\]
  where $\gnd^{-1}(C\theta) = C = C' \llor s \eq s'$,
  the literal $s\theta \eq s'\theta$
  is strictly $\succeq$-eligible \wrt\ $\GHSel$, and
  $s\theta$ and $s'\theta$ are of functional type.
  It follows that $s$ and $s'$ have either a functional
  or a polymorphic type.
  Let $\sigma$ be the most general substitution such that
  $s\sigma$ and $s'\sigma$ take $n$ arguments.
  By Lemma~\ref{lem:eligibility-lifting}, we may assume without loss of generality that 
  $s \noteq s'$ is strictly $\succsim$-eligible in $C$ \wrt\ $\sigma$ and $\HSel$.
  Hence the following inference $\iota'\in\HInf$ is applicable:
  \[\namedinference{ArgCong}{C' \llor s \eq s'}
  {C'\sigma \llor s\sigma\>\tuple{x}_n \eq s'\sigma\>\tuple{x}_n}\]
  Since $\sigma$ is the most general substitution
  that ensures well-typedness of the conclusion,
  $\iota$ is a ground instance of $\iota'$
  and is therefore liftable.

\medskip

\noindent
\infname{GExt}: The conclusion of a \infname{GExt} inference in $\GHInf$ is by definition a
ground instance of the conclusion of an \infname{Ext} inference in $\HInf$.
Hence, the \infname{GExt} inference is a ground instance of the \infname{Ext} inference.
Therefore it is liftable.
\qedhere
\end{proof}

Some of the \infname{Sup} inferences in $\GHInf$ are liftable as well:

\begin{lemmax}[Instances of green subterms]
Let $s$ be a $\lambda$-term in $\eta$-short $\beta$-normal form,
let $\sigma$ be a substitution, and
let $p$ be a green position of both $s$ and $\betanf{s\sigma}$.
Then $\betanf{(s|_p)\sigma}
= (\betanf{s\sigma})|_p$.
\label{lem:arg-subterm-instances}
\end{lemmax}
\begin{proof}
  By induction on $p$.
  If $p=\varepsilon$, then $\betanf{(s|_p)\sigma} = \betanf{s\sigma} = (\betanf{s\sigma})|_p$.
  If $p=i.p'$, then
  $%
  s = \cst{f}\typeargs{\tuple{\tau}} \> s_1 \dots s_n
  $ %
  and
  $%
  s\sigma = \cst{f}\typeargs{\tuple{\tau}\sigma} \> (s_1\sigma) \dots (s_n\sigma)
  $, %
  where $1 \leq i \leq n$ and
  $p'$ is a green position of $s_i$.
  Clearly,
  $\beta\eta$-normalization steps of $s\sigma$ can take place only in proper subterms.
  So
  $%
  \betanf{s\sigma}
  =
  \cst{f}\typeargs{\tuple{\tau}\sigma} \> (\betanf{s_1\sigma}) \dots (\betanf{s_n\sigma}).
  $ %
  Since $p = i.p'$ is a green position of $\betanf{s\sigma}$,
  $p'$ must be a green position of $\betanf{(s_i\sigma)}$.
  By the induction hypothesis,
  $\betanf{(s_i|_{p'})\sigma} = (\betanf{s_i\sigma})|_{p'}$.
  Therefore
  $\betanf{(s|_p)\sigma}
  = \betanf{(s|_{i.p'})\sigma}
  = \betanf{(s_i|_{p'})\sigma}
  = (\betanf{s_i\sigma})|_{p'}
  = (\betanf{s\sigma})|_p$.
  \qedhere
\end{proof}

\begin{lemmax}[Lifting of \infname{Sup}]
  Let $\iota\in\GHInf^\GHSel$ be a \infname{Sup} inference
  \[\namedinference{Sup}
  {\overbrace{D'\theta \llor t\theta \eq t'\theta}^{\vphantom{\cdot}\smash{D\theta}}
  \hypsep
  \overbrace{C'\theta \llor \greensubterm{s\theta}{t\theta}_p \doteq s'\theta}^{\vphantom{\cdot}\smash{C\theta}}}
  {D'\theta \llor C'\theta \llor \greensubterm{s\theta}{ t'\theta}_p \doteq s'\theta}
  \]
  where $\gnd^{-1}(D\theta) = D = D' \llor t \eq t'\in N$, $s\theta = \greensubterm{s\theta}{t\theta}_p$, and
  $\gnd^{-1}(C\theta) = C = C' \llor s \doteq s'\in N$.
  We assume that $s$, $t$, $s\theta$, and $t\theta$ are represented by $\lambda$-terms
  in $\eta$-short $\beta$-normal form.
  Let $p'$ be the longest prefix of $p$
  that is a green position of $s$.
  Since $\varepsilon$ is a green position of $s$,
  the longest prefix always exists.
  Let $u = s|_{p'}$.
  Suppose one of the following conditions applies:\
  {\upshape(i)} $u$ is a deeply occurring variable in $C$;
  {\upshape(ii)} $p = p'$ and the variable condition holds for $D$ and~$C$; or
  {\upshape(iii)} $p \neq p'$ and $u$ is not a variable.
  Then $\iota$ is liftable.
  \label{lem:lifting2}
\end{lemmax}
\begin{proof}
  The \infname{Sup} inference conditions for $\iota$ are that
  $t\theta\eq t'\theta$ is strictly $\succeq$-eligible,
  $s\theta\doteq s'\theta$ is strictly $\succeq$-eligible if positive
  and $\succeq$-eligible if negative,
  $D\theta \not\succsim C\theta$,
  $t\theta \not\precsim t'\theta$, and $s\theta \not\precsim s'\theta$.
  We assume that
  $s$, $t$, $s\theta$, and $t\theta$ are represented by $\lambda$-terms
  in $\eta$-short $\beta$-normal form.
  By\ Lemma~\ref{lem:arg-subterm-instances},
  $u\theta$ agrees with $s\theta|_{p'}$
  (considering both as terms rather than as $\lambda$-terms).

  \medskip\noindent
  \textsc{Case 1:}\enskip We have (a) $p = p'$, (b) $u$ is not fluid, and (c) $u$ is not a variable deeply occurring in $C$.
  Then $u\theta = s\theta|_{p'} = s\theta|_p = t\theta$.
  Since $\theta$ is a unifier of $u$ and $t$,
  there exists an idempotent $\sigma \in \csu(t,u)$
  such that for some substitution~$\rho$ and for all variables $x$ occurring in $D$ and $C$, 
  we have $x\sigma\rho = x\theta$.
  The inference conditions can be lifted:
  (Strict) eligibility of $t\theta\eq t'\theta$ and $s\theta\doteq s'\theta$ \wrt\ $\GHSel$
  implies (strict) eligibility of $t \eq t'$ and $s \doteq s'$ \wrt\ $\sigma$ and $\HSel$;
  $D\theta \not\succsim C\theta$ implies $D \not\succsim C$;
  $t\theta \not\precsim t'\theta$ implies $t \not\precsim t'$; and
  $s\theta \not\precsim s'\theta$ implies $s \not\precsim s'$.
  Moreover, by (a) and (c), condition~(ii) must hold and thus the variable condition holds for $D$ and~$C$.
  Hence there is the following \infname{Sup} inference $\iota'\in\HInf$:
  \[\namedinference{Sup}
  {D' \llor t \eq t' \hypsep C' \llor \greensubterm{s}{u}_p \doteq s'}
  {(D' \llor C' \llor \greensubterm{s}{t'}_p \doteq s')\sigma}
  \]
  Then $\iota$ is the $\sigma\rho$-ground instance
  of $\iota'$ and therefore liftable.

  \medskip\noindent
  \textsc{Case 2:}\enskip We have (a) $p \neq p'$, or (b) $u$ is fluid, or (c) $u$ is a variable deeply occurring in $C$.
  We will first show that (a) implies (b) or (c).
  Suppose (a) holds but neither (b) nor (c) holds.
  Then condition (iii) must hold---i.e., $u$ is not a variable.
  Moreover, since (b) does not hold, $u$ cannot have the form $y\>\tuple{u}_n$ for a variable $y$ and $n \geq 1$.
  If $u$ were of the form
  $\cst{f}\typeargs{\tuple{\tau}} \> s_1 \dots {s_n}$ with $n \geq 0$,
  $u\theta$ would have the form
  $\cst{f}\typeargs{\tuple{\tau}\theta} \> (s_1\theta)\dots (s_n\theta)$,
  but then there is some $1 \leq i \leq n$
  such that $p'.i$ is a prefix of $p$
  and $s|_{p'.i}$ is a green subterm of $s$,
  contradicting the maximality of $p'$.
  So $u$ must be a $\lambda$-expression,
  but since $t\theta$ is a proper green subterm of $u\theta$,
  $u\theta$ cannot be a $\lambda$-expression,
  yielding a contradiction. We may thus assume that (b) or (c) holds.

  Let $p = p'.p''$.
  Let $z$ be a fresh variable.
  Define a substitution $\theta'$ that maps this variable $z$ to
  $\lambda y.\> \greensubterm{(s\theta|_{p'})}{\, y}_{p''}$
  and any other variable $w$ to $w\theta$.
  Clearly,
  $(z \> t)\theta'
  = \greensubterm{(s\theta|_{p'})}{t\theta}_{p''}
  = s\theta|_{p'} = u\theta = u\theta'$.
  Since $\theta'$ is a unifier of $u$ and
  $z \> t$,
  there exists an idempotent $\sigma \in \csu(z \> t, u)$
  such that for some substitution~$\rho$, for $x=z$,
  and for all variables $x$ in $C$ and $D$, 
  we have $x\sigma\rho = x\theta'$.
  As in case 1,
  (strict) eligibility of the ground literals implies
  (strict) eligibility of the nonground literals.
  Moreover, by construction of $\theta'$,
  $t\theta' = t\theta \not= t'\theta = t'\theta'$ implies
  $(z \> t)\theta' \not= (z \> t')\theta'$,
  and thus
  $(z \> t)\sigma \not= (z \> t')\sigma$.
  Since we also have (b) or (c), there is the following inference $\iota'$:
  \[\namedinference{FluidSup}
  {D' \llor t \eq t' \hypsep C' \llor \greensubterm{s}{u}_{p'} \doteq s'}
  {(D' \llor C' \llor \greensubterm{s}{z \> t'}_{p'} \doteq s')\sigma}
  \]
  Then $\iota$ is the $\sigma\rho$-ground instance
  of $\iota'$ and therefore liftable. \qedhere
\end{proof}

The other \infname{Sup} inferences might not be liftable, but they are
redundant:

\begin{lemmax}\label{lem:nonliftable-sup-redundant}
  Let $\iota\in\GHInf^\GHSel$ be a \infname{Sup} inference from $\gnd(N)$ not covered by Lemma~\ref{lem:lifting2}. Then $\iota\in\GHRedI^\GHSel(\gnd(N))$.
\end{lemmax}
\begin{proof}
  Let $C\theta = C'\theta \lor s\theta \doteq s'\theta$ and
  $D\theta = D'\theta \lor t\theta \eq t'\theta$ be the premises of $\iota$,
  where $s\theta \doteq s'\theta$
  and $t\theta \eq t'\theta$ are the literals involved in the inference,
  $s\theta \succ s'\theta$, $t\theta \succ t'\theta$, and $C'$, $D'$, $s$,
  $s'$, $t$, $t'$ are the respective subclauses and terms in $C = \gnd^{-1}(C\theta)$ and $D = \gnd^{-1}(D\theta).$
  Then the inference $\iota$ has the form
  \[\namedinference{Sup}
  {{D'\theta \llor { t\theta \eq t'\theta}} \hypsep
  {C'\theta \llor s\theta\leftgreensubterm t\theta\rightgreensubterm \doteq s'\theta}}
  {D'\theta \llor C'\theta \llor s\theta\leftgreensubterm t'\theta\rightgreensubterm \doteq s'\theta}\]
  To show that $\iota \in \GHRedI^\GHSel(\gnd(N))$,
  it suffices to show
  $\{D\in\floor{\gnd(N)}\mid D\prec \floor{C\theta}\}\models\floor{\concl(\iota)}$.
  To this end, let $\III$ be an interpretation in $\GF$ such that
  $\III\models\{D\in\floor{\gnd(N)}\mid D\prec \floor{C\theta}\}$.
  We need to show that $\III\models \floor{\concl(\iota)}$.
  If $\floor{D'\theta}$ is true in $\III$, then obviously $\III\models \floor{\concl(\iota)}$.
  So we assume that $\floor{D'\theta}$ is false in $\III$.
  Since $C\theta \succ D\theta$ by the \infname{Sup} order
  conditions, it follows that $\III\models \floor{t\theta \eq t'\theta}$.
  Therefore, it suffices to show $\III\models \floor{C\theta}$.

  Let $p$ be the position in $s\theta$ where $\iota$ takes place
  and $p'$ be the longest prefix of $p$ that is a green subterm of $s$. Let $u = s|_{p'}$.
  Since Lemma~\ref{lem:lifting2} does not apply to $\iota$, $u$ is not a deeply occurring variable;
  if $p=p'$, the variable condition does not hold for $D$ and $C$; and
  if $p\neq p'$, $u$ is a variable.
  This means either
  the position $p$ does not exist in $s$, because it is below an unapplied variable that does not occur deeply in $C$,
  or
  $s|_p$ is an unapplied variable that does not occur deeply in $C$ and
  for which the variable condition does not hold.

  \medskip\noindent
  \textsc{Case 1:}\enskip The position $p$ does not exist in $s$ because it is below a variable $x$ that does not occur deeply in $C$.
  Then $t\theta$ is a green subterm of~$x\theta$
  and hence a green subterm of $x\theta \> \tuple{w}$ for any arguments~$\tuple{w}$.
  Let $v$ be the term that we obtain by replacing $t\theta$ by $t'\theta$ in $x\theta$
  at the relevant position.
  Since $\III\models \floor{t\theta\eq t'\theta}$,
  by congruence, $\III\models \floor{x\theta \> \tuple{w}\eq v \> \tuple{w}}$
  for any arguments $\tuple{w}$.
  Hence, $\III\models \floor{C\theta}$
  if and only if  $\III\models \floor{C\{x\mapsto v\}\theta}$ by congruence.
  Here, it is crucial that the variable does not occur deeply in $C$ because congruence does
  not hold in $\flooronly$-encoded terms below $\lambda$-binders.
  By the inference conditions, we have $t\theta \succ t'\theta$,
  which implies $\floor{C\theta} \succ \floor{C\{x\mapsto v\}\theta}$
  by compatibility with green contexts.
  Therefore, by the assumption about $\III$,
  we have $\III\models \floor{C\{x\mapsto v\}\theta}$
  and hence $\III\models \floor{C\theta}$.

  \medskip\noindent
  \textsc{Case 2:}\enskip The term $s|_p$ is a variable $x$ that does not occur deeply in $C$ and
  for which the variable condition does not hold.
  From this, we know that
  $C\theta \succeq C''\theta$,
  where $C'' = C\{x\mapsto t'\}$.
  We cannot have $C\theta = C''\theta$ because $x\theta = t\theta\neq t'\theta$ and $x$ occurs in $C$. Hence, we have $C\theta \succ C''\theta$.
  By the definition of $\III$, $C\theta \succ C''\theta$ implies $\III\models\floor{C''\theta}$. We will use equalities that are true in $\III$ to rewrite
  $\floor{C\theta}$ into $\floor{C''\theta}$, which implies $\III\models \floor{C\theta}$ by congruence.

  By saturation,
  every \infname{ArgCong} inference $\iota'$ from $D$ is in $\HRedI(N)$%
---i.e.,
  $\gnd(\concl(\iota')) \allowbreak\subseteq \gnd(N) \cup \GHRedC(\gnd(N))$.
  Hence, $D'\theta \lor t\theta \> \tuple{u} \eq t'\theta \> \tuple{u}$ is in $\gnd(N) \cup \GHRedC(\gnd(N))$
  for any %
  ground arguments $\tuple{u}$.

  We observe that whenever $t\theta \> \tuple{u}$ and $t'\theta \> \tuple{u}$
  are smaller than the $\succeq$-maximal term of $C\theta$ for some arguments $\tuple{u}$, we have
  \[\III\models \floor{t\theta \> \tuple{u}} \eq \floor{t'\theta \> \tuple{u}} \tag{$*$}\label{eq:congruences}  \]
  To show this, we assume that $t\theta \> \tuple{u}$ and $t'\theta \> \tuple{u}$
  are smaller than the $\succeq$-maximal term of $C\theta$ and we distinguish two cases:
  If $t\theta$ is smaller than the $\succeq$-maximal term of $C\theta$, all terms in $D'\theta$
  are smaller than the $\succeq$-maximal term of $C\theta$ and hence
  $D'\theta \lor t\theta \> \tuple{u} \eq t'\theta \> \tuple{u} \prec C\theta$.
  If, on the other hand, $t\theta$ is equal to the $\succeq$-maximal term of $C\theta$, then
  $t\theta \> \tuple{u}$ and $t'\theta \> \tuple{u}$ are smaller than $t\theta$.
  Hence $t\theta \> \tuple{u} \eq t'\theta \> \tuple{u} \prec t\theta \eq t'\theta$ and
  $D'\theta \lor t\theta \> \tuple{u} \eq t'\theta \> \tuple{u} \prec D\theta \prec C\theta$.
  In both cases, since $D'\theta$ is false in $\III$, by the definition of $\III$,
  we have (\ref{eq:congruences}).

  Next, we show the equivalence of $C\theta$ and $C''\theta$ via rewriting with equations
  of the form (\ref{eq:congruences})
  where $t\theta \> \tuple{u}$ and $t'\theta \> \tuple{u}$ are smaller than the $\succeq$-maximal term of $C\theta$.
  Since $x$ does not occur deeply in~$C$,
  every occurrence of $x$ in $C$ is not inside a $\lambda$-expression
  and not inside an argument of an applied variable.
  Therefore, all occurrences of $x$ in $C$ are in a green subterm of the form $x\>\tuple{v}$
  for some terms $\tuple{v}$ that do not contain $x$. Hence, every occurrence of $x$ in $C$ corresponds
  to a subterm $\floor{(x\>\tuple{v})\theta} = \floor{t\theta\>\tuple{v}\theta}$ in $\floor{C\theta}$
  and to a subterm
  $\floor{(x\>\tuple{v})\{x\mapsto t'\}\theta} = \floor{t'\theta\>\tuple{v}\{x\mapsto\nobreak t'\}\theta} = \floor{t'\theta\>\tuple{v}\theta}$ in $\floor{C''\theta}$.
  These are the only positions where $C\theta$ and $C''\theta$ differ.

  \looseness=-1
  To justify the necessary rewrite steps from $\floor{t\theta\>\tuple{v}\theta}$
  into $\floor{t'\theta\>\tuple{v}\theta}$ using (\ref{eq:congruences}),
  we must show that $\floor{t\theta\>\tuple{v}\theta}$ and $\floor{t'\theta\>\tuple{v}\theta}$ are smaller
  than the $\succeq$-maximal term in $\floor{C\theta}$ for the relevant $\tuple{v}$.
  If $\tuple{v}$ is the empty tuple, we do not need to show this because $\III \models \floor{t\theta \eq t'\theta}$
  follows from $\floor{D\theta}$'s being true and $\floor{D'\theta}$'s being false.
  If $\tuple{v}$ is nonempty, it suffices to show that $x\>\tuple{v}$ is not a $\succeq$-maximal term in $C$.
  Then $\floor{t\theta\>\tuple{v}\theta}$ and $\floor{t'\theta\>\tuple{v}\theta}$,
  which correspond to the term $x\>\tuple{v}$ in $C$, cannot be $\succeq$-maximal
  in $\floor{C\theta}$ and $\floor{C''\theta}$.
  Hence they must be smaller than the $\succeq$-maximal term in $\floor{C\theta}$
  because they are subterms of $\floor{C\theta}$
  and $\floor{C''\theta}\prec \floor{C\theta}$, respectively.

  To show that $x\>\tuple{v}$ is not a $\succeq$-maximal term in $C$,
  we make a case distinction on whether $s\theta \doteq s'\theta$ is selected in $C\theta$
  or $s\theta$ is the $\succeq$-maximal term in $C\theta$.
  One of these must hold because $s\theta \doteq s'\theta$ is $\succeq$-eligible in $C\theta$.
  If it is selected, by the selection restrictions, $x$ cannot be the head of a $\succeq$-maximal term of $C$.
  If $s\theta$ is the $\succeq$-maximal term in $C\theta$, we can argue that $x$ is a green subterm of $s$
  and, since $x$ does not occur deeply, $s$ cannot be of the form $x\>\tuple{v}$ for a nonempty $\tuple{v}$.
  This justifies the necessary rewrites between $\floor{C\theta}$ and $\floor{C''\theta}$ and it
  follows that $\III \models \floor{C\theta}$.
  \qedhere
\end{proof}

With these properties of our inference systems in place,
Theorem~\ref{thm:lifting-theorem}
guarantees static and dynamic refutational completeness
of $\HInf$ \wrt\ $\HRedI$. However, this theorem gives us refutational
completeness \wrt\ the Herbrand entailment $\Gmodels$,
defined as $N_1 \Gmodels N_2$ if $\gnd(N_1) \models \gnd(N_2)$,
whereas our semantics is Tarski entailment $\models$, defined as
$N_1 \models N_2$ if any model of $N_1$ is a model of $N_2$.
To repair this mismatch, we use the following lemma, which can
be proved along the lines of 
Lemma~4.16
of Bentkamp et al.~\cite{bentkamp-et-al-lfhosup-arxiv},
using Lemma~\ref{lem:subst-lemma-general} and Lemma~\ref{lem:apply-subst}.

\begin{lemmax}
  \label{lem:herbrand-tarski}
  For $N \subseteq \CHH$, we have $N \Gmodels \bot$ if and only if $N \models \bot$.
\end{lemmax}

\begin{theoremx}[Static refutational completeness]
  The inference system $\HInf$ is statically refutationally complete \wrt\ $(\HRedI, \HRedC)$.
  In other words, if $N \subseteq \CHH$ is a clause set saturated \wrt\ $\HInf$ and $\HRedI$,
  then we have $N \models \bot$ if and only if $\bot \in N$.
  \label{thm:static-refutational-completeness}
\end{theoremx}
\begin{proof}
  We apply Theorem~\ref{thm:lifting-theorem}.
  By Theorem~\ref{thm:GH-refutational-completeness},
  $\GHInf^\GHSel$ is statically refutationally complete for all $\GHSel\in\gnd(\HSel)$.
  By Lemmas~\ref{lem:lifting1}, \ref{lem:lifting2}, and~\ref{lem:nonliftable-sup-redundant},
  for every saturated $N\subseteq \CHH$, there exists a selection
  function $\GHSel\in\gnd(\HSel)$ such that
  all inferences $\iota\in\GHInf^\GHSel$ with $\prem(\iota)\in\gnd(N)$
  either are $\gnd^\GHSel$-ground instances of $\HInf$-inferences from $N$ or
  belong to $\smash{\GHRedI^\GHSel(\gnd(N))}$.

  Theorem~\ref{thm:lifting-theorem} implies that
  if $N \subseteq \CHH$ is a clause set saturated \wrt\ $\HInf$ and $\HRedI$,
  then $N \Gmodels \bot$ if and only if $\bot \in N$.
  By Lemma~\ref{lem:herbrand-tarski}, this also holds for the Tarski entailment $\models$.
  That is,
  if $N \subseteq \CHH$ is a clause set saturated \wrt\ $\HInf$ and $\HRedI$,
  then $N \models \bot$ if and only if $\bot \in N$.
 \qedhere
\end{proof}

From static refutational completeness, we can easily derive dynamic
refutational completeness.

\begin{theoremx}[Dynamic refutational completeness]
  The inference system $\HInf$ is dynamically refutationally complete \wrt\ $(\HRedI, \HRedC)$,
  as defined in 
  Definition~\ref{def:dyn-complete}.%
\label{thm:dynamic-refutational-completeness}
\end{theoremx}
\begin{proof}
  By
  Theorem~17 %
  of the saturation framework,
  this follows from Theorem~\ref{thm:static-refutational-completeness} and Lemma~\ref{lem:herbrand-tarski}.
  \qedhere
\end{proof}

\section{Extensions}
\label{sec:extensions}

The core calculus can be extended with various optional rules. Although
these are not necessary for refutational completeness, they can allow the
prover to find more direct proofs. Most of these rules are concerned with the
areas covered by the \infname{FluidSup} rule and the extensionality axiom. 

Two of the optional rules below rely on the notion of ``orange subterms.''
\begin{definitionx}
A $\lambda$-term $t$ is an \emph{orange subterm} of a $\lambda$-term $s$ if $s = t$;
or if
$s = \cst{f}\typeargs{\tuple{\tau}}\> \tuple{s}$ and $t$ is an orange subterm of $s_i$ for some~$i$;
or if
$s = x\> \tuple{s}$ and $t$ is an orange subterm of $s_i$ for some~$i$;
or if
$s = (\lambda x.\> u)$ and $t$ is an orange subterm of $u$.
\end{definitionx}
For example, in the term $\cst{f}\> (\cst{g}\> \cst{a})\> (y\> \cst{b})\>
(\lambda x.\> \cst{h}\> \cst{c}\> (\cst{g}\> x))$,
the orange subterms
are all the green subterms---$\cst{a}$, $\cst{g}\> \cst{a}$,
$y\> \cst{b}$, $\lambda x.\> \cst{h}\> \cst{c}\> (\cst{g}\> x)$ and the whole term---and
in addition
$\cst{b}$, $\cst{c}$, $x$, $\cst{g}\> x$, and $\cst{h}\>
\cst{c}\> (\cst{g}\> x)$.
Following Convention~\ref{conv:beta-eta-normal-form},
this notion is lifted to $\beta\eta$-equivalence classes via
representatives in $\eta$-short $\beta$-normal form.
We write $t = \orangesubterm{s}{\tuple{x}_n}{u}$ to
indicate that $u$ is an orange subterm of $t$, where $\tuple{x}_n$ are the variables
bound in the \emph{orange context} around $u$, from outermost to innermost.
If $n = 0$, we simply write $t = \yellowsubterm{s}{u}$.

Once a term $\orangesubterm{s}{\tuple{x}_n}{u}$ has been
introduced, we write $\orangesubtermeta{s}{\tuple{x}_n}{u'}$ to denote the same context with a
different subterm $u'$ at that position. The $\eta$ subscript is a reminder that $u'$ is
not necessarily an orange subterm of $\orangesubtermeta{s}{\tuple{x}_n}{u'}$ due to
potential applications of $\eta$-reduction.
For example, if $\orangesubterm{s}{x}{\cst{g}\>x\>x} = \cst{h}\>\cst{a}\>(\lambda x.\>\cst{g}\>x\>x)$,
then $\orangesubtermeta{s}{x}{\cst{f}\>x} = \cst{h}\>\cst{a}\>(\lambda x.\>\cst{f}\>x) = \cst{h}\>\cst{a}\>\cst{f}$.

Demodulation, which destructively rewrites using an equality $t \eq
t'$, is available at green positions. 
In addition, a variant of demodulation rewrites in orange contexts:
\[\namedsimp{\ensuremath{\lambda}DemodExt}{t \eq t' \hypsep
  \greensubterm{C}{\orangesubterm{s}{\tuple{x}}{t\sigma}}\phantom{\negvthinspace'_\eta} \hypsep
  \phantom{\orangesubterm{s}{\tuple{x}}{t\sigma} \eq \orangesubtermeta{s}{\tuple{x}}{t'\negvthinspace\sigma}}}
 {t \eq t' \hypsep
  \greensubterm{C}{\orangesubtermeta{s}{\tuple{x}}{t'\negvthinspace\sigma}} \hypsep
  \orangesubterm{s}{\tuple{x}}{t\sigma} \eq \orangesubtermeta{s}{\tuple{x}}{t'\negvthinspace\sigma}}\]
where the term $t\sigma$ may refer to the bound variables $\tuple{x}$.
The following side conditions apply:\
\begin{enumerate}
  \item[1.] $\betanf{\orangesubterm{s}{\tuple{x}}{t\sigma}}$ is a $\lambda$-expression or a term of the form $y\>\tuple{u}_n$ with $n>0$; %
  \item[2.] $\orangesubterm{s}{\tuple{x}}{t\sigma} \succ \orangesubtermeta{s}{\tuple{x}}{t'\negvthinspace\sigma}$;%
  \hfill 3.\enskip $\greensubterm{C}{\orangesubterm{s}{\tuple{x}}{t\sigma}} \succ \orangesubterm{s}{\tuple{x}}{t\sigma} \eq \orangesubtermeta{s}{\tuple{x}}{t'\negvthinspace\sigma}$
  \hfill\hbox{}
\end{enumerate}
Condition~3 ensures that the second premise is redundant \wrt\ the conclusions and may be removed.
The double bar indicates that the conclusions collectively make the premises
redundant and can replace them.

The third conclusion, which is entailed by $t \eq t'$ and (\infname{Ext}),
could be safely omitted if the corresponding (\infname{Ext}) instance
is
smaller than the second premise. But in general, the third
conclusion is necessary for the proof, and the variant of \infname{$\lambda$DemodExt}
that omits it---let us call it \infname{$\lambda$Demod}---might not preserve
refutational completeness.

An instance of \infname{$\lambda$DemodExt}, where $\cst{g}\>z$ is
rewritten to $\cst{f}\>z\>z$ under a $\lambda$-binder, follows:
\[\namedsimp{\ensuremath{\lambda}DemodExt}{
  \cst{g}\>x \eq \cst{f}\>x\>x
  \hypsep
  \cst{k}\>\rlap{\ensuremath{(\lambda z.\> \cst{h}\>(\cst{g}\>z))\eq \cst{c}}}
    \phantom{(\lambda z.\> \cst{h}\>(\cst{f}\>z\>z)\eq \cst{c})
    \hypsep (\lambda z.\> \cst{h}\>(\cst{g}\>z)) \eq (\lambda z.\> \cst{h}\>(\cst{f}\>z\>z))}}
 {\cst{g}\>x \eq \cst{f}\>x\>x
 \hypsep \cst{k}\>(\lambda z.\> \cst{h}\>(\cst{f}\>z\>z))\eq \cst{c}
 \hypsep (\lambda z.\> \cst{h}\>(\cst{g}\>z)) \eq (\lambda z.\> \cst{h}\>(\cst{f}\>z\>z))}\]

\begin{lemmax}
\infname{\ensuremath{\lambda}DemodExt} is sound and preserves refutational
completeness of the calculus.
\end{lemmax}

\begin{proof}
  Soundness of the first conclusion is obvious.
  Soundness of the second and third conclusion follows from congruence and extensionality using the
  premises.
  Preservation of completeness is justified by redundancy.
  Specifically, we justify the deletion of the second premise
  by showing that it is redundant \wrt\ the conclusions.
  By definition, it is redundant if for every ground instance
  $\greensubterm{C}{\orangesubterm{s}{\tuple{x}}{t\sigma}}\theta \in \gnd(\greensubterm{C}{\orangesubterm{s}{\tuple{x}}{t\sigma}})$,
  its encoding $\floor{\greensubterm{C}{\orangesubterm{s}{\tuple{x}}{t\sigma}}\theta}$ is entailed by
  $\floor{\gnd(N)}$, where $N$ are the conclusions of $\infname{\ensuremath{\lambda}DemodExt}$.
  The first conclusion cannot help us prove redundancy because
  $\betanf{\orangesubterm{s}{\tuple{x}}{t\sigma}\theta}$
  might be a $\lambda$-expression and then $\floor{\orangesubterm{s}{\tuple{x}}{t\sigma}\theta}$ is a symbol that is
  unrelated to $\floor{t\sigma\theta}$.
  Instead, we use the $\theta$-instances of the last two conclusions.
  By Lemma~\ref{lem:subterm-correspondence1},
  $\floor{\greensubterm{C}{\orangesubtermeta{s}{\tuple{x}}{t'\negvthinspace\sigma}}\theta}$
  has $\floor{\orangesubtermeta{s}{\tuple{x}}{t'\negvthinspace\sigma}\theta}$ as a subterm.
  If this subterm is replaced by $\floor{\orangesubterm{s}{\tuple{x}}{t\sigma}\theta}$,
  we obtain $\floor{\greensubterm{C}{\orangesubterm{s}{\tuple{x}}{t\sigma}}\theta}$.
  Hence, the $\flooronly$-encodings of the $\theta$-instances of the last two conclusions
  entail the $\flooronly$-encoding of the $\theta$-instance of the second premise by congruence.
  Due to the side condition that the second premise is larger than the second and third conclusion,
  by stability under grounding substitutions,
  the $\theta$-instances of the last two conclusions
  must be smaller than the $\theta$-instance of the second premise.
  Thus, the second premise is redundant. \qedhere
\end{proof}

The next simplification rule can be used to prune arguments of applied variables
if the arguments can be expressed as functions of the remaining arguments.
For example, the clause
$\subterm{C}{\, y\>\cst{a}\>\cst{b}\>(\cst{f}\>\cst{b}\>\cst{a}){,}\allowbreak\;
 y\>\cst{b}\>\cst{d}\>(\cst{f}\>\cst{d}\>\cst{b})}$, in which $y$
occurs twice, can be simplified to
$\subterm{C}{\, y'\>\cst{a}\>\cst{b}{,}\; y'\>\cst{b}\>\cst{d}}$.
Here, for each occurrence of $y$, the third argument can be computed by
applying $\cst{f}$ to the second and first arguments.
The rule can also be used to remove the repeated arguments in
$y\>\cst{b}\>\cst{b} \noteq y\>\cst{a}\>\cst{a}$, the static
argument~$\cst{a}$ in $y\>\cst{a}\>\cst{c} \noteq y\>\cst{a}\>\cst{b}$, and
all four arguments in $y\>\cst{a}\>\cst{b} \noteq z\>\cst{b}\>\cst{d}$.
It is stated as
\[\namedsimp{PruneArg}{C}{C\sigma}\]
where the following conditions apply:
\begin{enumerate}
  \item[1.] $\sigma = \{y \mapsto \lambda\tuple{x}_{\!j}.\> y'\> \tuple{x}_{\!j-1}\}$;%
  \hfill 2.\enskip $y'$ is a fresh variable;%
  \hfill 3.\enskip $C\sqsupset C\sigma$;
  \hfill\hbox{}%
  \item[4.] the minimum number~$k$ of arguments passed to any
  occurrence of $y$ in the clause $C$ is at least $j$;%
  \item[5.] there exists
  a term $t$ containing no variables bound in the clause such that for all terms of the form $y\>\tuple{s}_k$
  occurring in the clause we have $s_{\!j} = t\>
  \tuple{s}_{\!j-1}\>s_{\!j+1}\ldots s_k$.
\end{enumerate}

Clauses with a static argument correspond to the case $t :=
(\lambda \bar{x}_{\!j-1}\> x_{\!j+1} \ldots x_k.\; u)$, where $u$ is the
static argument (containing no variables bound in $t$) and $j$ is its index
in $y$'s argument list.
The repeated argument case corresponds to
$t := (\lambda \bar{x}_{\!j-1} \> x_{\!j+1} \ldots x_k.\; x_i)$,
where $i$ is the index of the repeated argument's mate.
\begin{lemmax}
\infname{PruneArg} is sound and preserves refutational completeness of the calculus.
\end{lemmax}

\begin{proof}
The rule is sound because it simply applies a substitution to $C$.
It preserves completeness
because the premise $C$ is redundant \wrt\ the conclusion $C\sigma$.
This is because the sets of ground instances of $C$ and $C\sigma$ are the
same and $C \sqsupset C\sigma$. Clearly $C\sigma$ is an instance of $C$. We will show the inverse:\ that
$C$ is an instance of $C\sigma$. Let
$\rho =
\{y' \mapsto \lambda\tuple{x}_{\!j-1}\> x_{\!j+1} \ldots x_k.\;y\>\tuple{x}_{\!j-1}\allowbreak\>
(t\>\tuple{x}_{\!j-1}\> x_{\!j+1} \ldots x_k)\> x_{\!j+1} \ldots x_k\}$.
We show $C\sigma\rho = C$.
Consider an occurrence of $y$ in $C$. By the side conditions,
it will have the form $y\>\tuple{s}_k\>\tuple{u}$,
where $s_{\!j} = t\> \tuple{s}_{\!j-1}\>s_{\!j+1}\ldots s_k$. Hence,
$(y\>\tuple{s}_k)\sigma\rho
  = (y'\>\tuple{s}_{\!j-1}\>s_{\!j+1} \ldots s_k)\rho
  = y\>\tuple{s}_{\!j-1}\>(t\> \tuple{s}_{\!j-1}\>s_{\!j+1}\ldots s_k)\>s_{\!j+1} \ldots s_k
  = y\>\tuple{s}_k$.
Thus, $C\sigma\rho = C$. \qedhere
\end{proof}

We designed an algorithm that efficiently computes the subterm $u$ of the term
$t = (\lambda x_1 \ldots \, x_{\!j-1} \, x_{\!j+1} \ldots \, x_k.\allowbreak\; u)$ occurring in
the side conditions of $\infname{PruneArg}$. The algorithm is incomplete, but
our tests suggest that it discovers most cases of prunable arguments that occur
in practice.
The algorithm works by maintaining a mapping of pairs $(y, i)$ of functional
variables $y$ and indices $i$ of their arguments to a set of candidate terms
for $u$. For an occurrence $y \>
\tuple{s}_{n}$ of~$y$ and for an argument $s_{\!j}$, the algorithm
approximates this set by computing all possible ways in which subterms of
$s_{\!j}$ that are equal to any other $s_i$ can be replaced with the variable
$x_i$ corresponding to the $i$th argument of $y$. The candidate sets for
all occurrences of $y$ are then intersected. An arbitrary element of the
final intersection is returned as the term~$u$.

For example, suppose that $y\>\cst{a}\>(\cst{f} \> \cst{a})\>\cst{b}$ and
$y\>z\>(\cst{f} \>z)\>\cst{b}$ are the only occurrences of $y$ in the clause
$C$. The initial mapping is
$\{1 \mapsto \THH{,}\;
   2 \mapsto \THH{,}\;
   3 \mapsto \THH\}$.
After computing the ways in which each argument can be expressed using the
remaining ones for the first occurrence and intersecting the sets, we get
$\{1 \mapsto \{\cst{a}\}{,}\;
   2 \mapsto \{\cst{f}\>\cst{a}{,}\; \cst{f}\>x_1\}{,}\;
   3 \mapsto \{\cst{b}\}\}$,
where $x_1$ represents $y$'s first argument. Finally, after computing the
corresponding sets for the second occurrence of $y$ and intersecting them with
the previous candidate sets, we get
$\{1 \mapsto \emptyset{,}\;
   2 \mapsto \{\cst{f}\>x_1\}{,}\;
   3 \mapsto \{\cst{b}\}\}.$
The final mapping shows that we can remove the second argument,
since it can be expressed as a function of the first argument:
$t = (\lambda x_1 \, x_3.\; \cst{f}\> x_1\> x_3)$.
We can also remove the third argument, since its value is fixed:
$t = (\lambda x_1 \, x_3.\; \cst{b})$.
An example where our procedure fails is the pair of occurrences
$y\>(\lambda x.\> \cst{a})\>(\cst{f}\>\cst{a})\>\cst{c}$
and
$y\>(\lambda x.\> \cst{b})\>(\cst{f}\>\cst{b})\>\cst{d}$.
\infname{PruneArg} can be used to eliminate the second argument by taking $t :=
(\lambda x_1\>x_3.\; \cst{f}\>(x_1\>x_3))$, but our algorithm will not detect
this.

Following the literature \cite{gupta-et-al-2014,steen-benzmueller-2018}, we
provide a rule for negative extensionality:
\[\namedinference{NegExt}{C' \llor s \noteq s'}
  {C' \llor s\>(\cst{sk}\typeargs{\tuple{\alpha}}\>\tuple{y})
  \noteq s'\>(\cst{sk}\typeargs{\tuple{\alpha}}\>\tuple{y})}\]
where the following conditions apply:
\begin{enumerate}
  \item[1.] $\cst{sk}$ is a fresh Skolem symbol;%
  \hfill 2.\enskip $s \noteq s'$ is $\succsim$-eligible in the premise;%
  \hfill\hbox{}%
  \item[3.] $\tuple{\alpha}$ and $\tuple{y}$ are
  the type and term variables occurring free in the literal $s \noteq s'$.%
\end{enumerate}
Negative extensionality can be applied as an inference rule at any time 
or as a simplification rule during preprocessing of the initial problem. The
rule uses Skolem terms $\cst{sk}\>\tuple{y}$ rather than $\diff\> s\> s'$
because they tend to be more compact.

\begin{lemmax}[\infname{NegExt}'s satisfiability preservation]
Let $N\subseteq\CHH$ and let $E$ be the conclusion of a \infname{NegExt} inference from $N.$
If $N \ccup \{\text{\upshape(\infname{Ext})}\}$ is satisfiable, then $N \ccup \{\text{\upshape(\infname{Ext})}, E\}$ is satisfiable.
\end{lemmax}
\begin{proof}
Let $\III$ be a model of $N \ccup \{\text{\upshape(\infname{Ext})}\}.$
We need to construct a model of $N \ccup \{\text{\upshape(\infname{Ext})}, E\}.$
Since (\infname{Ext}) holds in~$\III$, so does its
instance
$s\>(\diff\> s\> s')\noteq s'\>(\diff\> s\> s') \llor s \eq s'$.
We extend the model $\III$ to a model $\III'$, interpreting $\cst{sk}$
such that $\III' \models \cst{sk}\typeargs{\tuple{\alpha}}\>\tuple{y} \eq \diff\> s\> s'$.
The Skolem symbol $\cst{sk}$ takes the free type and term variables of $s
\noteq s'$ as arguments, which include all the free variables of
$\diff\> s\> s'$, allowing us to extend $\III$ in this way.

By assumption, the premise $C' \llor s \noteq s'$
is true in $\III$ and hence in $\III'$. Since the above instance of
(\infname{Ext}) holds in $\III$, it also holds in $\III'$. Hence, the conclusion
$C' \llor s\>(\cst{sk}\typeargs{\tuple{\alpha}_m}\>\tuple{y}_n)
\noteq s'\>(\cst{sk}\typeargs{\tuple{\alpha}_m}\>\tuple{y}_n)$
also holds, which can be seen by resolving the
premise against the (\infname{Ext}) instance and unfolding the defining
equation of~$\cst{sk}$.
\qedhere
\end{proof}

\looseness=-1
One reason why the extensionality axiom is so prolific is that both sides of its
maximal literal, $y\>(\diff\> y\> z) \noteq
z\>(\diff\> y\> z)$, are fluid. As a pragmatic alternative to the axiom,
we introduce the ``abstracting'' rules \infname{AbsSup}, \infname{AbsERes}, and
\infname{AbsEFact} with the same premises as the core \infname{Sup},
\infname{ERes}, and \infname{EFact}, respectively. We call these
rules collectively \infname{Abs}.
Each new rule shares all the side conditions of the
corresponding core rule except that of the form $\sigma\in\csu(s,t)$. Instead, it lets
$\sigma$ be the most general unifier of $s$ and $t$'s types and adds this condition:
Let $\greensubterm{v}{s_1, \ldots, s_n} = s\sigma$ and
$\greensubterm{v}{t_1,\ldots,t_n} = t\sigma$, where $\greensubterm{v}{\phantom{.}}$ is the largest
common green context of $s\sigma$ and $t\sigma$.
If any $s_i$ is of functional type and the core rule has conclusion $E\sigma$,
the new rule has conclusion $E\sigma \llor s_1 \noteq t_1 \llor \cdots \llor s_n \noteq t_n$.
The \infname{NegExt} rule can then be applied to those literals $s_i \noteq
t_i$ whose sides have functional type. Essentially the same idea was proposed
by Bhayat and Reger as \emph{unification with abstraction} in the context of
combinatory superposition \cite[\Section~3.1]{bhayat-reger-2020-combsup}.
The approach regrettably does not fully eliminate the need for
axiom~(\infname{Ext}), as Visa Nummelin demonstrated via the following example.

\begin{examplex}
Consider the unsatisfiable clause set consisting of
$\cst{h}\>x \eq \cst{f}\>x$,
$\cst{k}\>\cst{h} \eq \cst{k}\;\cst{g}$, and
$\cst{k}\>\cst{g} \noteq \cst{k}\>\cst{f}$,
where $\cst{k}$ takes at most one argument and
$\cst{h} \succ \cst{g} \succ \cst{f}$.
The only nonredundant \infname{Abs} inference applicable is
\infname{AbsERes} on the third clause, resulting in $\cst{g} \noteq \cst{f}$.
Applying \infname{ExtNeg} further produces $\cst{g}\>\cst{sk} \noteq
\cst{f}\>\cst{sk}$. The set consisting of all five clauses is saturated.
\end{examplex}

A different approach is to instantiate the extensionality axiom with arbitrary
terms $s, s'$ of the same functional type:\strut
\[
\namedinference{ExtInst}{}
{s\>(\diff\> s\> s') \noteq
 s'\>(\diff\> s\> s') \llor
s \eq s'}
\]
We would typically choose $s, s'$ among the green subterms occurring in the
current clause set.
Intuitively, if we think in terms of eligibility, \infname{ExtInst}
demands $s\>(\diff\> s\> s') \eq s'\>(\diff\> s\> s')$ to be proved before $s \eq s'$
can be used. This can be advantageous because simplifying inferences (based on
matching) will often be able to rewrite the applied terms $s\>(\diff\> s\> s')$
and $s'\>(\diff\> s\> s')$. In contrast, \infname{Abs} assume $s \eq s'$ and
delay the proof obligation that $s\>(\diff\> s\> s') \eq s'\>(\diff\> s\> s')$.
This can create many long clauses, which will be subject to expensive
generating inferences (based on full unification).

Superposition can be generalized to orange subterms as follows:
\[\namedinference{\ensuremath{\lambda}Sup}
{D' \llor { t \eq t'} \hypsep
 C' \llor \orangesubterm{s}{\tuple{x}}{u} \doteq s'}
{(D' \llor C' \llor \orangesubtermeta{s}{\tuple{x}}{t'} \doteq s')
\sigma\rho}\]
where the substitution $\rho$ is defined as follows:
  Let $P_y = \{y\}$ for all type and term variables $y \not\in\tuple{x}$.
  For each $i$, let $P_{x_i}$ be
  recursively defined as
  the union of all $P_y$ such that $y$ occurs free in the $\lambda$-expression that
  binds $x_i$ in $\orangesubterm{s}{\tuple{x}}{u}\sigma$ or
  that occurs free in the corresponding subterm of $\smash{\orangesubtermeta{s}{\tuple{x}}{t'}\sigma}$.
  Then $\rho$ is defined as
  $\{x_i \mapsto \cst{sk}_i\typeargs{\tuple{\alpha}_i}\>\tuple{y}_i\text{ for each $i$}\}$,
  where $\tuple{y}_i$ are the term variables in $P_{x_i}$ and $\tuple{\alpha}_i$
  are the type variables in $P_{x_i}$
  and the type variables occurring in the type of the $\lambda$-expression binding $x_i$.
In addition, \infname{Sup}'s side conditions and the following conditions apply:
\begin{enumerate}
  \item[10.] $\tuple{x}$ has length $n > 0$;%
  \hfill 11.\enskip $\tuple{x}\sigma = \tuple{x}$;%
  \hfill\hbox{}%
  \item[12.] the variables $\tuple{x}$ do not occur in $y\sigma$ for all variables $y$ in $u$.
\end{enumerate}

The substitution $\rho$ introduces Skolem terms to represent bound variables that would
otherwise escape their binders.
The rule can be justified in terms of paramodulation and extensionality,
with the Skolem terms standing for $\diff$ terms. 
We can shorten the derivation of Example~\ref{ex:prod-div} by applying this rule to
the clauses $C_{\text{div}}$ and $C_{\text{conj}}$ as follows:
\[\namedinference{\ensuremath{\lambda}Sup}
{n \eq \cst{zero} \llor \cst{div}\;n\;n \eq \cst{one} \hypsep
 \cst{prod}\; K\;(\lambda k.\> \cst{div}\; (\cst{succ}\; k)\; (\cst{succ}\; k)) \noteq \cst{one}}
{\cst{succ}\; \cst{sk} \eq \cst{zero} \llor
 \cst{prod}\; K\;(\lambda k.\> \cst{one}) \noteq \cst{one}}\]
From this conclusion, $\bot$ can be derived using only \infname{Sup} and \infname{EqRes} inferences.
We thus avoid both \infname{FluidSup} and (\infname{Ext}).

\begin{lemmax}[\infname{\ensuremath{\lambda}Sup}'s satisfiability preservation]
  Let $N\subseteq\CHH$ and let $E$ be the conclusion of a \infname{\ensuremath{\lambda}Sup} inference from $N.$
  If $N \ccup \{\text{\upshape(\infname{Ext})}\}$ is satisfiable, then $N \ccup \{\text{\upshape(\infname{Ext})}, E\}$ is satisfiable.
\end{lemmax}

\begin{proof}
Let $\III$ be a model of $N \ccup \{\text{\upshape(\infname{Ext})}\}.$ We need to construct a model of $N \ccup \{\text{\upshape(\infname{Ext})}, E\}.$
For each $i$, let $v_i$ be the $\lambda$-expression binding $x_i$ in the term
$\orangesubterm{s}{\tuple{x}}{u}\sigma$ in
the rule. Let $v'_i$ be the variant of $v_i$ in which the relevant occurrence of
$u\sigma$ is replaced by $t'\sigma$. We define a substitution $\pi$ recursively
by $x_i\pi = \diff\> (v_i\pi)\> (v'_i\pi)$
for all $i$. This definition is well founded because the variables $x_{\!j}$ with
$j \geq i$ do not occur freely in $v_i$ and $v_i'$.
We extend the model $\III$ to a model $\III'$, interpreting $\cst{sk}_i$
such that $\III' \models \cst{sk}_i\typeargs{\tuple{\alpha}_i}\>\tuple{y}_i \eq \diff\> (v_i\pi)\allowbreak\> (v'_i\pi)$
for each $i$. Since the free type and term variables of any $x_i\pi$ are
necessarily contained in $P_{x_i}$, the arguments of $\cst{sk}_i$ include the
free %
variables of $\diff\> (v_i\pi)\> (v'_i\pi)$, allowing us to extend $\III$ in this way.

By assumption, the premises of the \infname{\ensuremath{\lambda}Sup} inference
are true in $\III$ and hence in $\III'$.
We need to show that
the conclusion $(D' \llor C' \llor \orangesubtermeta{s}{\tuple{x}}{t'} \doteq s')\sigma\rho$
is also true in $\III'$.
Let $\xi$ be a valuation.
If $\III',\xi \models (D' \llor C')\sigma\rho$, we are done.
So we assume that $D'\sigma\rho$ and $C'\sigma\rho$ are false in $\III'$ under $\xi$.
In the following, we omit `$\III',\xi\models$', but all equations ($\eq$) are meant to be true in $\III'$ under $\xi$.
Assuming $D'\sigma\rho$ and $C'\sigma\rho$ are false,
we will show inductively that  $v_i\pi \eq v'_i\pi$ for all $i = k, \dots, 1$.
By this assumption, the premises imply that $t\sigma\rho \eq t'\sigma\rho$ and
$\orangesubterm{s}{\tuple{x}}{u}\sigma\rho \doteq s'\sigma\rho$.
Due to the way we constructed $\III'$, we have $w\pi \eq w\rho$ for any term $w$.
Hence, we have $t\sigma\pi \eq t'\sigma\pi$.
The terms $v_k\pi\>(\diff\> (v_k\pi)\> (v'_k\pi))$ and
$v_k'\pi\>(\diff\> (v_k\pi)\> (v'_k\pi))$ are the respective
result of applying $\pi$ to the body of the $\lambda$-expressions $v_k$ and $v'_k$.
Therefore, by congruence, $t\sigma\pi \eq t'\sigma\pi$ and $t\sigma = u\sigma$ imply that
$v_k\pi\>(\diff\> (v_k\pi)\> (v'_k\pi)) \eq v'_k\pi\>(\diff\> (v_k\pi)\> (v'_k\pi)).$
The extensionality axiom then implies $v_k\pi \eq v'_k\pi$.

It follows directly from the definition of $\pi$ that for all $i$,
$v_i\pi\>(\diff\> (v_i\pi)\> (v'_i\pi)) = \yellowsubterm{s_i}{v_{i+1}\pi}$ and
$v'_i\pi\>(\diff\> (v_i\pi)\> (v'_i\pi)) = \yellowsubterm{s_i}{v'_{i+1}\pi}$
for some context $\yellowsubterm{s_i}{\phantom{\cdot}}$.
The subterms $v_{i+1}\pi$ of $\yellowsubterm{s_i}{v_{i+1}\pi}$ and $v_{i+1}'\pi$ of $\yellowsubterm{s_i}{v_{i+1}'\pi}$ may be below
applied variables but not below $\lambda$s.
Since substitutions avoid capture, in $v_i$ and $v_i'$, $\pi$ only substitutes $x_{\!j}$ with $j<i$, but
in $v_{i+1}$ and $v_{i+1}'$, it substitutes all
$x_{\!j}$ with $j\leq i$.
By an induction using these equations, congruence, and the extensionality axiom,
we can derive from $v_k\pi \eq v'_k\pi$ that
$v_1\pi \eq v'_1\pi.$
Since $\III' \models w\pi \eq w\rho$ for any term $w$, we have $v_1\rho \eq v'_1\rho.$
By congruence, it follows that $\orangesubterm{s}{\tuple{x}}{u}\sigma\rho \eq \orangesubtermeta{s}{\tuple{x}}{t'}\sigma\rho.$
With $\orangesubterm{s}{\tuple{x}}{u}\sigma\rho \doteq s'\sigma\rho,$ it follows that
$(\orangesubtermeta{s}{\tuple{x}}{t'} \doteq s')\sigma\rho.$
Hence, the conclusion of the \infname{\ensuremath{\lambda}Sup} inference is true in $\III'$.
\qedhere
\end{proof}

The next rule, \emph{duplicating flex subterm superposition}, is a lightweight
alternative to \infname{FluidSup}:
\[
\namedinference{DupSup}{D'\lor t\eq t' \quad
C'\lor \greensubterm{s}{\, y\>\tuple{u}_n} \doteq s'}
{(D'\lor C'\lor \greensubterm{s}{z\>\tuple{u}_n\>t'}\doteq s')\rho\sigma}
\]
where $n > 0$, $\rho = \{y \mapsto
\lambda\tuple{x}_n.\>z\>\tuple{x}_n\>(w\>\tuple{x}_n)\}$, and $\sigma \in \csu(t{,}\>
w\>(\tuple{u}_n\rho))$ for fresh variables $w, z$. The order and eligibility
restrictions are as for \infname{Sup}. The rule can be understood as the
composition of an inference that applies the substitution~$\rho$ and of a
paramodulation inference into the subterm
$w\>(\tuple{u}_n\rho)$ of $\greensubterm{s}{z\>(\tuple{u}_n\rho)\>(w\>(\tuple{u}_n\rho))}$.
\infname{DupSup} is general enough to replace \infname{FluidSup} in Examples
\ref{ex:wsup-1}~and~\ref{ex:wsup-2} but not in Example~\ref{ex:wsup-3}.
On the other hand, \infname{FluidSup}'s unification problem is usually a
flex--flex pair, whereas \infname{DupSup} yields a less explosive flex--rigid
pair unless $t$ is variable-headed.

\begin{notyet}
Let us call a $\lambda$-term $t$ \emph{second-order-like} if $t =
\cst{f}\typeargs{\tuple{\tau}}\> \tuple{s}$ and each $s_i$ is
second-order-like; or if $s = x\> \tuple{s}$, each
$s_i$'s type is not functional or a type variable, and each $s_i$ is
second-order-like; or if $\betanf{t}$ is a $\lambda$-expression.
The \infname{DupSup} rule, in conjunction with an extended \infname{Sup} rule
that rewrites into the yellow subterms, constitutes a complete alternative to
the explosive \infname{FluidSup} rule for second-order-like fluid subterms of
the form $y\>\tuple{u}_n$.

\begin{theoremx}[Static refutational completeness with \infname{DupSup}]
  Let $\HInf'$ be the inference system obtained by removing
  \infname{FluidSup} inferences into second-order-like fluid subterms of
  the form $y\>\tuple{u}_n$, where $n > 0$, by adding \infname{DupSup},
  and by extending \infname{Sup} so that it rewrites into yellow subterms. Then
  $\HInf'$ is statically refutationally complete.
  \label{thm:static-refutational-completeness-with-dupsup}
\end{theoremx}
\begin{proof}
  The proof is essentially as for
  Theorem~\ref{thm:static-refutational-completeness}. The only necessary change is
  in case~2 of the proof of Lemma~\ref{lem:lifting2}, subcase~(b). We reuse the
  notations from that proof.

  If $u$ is not of the form $y\>\tuple{u}_n$ or is not second-order-like,
  $\HInf'$ contains an applicable \infname{FluidSup} inference that lifts the
  ground inference~$\iota$.

  Otherwise, let $p = p'.p''$.
\qedhere
\end{proof}
\end{notyet}

The last rule, \emph{flex subterm superposition}, is an even more lightweight
alternative to \infname{Fluid\-Sup}:
\[
\namedinference{FlexSup}{D'\lor t\eq t' \quad
C'\lor \greensubterm{s}{\, y\>\tuple{u}_n} \doteq s'}
{(D'\lor C'\lor \greensubterm{s}{t'}\doteq s')\sigma}
\]
where $n > 0$ and $\sigma \in \csu(t{,}\> y\>\tuple{u}_n)$. The order and
eligibility restrictions are as for \infname{Sup}.

\section{Implementation}
\label{sec:implementation}

Zipperposition \cite{cruanes-2015,cruanes-2017} is an open source
superposition prover written in OCaml.%
\footnote{\url{https://github.com/sneeuwballen/zipperposition}} Originally
designed for polymorphic first-order logic (TF1
\cite{blanchette-paskevich-2013}), it was later extended with an incomplete
higher-order mode based on pattern unification \cite{miller-1991}. Bentkamp et
al.~\cite{bentkamp-et-al-2018} extended it further with a complete
$\lambda$-free clausal higher-order mode. We have now
implemented a clausal higher-order mode based on our calculus.
We use the order $\lsucc$ (\Section~\ref{ssec:a-derived-term-order})
derived from the Knuth--Bendix order \cite{knuth-bendix-1970} and the lexicographic path order \cite{kamin-levy-1980-cannotfind}.
We currently use the corresponding nonstrict order $\lsucceq$ as~$\succsim$.

Except for \infname{FluidSup}, the core calculus rules already existed in
Zipperposition in a similar form. To improve efficiency, we extended
the prover to use a higher-order generalization \cite{vukmirovic-et-al-2020-unif} of fingerprint indices
\cite{schulz-fingerprint-2012} to find inference
partners for all new binary inference rules.
To speed up the computation of the \infname{Sup} conditions,
we omit the condition $C\sigma \not\precsim D\sigma$ in the implementation,
at the cost of performing some additional inferences.
Among the optional rules, we implemented \infname{$\lambda$Demod},
\infname{PruneArg}, \infname{NegExt}, \infname{Abs}, \infname{ExtInst},
\infname{$\lambda$Sup}, \infname{DupSup}, and \infname{FlexSup}.
For \infname{$\lambda$Demod} and \infname{$\lambda$Sup}, demodulation,
subsumption, and other standard simplification rules (as implemented in
E~\cite{schulz-et-al-2019}), we use pattern unification.
For generating inference rules
that require enumerations of complete sets of unifiers, we use the complete
procedure of Vukmirovi\'c et al.\ \cite{vukmirovic-et-al-2020-unif}.
It has better termination behavior, produces fewer redundant unifiers,
and can be implemented more efficiently than procedures such as Jensen
and Pietrzykowski's \cite{jensen-pietrzykowski-1976} and Snyder and Gallier's
\cite{snyder-gallier-1989}.
The set of fluid terms is overapproximated in the implementation by 
the set of terms that are either nonground $\lambda$-expressions
or terms of the form $y\>\tuple{u}_n$ with $n>0$.
To efficiently retrieve candidates for \infname{Abs} inferences without slowing down
superposition term indexing structures, we implemented dedicated indexing for clauses
that are eligible for \infname{Abs} inferences \cite[\Section~3.3]{vukmirovic-nummelin-2020-boolean}.

Zipperposition implements a DISCOUNT-style given clause
procedure \cite{avenhaus-et-al-1995}. The proof
state is represented by a set $A$ of active clauses
and a set $P$ of passive clauses. To interleave nonterminating unification with other
computation, we added a set $T$ containing possibly infinite sequences of
scheduled inferences.
These sequences are stored as finite instructions of how to compute the inferences.
Initially, all clauses are in $P$. At each iteration of the
main loop, the prover heuristically selects a \emph{given clause} $C$ from
$P$. If $P$ is empty, sequences from $T$ are evaluated to generate more clauses into $P$;
if no clause can be produced in this way, $A$ is saturated and the prover stops.
Assuming a given clause $C$ could be selected, it is first simplified using $A$.
Clauses in $A$ are then simplified \wrt\ $C$, and any simplified clause is moved to $P$.
Then $C$ is added to $A$ and all sequences representing nonredundant inferences between $C$ and $A$ are added to $T$.
This maintains the invariant that all nonredundant inferences
between clauses in $A$ have been scheduled or performed.
Then some of the scheduled inferences in $T$ are performed and the
conclusions are put into $P$.

We can view the above loop as an instance of the abstract Zipperposition loop prover
\textsf{ZL} of Waldmann et
al.~\cite[Example~34]{waldmann-et-al-2020-saturation}. %
Their Theorem~32 %
allows us to
obtain dynamic completeness for this prover architecture from
our static completeness result (Theorem~54). %
This requires that the sequences in $T$ are visited fairly, that clauses in $P$
are chosen fairly, and that simplification terminates, all of which are
guaranteed by our implementation.

The unification procedure we use returns a sequence of either singleton sets
containing the unifier or an empty set signaling that a unifier is still not
found. Empty sets are returned to give back control to the caller of unification
procedure and avoid getting stuck on nonterminating problems. These sequences
of unifier subsingletons are converted into sequences containing subsingletons of
clauses representing inference conclusions.

\section{Evaluation}
\label{sec:evaluation}

\newcommand\HEAD[1]{\hbox to \wd\mybox{\hfill\hbox{#1}\hfill}}
\newcommand\Z{\phantom{0}}
\newcommand\MIDLINE{\\[.25ex]\hline\rule{0pt}{3ex}}

We evaluated our prototype implementation of the calculus in Zipperposition
with other higher-order provers and with Zipperposition's modes for less
expressive logics. All of the experiments were
performed on StarExec nodes equipped with Intel Xeon E5-2609$\>$0 CPUs clocked
at 2.40\,GHz. Following CASC 2019,\footnote{\url{http://tptp.cs.miami.edu/CASC/27/}}
we use 180\,s as the CPU time limit.

We used both standard TPTP benchmarks~\cite{sutcliffe-2017-tptp} and
Sledgehammer-generated benchmarks \cite{meng-paulson-2008-trans}. From the
TPTP, version~7.2.0, we used 1000 randomly selected first-order (FO) problems in
CNF, FOF, or TFF syntax without arithmetic and all 499 mono\-morphic higher-order
theorems in TH0 syntax without interpreted Booleans and arithmetic. We
partitioned the TH0 problems into those containing no $\lambda$-expressions
(\relax{TH0$\lambda$f}, 452~problems) and those containing
$\lambda$-expressions (\relax{TH0$\lambda$}, 47~problems).
The Sledgehammer benchmarks, corresponding to Isabelle's Judgment Day suite
\cite{boehme-nipkow-2010}, were regenerated to target clausal higher-order
logic. They comprise 2506~problems, divided in two groups:
\relax{SH-$\lambda$} preserves
$\lambda$-expressions, whereas \relax{SH-ll} encodes them as $\lambda$-lifted
supercombinators \cite{meng-paulson-2008-trans} to make the problems accessible
to $\lambda$-free clausal higher-order provers. Each group of problems is
generated from 256 Isabelle facts (definitions and lemmas). Our results are
publicly available.%
\footnote{\url{https://doi.org/10.5281/zenodo.4032969}}

\ourparagraph{Evaluation of Extensions}
To assess the usefulness of the extensions described in
\Section~\ref{sec:extensions}, we fixed a \emph{base} configuration
of Zipperposition parameters. For each extension, we then changed the
corresponding parameters and observed the effect on the success rate. The base
configuration uses the complete variant of the unification procedure
of Vukmirovi\'c et al.\ \cite{vukmirovic-et-al-2020-unif}.
It also includes the optional rules \infname{NegExt} and
\infname{PruneArg}, substitutes \infname{FlexSup} for the highly explosive
\infname{FluidSup},
and excludes axiom~(\infname{Ext}). The base configuration is not refutationally complete.

\newcommand{\mainevalnumbers}{
  \newbox\mybox
  \setbox\mybox=\hbox{\small SH256-$\lambda$}
	\def\arraystretch{1.1}%
  \relax{\begin{tabular}{@{}l@{\hskip 0.5em}c@{\hskip 1em}c@{\hskip 1em}c@{\hskip 1em}c@{\hskip 1em}c@{\hskip 1em}c@{\hskip 1em}c@{}}
    \strut                   &  \HEAD{FO}  & \HEAD{TH0$\lambda$f}  & \HEAD{TH0$\lambda$} & \HEAD{SH-ll} & \HEAD{SH-$\lambda$}
    \MIDLINE
    CVC4                     & 539         &  424                        &  31                       & 696             & 650            \\
    Ehoh                     & 681         &  418                        &  --                       & 691             & --              \\
    Leo-III-uncoop           & 198         &  389                        &  42                       & 226             & 234            \\
    Leo-III-coop             & 582         &  {\bf438}                   &  43                       & 683             & 674         \\
    Satallax-uncoop          & --          &  398                        &  43                       & 489             & 507 \\
    Satallax-coop            & --          &  432                        &  43                       & 602             & 616            \\
    Vampire                  & {\bf 729}   &  432                        &  42                       & \textbf{718}    & 707            \\[\jot]
    FOZip                    & 399         &  --                         &  --                       &  --             & --                \\
    @+FOZip                  & 363         &  400                        &  --                       & 478             & --                \\
    $\lambda$freeZip         & 395         &  398                        &  --                       & {538}           & --            \\
    $\lambda$Zip-base        & 388         &  {408}                      &  39                       & 420             & 436        \\
    $\lambda$Zip-pragmatic   & 396         &  411                        &  33                       & 496             & 503        \\
    $\lambda$Zip-full        & 177         &  339                        &  34                       & 353             & 361        \\
    Zip-uncoop               & 514         &  426                        &  {\bf 46}                 & 661             & 677        \\
    Zip-coop                 & 625         &  434                        &  {\bf 46}                 & 710             & \textbf{717}
    \end{tabular}}
	\caption{Number of problems proved by the different provers}
  \label{fig:res}
}

\begin{figure}
  \def\arraystretch{1.1}%
  \relax{\begin{tabular}{@{}l@{\hskip 2em}c@{\hskip 2em}c@{\hskip 2em}c@{\hskip 2em}c}
    \strut                   & $-$NE,$-$PA  & $-$NE     & $-$PA  & Base
    \MIDLINE
    TH0                      &  446 (0)      &  446 (0)   &  447 (0)   & 447 (0) \\
    SH-$\lambda$             &  431 (0)      &  433 (0)   &  433 (0)   & 436 (1)
    \end{tabular}}
  \caption{Number of problems proved without rules included in the base configuration}
  \label{fig:neg-ext-prune-arg}

  \vspace*{\floatsep}

  \def\arraystretch{1.1}%
  \relax{\begin{tabular}{@{}l@{\hskip 2em}c@{\hskip 2em}c@{\hskip 2em}c@{\hskip 2em}c@{\hskip 2em}c@{\hskip 2em}c@{\hskip 2em}c@{\hskip 2em}c}
    \strut                  & Base          & $+\lambda$\infname{D}     & $+\lambda$S0  & $+\lambda$S1 & $+\lambda$S2 & $+\lambda$S4 & $+\lambda$S8 & $+\lambda$S1024
    \MIDLINE
    TH0                     & 447 (0)        &  448 (0)  & 449 (0) & 449 (0) & 449 (0)  & 449 (0)  & 449 (0)  & 449 (0) \\
    SH-$\lambda$            & 436 (1)        &  435 (4)  & 430 (1) & 429 (0) & 429 (0)  & 429 (0)  & 429 (0)  & 429 (0)
    \end{tabular}}
  \caption{Number of problems proved using rules that perform rewriting under $\lambda$-binders}
  \label{fig:lambda-rules}

  \vspace*{\floatsep}

  \def\arraystretch{1.1}%
  \relax{\begin{tabular}{@{}l@{\hskip 2em}c@{\hskip 2em}c@{\hskip 2em}c@{\hskip 2em}c}
    \strut                  & Base                   & $+$\infname{Abs}       & $+$\infname{ExtInst}      & $+$(\infname{Ext})
    \MIDLINE
    TH0                     & 447 (0)\phantom{0}     &  450 (1)\phantom{0}    &  450 (1)                  &  376 (0)               \\
    SH-$\lambda$            & 436 (11)               &  430 (11)              &  402 (1)                  &  365 (2)
    \end{tabular}}
    \caption{Number of problems proved using rules that perform extensionality reasoning}
  \label{fig:ext-rules}

  \vspace*{\floatsep}

  \def\arraystretch{1.1}%
  \relax{\begin{tabular}{@{}l@{\hskip 2em}c@{\hskip 2em}c@{\hskip 2em}c@{\hskip 2em}c}
    \strut                  & $-$\infname{FlexSup} &  Base        & $-$\infname{FlexSup},$+$\infname{DupSup}   & $-$\infname{FlexSup},$+$\infname{FluidSup}
    \MIDLINE
    TH0                     &  446 (0)\phantom{0}  & 447 (0)  &  448 (1)    &  447 (0)                   \\
    SH-$\lambda$            &  469 (10) & 436 (4)  &  451 (3)    &  461 (7)
    \end{tabular}}
  \caption{Number of problems proved with rules that perform superposition into fluid terms}
  \label{fig:fluid-rules}

\end{figure}

The rules \infname{NegExt} (NE) and \infname{PruneArg} (PA) were added to
the base configuration because our informal experiments showed that they
usually help. Fig.~\ref{fig:neg-ext-prune-arg} confirms this, although
the effect is small. In all tables, $+R$ denotes the inclusion of a
rule~$R$ not present in the base, and $-R$ denotes the exclusion of a
rule~$R$ present in the base. Numbers given in parentheses denote the number of
problems that are solved only by the given configuration and no other
configuration in the same table.

The rules \infname{$\lambda$Demod} ($\lambda$D) and \infname{$\lambda$Sup} extend the
calculus to perform some rewriting under \hbox{$\lambda$-binders}. While
experimenting with the calculus, we noticed that, in some configurations,
\infname{$\lambda$Sup} performs better when the number of fresh Skolem symbols
it introduces overall is bounded by some parameter $n$. As
Fig.~\ref{fig:lambda-rules} shows, inclusion of these rules has different
effect on the two benchmark sets. Different choices of $n$ for
\infname{$\lambda$Sup} (denoted by $\lambda$S$n$) do not seem to influence
the success rate much.

  The evaluation of the \infname{Abs} and \infname{ExtInst} rules and
  axiom~(\infname{Ext}), presented in Fig.~\ref{fig:ext-rules}, confirms our intuition
  that including the extensionality axiom is severely detrimental to
  performance.
  The $+$(\infname{Ext}) configuration
  solved two unique problems on SH-$\lambda$ benchmarks,
  but the success of the $+$(\infname{Ext}) configuration
  on these problems appears to be due to a coincidental
  influence of the axiom on heuristics---the axiom is not referenced in the
  generated proofs.

  The \infname{FlexSup} rule included in the base configuration
  did not perform as well as we expected.
  Even the \infname{FluidSup} and \infname{DupSup} rules
  outperformed \infname{FlexSup},
  as shown in Fig.~\ref{fig:fluid-rules}. This effect is especially visible on
  SH-$\lambda$ benchmarks. On TPTP, the differences are negligible.

Most of the extensions had a stronger effect on SH-$\lambda$ than on TH0.
A possible explanation is that the Boolean-free TH0 benchmark subset consists
mostly of problems that are simple to solve using most prover parameters.
On the other hand, SH-$\lambda$ benchmarks are of varying difficulty and
can thus benefit more from changing prover parameters.

\ourparagraph{Main Evaluation}
We selected all contenders in the THF division of CASC 2019 as representatives of
the state of the art:\ CVC4 1.8 prerelease \cite{barbosa-et-al-2019}, Leo-III 1.4
\cite{steen-benzmueller-2018}, Satallax 3.4 \cite{brown-2012-ijcar}, and Vampire
4.4 \cite{bhayat-reger-2019-restricted}. We also included Ehoh
\cite{vukmirovic-et-al-2019}, the $\lambda$-free clausal higher-order mode of
E~2.4. Leo-III and Satallax are cooperative higher-order provers that can be set
up to regularly invoke first-order provers as terminal proof procedures. To
assess the performance of their core calculi, we evaluated them with first-order
backends disabled. We denote these ``uncooperative'' configurations by
Leo-III-uncoop and Satallax-uncoop respectively, as opposed to the standard
versions Leo-III-coop and Satallax-coop.

\begin{figure}[b]
  \mainevalnumbers
\end{figure}

To evaluate the overhead our calculus incurs on first-order or $\lambda$-free
higher-order problems, we ran Zipperposition in first-order (FOZip) and
$\lambda$-free ($\lambda$freeZip) modes, as well as in a mode that encodes
curried applications using a distinguished binary symbol $\cst{@}$
before using first-order Zipperposition (@+FOZip). We
evaluated the implementation of our calculus in Zipperposition ($\lambda$Zip)
in three configurations:\ base, pragmatic, and full.
Pragmatic builds on base by disabling \infname{FlexSup} and replacing complete
unification with the pragmatic variant procedure pv$^2_{1121}$ of
Vukmirovi\'c et al. Full is a refutationally complete
extension of base that substitutes \infname{Fluid\-Sup} for \infname{FlexSup}
and includes axiom~(\infname{Ext}).
Finally, we evaluated Zipperposition in a portfolio mode that runs the prover in
various configurations (Zip-uncoop). We also evaluated a cooperative version of
the portfolio which, in some configurations,
after a predefined time invokes Ehoh as backend
on higher-order problems (Zip-coop). In this version, Zipperposition encodes
heuristically selected clauses from the current proof state to lambda-free
higher-order logic supported by Ehoh \cite{vukmirovic-et-al-2019}. On
first-order problems, we ran Ehoh, Vampire, and Zip-uncoop using the provers'
respective first-order modes.

A summary of these experiments is presented in \figurename~\ref{fig:res}.
In the pragmatic configuration, our calculus
outperformed $\lambda$freeZip on TH0$\lambda$f problems and incurred less than 1\%
overhead compared with FOZip, but fell behind $\lambda$freeZip on SH-ll problems.
The full configuration suffers greatly from the explosive extensionality axiom
and \infname{FluidSup} rule.

Except on TH0$\lambda$ problems, both base and pragmatic configurations outperformed Leo-III-uncoop, which runs
a fixed configuration, by a substantial margin. Zip-uncoop outperformed Satallax-uncoop, which
uses a portfolio. Our most competitive configuration, Zip-coop, emerges as the
winner on both problem sets containing \hbox{$\lambda$-expressions}.

On higher-order TPTP benchmarks
this configuration \OK{does not} solve any problems that no other (cooperative) higher-order prover solves.
By contrast, on SH-ll benchmarks Zip-coop solves \OK{21} problems no other higher-order prover solves,
and on SH-$\lambda$ benchmarks, it uniquely solves \OK{27} problems.

\section{Discussion and Related Work}
\label{sec:discussion-and-related-work}

Bentkamp et al.~\cite{bentkamp-et-al-2018} 
introduced four calculi for
$\lambda$-free clausal higher-order logic organized along two axes:\
\emph{intensional} versus \emph{extensional}, and \emph{nonpurifying} versus
\emph{purifying}. The purifying calculi flatten the clauses containing applied
variables, thereby eliminating the need for superposition into variables.
As we extended their work to support $\lambda$-expressions, we found the purification
approach problematic and gave it up because it needs $x$ to be smaller
than $x\;t$, which is impossible to achieve with a term order on
$\beta\eta$-equivalence classes.
We also quickly gave up our attempt at supporting intensional higher-order logic.
Extensionality is the norm for higher-order unification
\cite{dowek-2001} and is mandated by the TPTP THF format
\cite{sutcliffe-et-al-2009} and in proof assistants such as HOL4, HOL Light,
Isabelle/HOL, Lean, Nuprl, and PVS.

Bentkamp et al.\ viewed their approach as ``a stepping stone towards full
higher-order logic.'' It already included a notion analogous to green subterms
and an \infname{ArgCong} rule, which help cope with the complications
occasioned by $\beta$-reduction.

Our Boolean-free $\lambda$-superposition calculus joins the family of proof systems
for higher-order logic. It is related to Andrews's higher-order resolution
\cite{andrews-1971}, Huet's constrained resolution \cite{huet-1973}, Jensen
and Pietrzykowski's $\omega$-resolution \cite{jensen-pietrzykowski-1976},
Snyder's higher-order $E$-resolution \cite{snyder-1990}, Benz\-m\"uller and
Kohlhase's extensional higher-order resolution
\cite{benzmueller-kohlhase-1998}, Benzm\"uller's higher-order
unordered paramodulation and RUE resolution \cite{benzmueller-1999}, and
Bhayat and Reger's combinatory superposition \cite{bhayat-reger-2020-combsup}.
A noteworthy variant of higher-order unordered paramodulation is Steen and
Benzm\"uller's higher-order ordered paramodulation
\cite{steen-benzmueller-2018}, whose order restrictions undermine refutational
completeness but yield better empirical results.
Other approaches are based on analytic tableaux
\cite{robinson-1969,kohlhase-1995,konrad-1998,backes-brown-2011}, connections
\cite{andrews-1989}, sequents \cite{lindblad-2014}, and satisfiability modulo
theories (SMT) \cite{barbosa-et-al-2019}. Andrews \cite{andrews-2001} and
Benzm\"uller and Miller \cite{benzmueller-miller-2014} provide excellent
surveys of higher-order automation.

Combinatory superposition 
was developed shortly after $\lambda$-superposition
and is closely related. It is modeled on the intensional nonpurifying calculus
by Bentkamp et al.\ and targets extensional polymorphic clausal
higher-order logic. Both combinatory and $\lambda$-superposition gracefully
generalize the highly successful first-order superposition rules without
sacrificing refutational completeness, and both are equipped with a redundancy
criterion, which earlier refutationally complete higher-order calculi lack.
In particular, \infname{PruneArg} is a versatile simplification rule that could
be useful in other provers.
Combinatory superposition's distinguishing feature is that it uses $\cst{SKBCI}$
combinators to represent $\lambda$-expressions. Combinators can be implemented more
easily starting from a first-order prover; $\beta$-reduction amounts to
demodulation.
However, according to its developers
\cite{bhayat-reger-2020-combsup}, ``Narrowing terms with combinator axioms is
still explosive and results in redundant clauses. It is also never likely to be
competitive with higher-order unification in finding complex unifiers.''
Among the drawbacks of $\lambda$-superposition are the need to solve flex--flex
pairs eagerly and the explosion caused by the extensionality axiom. We believe
that this is a reasonable trade-off, especially for large problems with a
substantial first-order component.

Our prototype Zipperposition joins the league of automatic theorem provers for
higher-order logic. We list some of its rivals. TPS
\cite{andrews-et-al-1996} is based on the connection method and expansion
proofs. LEO \cite{benzmueller-kohlhase-1998} and \textsc{Leo}-II
\cite{benzmuller-2015-leo2} implement variants of RUE resolution.
Leo-III \cite{steen-benzmueller-2018} is based on higher-order paramodulation.
Satallax \cite{brown-2012-ijcar} implements a higher-order tableau calculus
guided by a SAT solver.
\textsc{Leo}-II, Leo-III, and Satallax integrate first-order provers as
terminal procedures. AgsyHOL \cite{lindblad-2014} is based on a focused sequent
calculus guided by narrowing.
The SMT solvers CVC4 and veriT have recently
been extended to higher-order logic
\cite{barbosa-et-al-2019}.
Vampire now implements both combinatory superposition and a version of standard
superposition with first-order unification replaced by restricted combinatory
unification \cite{bhayat-reger-2019-restricted}.

Half a century ago, Robinson \cite{robinson-1970} proposed to reduce
higher-order logic to first-order logic via a translation. ``Hammer'' tools such as
Sledgehammer \cite{paulson-blanchette-2010}, Miz$\mathbb{AR}$
\cite{urban-et-al-2013}, HOLyHammer \cite{kaliszyk-urban-2015}, and CoqHammer
\cite{czajka-kaliszyk-2018} have since popularized this approach in proof
assistants. The translation must eliminate the $\lambda$-expressions,
typically using $\cst{SKBCI}$ combinators or $\lambda$-lifting
\cite{meng-paulson-2008-trans}, and encode typing information
\cite{blanchette-et-al-2016-types}.
\section{Conclusion}
\label{sec:conclusion}

We presented the Boolean-free $\lambda$-superposition calculus, which targets a
clausal fragment of extensional polymorphic higher-order logic.
With the exception of a functional extensionality axiom, it gracefully
generalizes standard superposition. Our prototype prover Zipperposition shows
promising results on TPTP and Isabelle benchmarks. 
In future work, we plan to
pursue five main avenues of investigation.

We first plan to \emph{extend the calculus to support Booleans and Hilbert
choice.} Booleans are notoriously explosive. We want to experiment with both
axiomatizations and native support in the calculus. Native support would
likely take the form of a primitive substitution rule that enumerates
predicate instantiations \cite{andrews-1989}, delayed clausification rules
\cite{ganzinger-stuber-2005}, and rules for reasoning about Hilbert choice.

We want to investigate techniques to \emph{curb the explosion caused by
functional extensionality.} The extensionality axiom reintroduces the search
space explosion that the calculus's order restrictions aim at avoiding.
Maybe
we can replace it by more restricted inference rules without compromising
refutational completeness.

We will also look into approaches to \emph{curb the explosion caused by
higher-order unification.} Our calculus suffers from the need to solve
flex--flex pairs. Existing procedures
\cite{jensen-pietrzykowski-1976,snyder-gallier-1989,vukmirovic-et-al-2019}
enumerate redundant unifiers. This can probably be avoided to some extent. It
could also be useful to investigate unification procedures that would delay
imitation/projection choices via special schematic variables, inspired by
Libal's representation of regular unifiers \cite{libal-2015}.

We clearly need to \emph{fine-tune and develop heuristics.}
We expect heuristics to be a fruitful area for future research in
higher-order reasoning. Proof assistants are an inexhaustible source of
easy-looking benchmarks that are beyond the power of today's provers. Whereas
``hard higher-order'' may remain forever out of reach, we believe that there is
a substantial ``easy higher-order'' fragment that awaits automation.

Finally, we plan to \emph{implement the calculus in a state-of-the-art prover.}
A suitable basis for an optimized implementation of the calculus would be
Ehoh, the $\lambda$-free clausal higher-order version of E developed by
Vukmirovi\'c, Blanchette, Cruanes, and Schulz\ \cite{vukmirovic-et-al-2019}.

\def\ackname{Acknowledgment} %
\begin{acknowledgements}
        Simon Cruanes patiently explained Zipperposition's internals and allowed us to
        continue the development of his prover.
        Christoph Benzm\"uller and Alexander Steen shared insights and examples with
        us, guiding us through the literature and clarifying how the Leos work.
        Maria Paola Bonacina and Nicolas Peltier gave us some ideas on how to treat
        the extensionality axiom as a theory axiom, ideas we have yet to explore.
        Mathias Fleury helped us set up regression tests for Zipperposition.
        Ahmed Bhayat, Tomer Libal, and Enrico Tassi shared their insights on
        higher-order unification.
        Andrei Popescu and Dmitriy Traytel explained the terminology surrounding the
        $\lambda$-calculus.
        Haniel Barbosa, Daniel El Ouraoui, Pascal Fontaine, Visa Nummelin, and Hans-J\"org Schurr
        were involved in many stimulating discussions.
        Christoph Weidenbach made this collaboration possible.
        Ahmed Bhayat, Wan Fokkink, Mark Summerfield, and the anonymous reviewers suggested several
        textual improvements.
        The maintainers of StarExec let us use their service for the evaluation.
        We thank them all.

        Bentkamp, Blanchette, and Vukmirovi\'c's research has received funding from
        the European Research Council (ERC) under the European Union's Horizon 2020
        research and innovation program (grant agreement No.\ 713999, Matryoshka).
        Bentkamp and Blanchette also benefited from the Netherlands Organization for
        Scientific Research (NWO) Incidental Financial Support scheme.
        Blanchette has received funding from the NWO under the Vidi program (project
        No.\ 016.Vidi.189.037, Lean Forward).
\end{acknowledgements}

\bibliographystyle{spmpsci}
\bibliography{ms}

\end{document}